%% file: Spinors_revision.tex
\documentclass{article}
\pdfoutput=1
\usepackage{wrapfig}
\usepackage{amssymb}
\usepackage{amsbsy}
\usepackage[usenames]{color}        
\usepackage{graphicx}

\usepackage{verbatim}
\def\K{\mathcal K}

\def\P{{\boldsymbol  \Pi}}

\def \fK{{\mathfrak K}}

\def\wt{\widetilde}

\def\X{{\mathbf X}}
\def\Y{{\mathbf Y}}
\def\wh{\widehat}

\def\ds{\displaystyle}

\def\res{\mathop{\mathrm{res}}\limits_}

\makeatletter
\@addtoreset{equation}{section}
\makeatother

\def\le{\left}
\def\bc{\begin{corollary}}
\def\ec{\end{corollary}}
\def\&{&{\hskip -20pt}}

\def \s{\mathfrak s}

\def\br{\begin{remark}\rm\small}
\def\1{{\bf 1}}
\def\er{\end{remark}}
\def\bt{\begin{theorem}}
\def\et{\end{theorem}}

\def\bx{\begin{example}}
\def\ex{\end{example}}
\def\bd{\begin{definition}}
\def\ed{\end{definition}}
\def\bp{\begin{proposition}\rm}
\def\bl{\begin{lemma}\em}
\def\el{\end{lemma}}
\def\ep{\end{proposition}}
\def\be{\begin{equation}}
\def\ee{\end{equation}}
\def\bea{\begin{eqnarray}}
\def\eea{\end{eqnarray}}
\def\beaq{\begin{eqnarray}}
\def\eeaq{\end{eqnarray}}
\def \pa{\partial}
\def\C{{\mathbb C}}
\def\R{{\mathbb R}}
\def\N{{\mathbb N}}
\def\Z{{\mathbb Z}}

\def\ds{\displaystyle}
\def\res{\mathop{\mathrm {res}}\limits_}

\newtheorem{theorem}{Theorem}[section]
\newtheorem{examp}{Example}[section]
\newtheorem{coroll}{Corollary}[section]
\newtheorem{examps}{Examples}[section]

\newtheorem{lemma}{Lemma}[section]
\newtheorem{remark}{Remark}[section]
\newtheorem{remarks}[remark]{Remarks}
\newtheorem{proposition}{Proposition}[section]
\newtheorem{definition}{Definition}[section]
\def\ri{\right}
\def\le{\left}
\def\Amat{\mathbb A}
\def\br{\begin{remark}}
\def\1{{\bf 1}}
\def\er{\end{remark}}
\def\bt{\begin{theorem}}
\def\et{\end{theorem}}
\def\bc{\begin{coroll}}
\def\rd{{\rm d}}
\def\ec{\end{coroll}}
\def\brs{\begin{remarks}.\\ \rm\
\begin{enumerate}}
\def\ers{\end{enumerate}\end{remarks}}
\def\bx{\begin{examp}\small}
\def\ex{\end{examp}}
\def\bl{\begin{lemma}}
\def\el{\end{lemma}}
\def\bxs{\begin{examps}. \rm\begin{enumerate}}
\def\exs{\end{enumerate}\end{examps}}
\def\bd{\begin{definition}}
\def\ed{\end{definition}}
\def\bp{\begin{proposition}}
\def\ep{\end{proposition}}
\def\be{\begin{equation}}
\def\ee{\end{equation}}
\def\bes{$$}
\def\ees{$$}
\def\bea{\begin{eqnarray}}
\def\eea{\end{eqnarray}}
\def\beas{\begin{eqnarray*}}
\def\eeas{\end{eqnarray*}}
\def \pa{\partial}

\def\C{{\mathbb C}}
\def\A{\mathop{\mathbf a}}

\def\R{{\mathbb R}}
\def\N{{\mathbb N}}

\def\Z{{\mathbb Z}}
\def\a{{\alpha}}
\def \z{\mathbf z}

\usepackage[colorlinks=true, pdfstartview=FitV, linkcolor=blue, citecolor=blue, urlcolor=blue]{hyperref}

\date{}
\begin{document}
%

\baselineskip 16pt plus 1pt minus 1pt
\begin{titlepage}
\begin{flushright}
math-ph/0605043\\
CRM-3217 (2006)
\end{flushright}
\vspace{0.2cm}
\begin{center}
\begin{Large}
\textbf{Commuting difference operators, spinor bundles and the asymptotics of orthogonal polynomials with respect to varying complex weights  }
\end{Large}\\
\bigskip
\begin{large} {M.
Bertola}$^{\dagger\ddagger}$\footnote{Work supported in part by the Natural
    Sciences and Engineering Research Council of Canada (NSERC),
    Grant. No. 261229-03 and by the Fonds FCAR du
    Qu\'ebec No. 88353.}\footnote{bertola@crm.umontreal.ca}
    M.Y. Mo
\end{large}
\\
\bigskip
\begin{small}
$^{\dagger}$ {\em Centre de recherches math\'ematiques,
Universit\'e de Montr\'eal\\ C.~P.~6128, succ. centre ville, Montr\'eal,
Qu\'ebec, Canada H3C 3J7} \\
\smallskip
$^{\ddagger}$ {\em Department of Mathematics and
Statistics, Concordia University\\ 7141 Sherbrooke W., Montr\'eal, Qu\'ebec,
Canada H4B 1R6} \\
\end{small}
\end{center}
\bigskip
\begin{center}{\bf Abstract}\\
\end{center}
The paper has three parts.
In the first part we apply the theory of commuting pairs of (pseudo) difference operators to the (formal) asymptotics of orthogonal polynomials: using purely geometrical arguments we show heuristically  that the asymptotics,  for large degrees, of orthogonal polynomial with respect to varying weights is intimately related to certain spinor bundles on a hyperelliptic algebraic curve reproducing  formul\ae\ appearing in the works of Deift et al. on the subject.

In the second part we show that given an arbitrary nodal hyperelliptic curve satisfying certain conditions of admissibility we can reconstruct a sequence of polynomials orthogonal with respect to semiclassical complex varying  weights supported on several curves in the complex plane. The strong asymptotics of these polynomials will be shown to be given by the spinors introduced in the first part using a Riemann--Hilbert analysis.

In the third part we use Strebel  theory of quadratic differentials and the procedure of welding to reconstruct arbitrary admissible hyperelliptic curves.
As a result we can obtain orthogonal polynomials whose  zeroes may become dense  on a collection of Jordan arcs forming an arbitrary forest of trivalent loop-free trees.
\medskip
\bigskip
\bigskip
\bigskip
\bigskip

\end{titlepage}
\tableofcontents
\section{Introduction and summary}
The present paper deals with  the asymptotics of certain
(pseudo--)orthogonal polynomials, its formal properties and
connections with algebraic geometry. In order to explain the
framework, let us recall the main results for ordinary orthogonal
polynomials \cite{DKV} in a simple exemplifying case. Let $V(x)$
be an even-degree real polynomial bounded from below and consider
the Hilbert space $L^2(\R, {\rm e}^{-N V(x)} \rd x)$. Let $p_n(x)$
be the (real) orthogonal polynomials (OP) for this measure.

One of the main goals of modern asymptotic  analysis is to
describe their strong asymptotic as we let $n\to
\infty$ while $N\to \infty$ at the
same rate.

This problem has been brilliantly solved in \cite{BlIt,BlIt2,DKV,DKV2} using
the associated Riemann--Hilbert problem (RHP); indeed the matrix
\be Y(x):= \le[
\begin{array}{cc}
p_n(x) & \phi_n(x)\\
p_{n-1}(x) & \phi_{n-1}(x)
\end{array}
\ri]\ ,\qquad \phi_n(x)  = \frac {1}{2i\pi} \int_\R
\frac {{\rm e}^{-NV(s)}p_n(s)}{s-x}\rd s \ ,z\in \C\setminus \R
\label{1-1}
 \ee
solves a RHP with  jumps on the real axis and rather simple
asymptotics at $z=\infty$ \cite{FIK}. More importantly the solution
of this RHP {\bf characterizes} the orthogonal polynomials;
therefore if one could solve the RHP, then he/she would immediately
have access to all information on the corresponding OP.

Without entering now into the details it suffices to recall that the large parameter $N$ and the degree $n$ of the polynomials enter explicitly and in a simple way the Riemann--Hilbert data and the asymptotics at $x=\infty$. Therefore all the ``complication'' of the asymptotics is controlled by the (highly transcendental) solution of the RHP.

The technique developed in the last decade of the last millennium, called ``nonlinear steepest descent'' method, consists in transforming this RHP into a simpler, ``asymptotic'' one which differs from the exact one by small controllable errors, with suitable uniformly small bounds as $N\to \infty$.

The main character of the method is the so--called $G$-function (which we will discuss at length in the paper). Here we only indicate that it should satisfy some general properties which guarantee the amenability of the Deift--Zhou steepest descent method.

The main logic of approach in most literature is then:
\begin{itemize}
\item fix the potential $V$;
\item try to find an appropriate $G$-function for the given potential;
\item implement the steepest
descent method
\end{itemize}

In a certain sense this point of view mixes a ``forward problem'' (finding the asymptotics of
orthogonal polynomials for a given potential) with an ``inverse problem'' (reconstructing the
matrix solution of a monodromy/jump  problem).

Our approach is ``purely inverse'';
 namely we assign certain asymptotic data in the large $N$-limit
 (which --of course-- satisfy some consistency conditions) and remount the (class of)
 orthogonal polynomials and potentials whose asymptotics matches our given data.
In a subsequent paper \cite{BertoBoutroux} it is  shown that for any (polynomial) potential $V$ (possibly complex) and any ``Stokes' data'' for the orthogonal polynomials, it is possible to construct an appropriate $G$-function. Together, the present paper and loc. cit. completely solve the problem of the large--$N$ asymptotics for these {\em semiclassical} orthogonal polynomials.
\subsection{Asymptotics of generalized pseudo-orthogonal polynomials}

In order to explain the main theorem of the asymptotic analysis we need to introduce some notations.
The setting in which we move is rather algebro--geometric;
the main piece of data is a hyperelliptic (nodal) curve $\mathcal L$ of genus $g$
\be
y^2 = M^2(x) \prod_{j=1}^{2g+2} (x-\alpha_j)\label{1-2}
\ee
 satisfying the conditions of {\bf Boutroux} (we adopt the terminology of \cite{kapaev}) and {\bf admissibility} (to be explained presently).
Such curve is the {\bf asymptotic spectral curve}; the reader acquainted with the asymptotic analysis of ordinary OPs may think of this curve as the hyperelliptic curve associated with the equilibrium measure.
The Boutroux condition is a transcendental reality  condition
\be
\oint_\gamma y\rd x \in i\R\label{1-3}
\ee
where $\gamma$ is any closed loop on the  spectral curve $\mathcal L$.
The Boutroux condition implies that (see Section \ref{se:relat}) the set
\bes
\mathfrak  H_0:=  \le\{\Re\int_{\alpha_1}^x y\rd x = 0\ri\}
\ees
is well-defined independently of the branch-point $\alpha_j$ used in the integration; it also follows that it
consists of a {\bf forest of (open) trivalent trees} whose branches are Jordan arcs (possibly of infinite length).

\begin{wrapfigure}{r}{0.4\textwidth}
\resizebox{0.4\textwidth}{!}{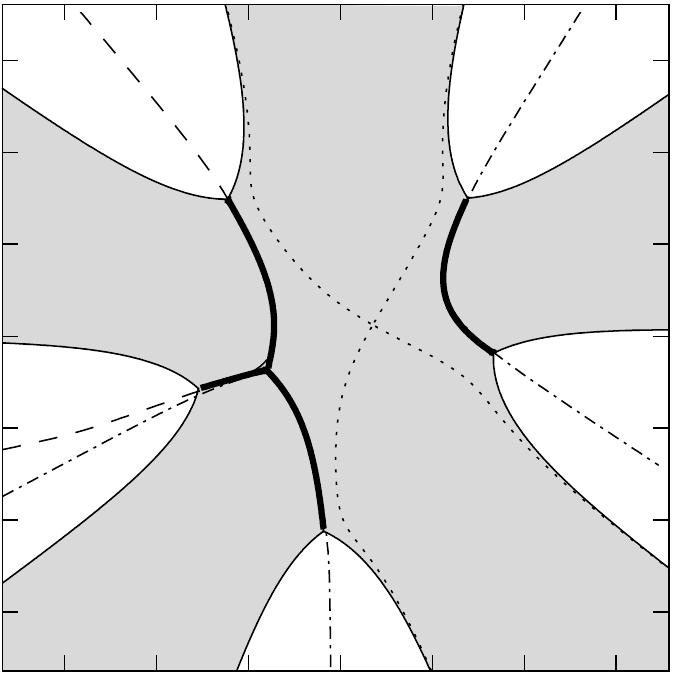}
\caption{An example of admissible Boutroux curve, with the set $\mathfrak H_0$ in evidence (solid-black contours). The branchcuts defined by $\mathfrak H_0$ are the thick arcs. The domains of negativity of $h$ are the shaded regions.
In this case the potential has degree $5$ (see Sec. \ref{se:reconstr} for an explanation of why). The contours in texture represent choices of the contours $\Gamma_j$. In this case there are three inequivalent choices. For example the dashed contour comes from the left, follows two arms of the cross of thick arcs and goes to infinity downwards. The dotted lines are critical levelsets of $h$ passing through a saddle--point.}
\label{introfig}
\end{wrapfigure}

It is then proved in Sec. \ref{se:relat} that this set uniquely
defines a collection $\mathcal B$ of {\bf finite} Jordan arcs
joining the branchpoints $\alpha_j$ which can be used as branch-cuts
for the algebraic function $y(x)$; one can then define uniquely (up
to overall sign) the function \bes h(x):= \Re \int_{\alpha_1}^x y\d
x \ees in such a way that it is {\bf continuous} on $\C$ and {\bf
harmonic} on $\C \setminus \mathcal B$ (see figure \ref{introfig}).

The {\bf admissibility} condition requires that the domains of negativity of (one of the two branches of) $h$ be such that $h<0$ on both sides of each branch-cut (and also that the zeroes of $M(x)$ do not belong to $\mathfrak H_0$).

An admissible Boutroux curve {\bf defines} a {\bf potential} via
\bes
V'(x) =2 (y(x))_{pol}
\ees
where the subscript means the polynomial part and the branch is such that $h(x)\sim \frac 12 \Re V(x)$ at $x=\infty$.

\subsubsection{Connection to the asymptotics of OPs}
Such an admissible Boutroux curve is connected to  the asymptotics of generalized OPs for the potential $V(x)$ above, in the following way (we omit some details for the sake of clarity; the full discussion is in Sec. \ref{se:relat}).

Given an admissible Boutroux curve, we can define some contours $\Gamma_j$ satisfying the properties
\begin{itemize}
\item They extend from $\infty$ to $\infty$ and approach it along directions for which $\Re V(x)\to +\infty$;
\item The (admissible branch of the) function $h(x)$ is nonnegative on these contours.
\end{itemize}
In a figurative way each contour must extend from $\infty$ to $\infty$ along different asymptotic directions and in order to do so is obliged to pass through the branchcuts $\mathcal B$ in order not to violate the second condition above.

To each contour we associate an arbitrary\footnote{These numbers must be sufficiently generic.}  complex number $\varkappa_j$ and then we can consider the generalized OPs as in \cite{BEH1}
\be
\label{eq:orthop}
\sum\varkappa_j \int_{\Gamma_j} p_n(x) p_m(x) {\rm e}^{-N V(x)}{\rm d}x = \delta_{nm}.
\ee

It should be clear that if we fix $N$, the contours $\Gamma_j$ can actually be deformed quite arbitrarily in the finite region of the complex plane; it is only in the large $N$-limit that we should require $h\leq 0$ on these contours (this is the {\em steepest descent condition}).

The main result of this part of the paper is Thm. \ref{stron} which provides strong asymptotic results for $p_{N+r}$ (where $r$ is an arbitrary fixed integer) in terms of the solution of a {\em model Riemann Hilbert problem} (eqs. \ref{eq:jumpcond} in Lemma \ref{le:monspinor});
This is the same logical approach used in \cite{DKV, DKV2}.

Such a RH problem is solved explicitly in terms of Theta functions and spinors on the Boutroux curve: this solution is essentially contained in \cite{DKV2} and is a special case of the more general approach in \cite{korotkin}.
However our approach is yet different and contained in Sec. \ref{se:Jacobi} where we solve rather an {\bf inverse spectral problem} (with a condition of Serre duality) and we find that the solution solves the desired RHP (see the next section in this introduction for the conceptual approach).
Note that the formul\ae\ that we obtain are significantly more streamlined compared to those appearing in the seminal work \cite{DKV2} and also ``explain'' the natural appearance of spinor bundles in terms of Serre-duality.

The main point of this inverse asymptotic study is that Boutroux and admissibility conditions are sufficient (and also necessary as follows from the forward problem) for the amenability of the Deift--Zhou steepest descent method\footnote{
The nonlinear steepest descent method was suggested by Deift and Zhou in 1993 for oscillatory Riemann--Hilbert problems appearing in soliton theory. Some elements of the method appeared in earlier works of A. Its (1980-1982) \cite{itsmatveev} and also in \cite{FIK}. An essential  extension of the method, namely the introduction of the concept of $G$--function, allowed to include into the scheme in particular the RHP for orthogonal polynomials and it is due to Deift, Venakides and Zhou \cite{DKV, DKV2}. }

A consequence and feature of this inverse asymptotic study is that we can construct very easily very complicated and somewhat ``surprising'' examples:
 we mention that a consequence of the uniform asymptotics given by the steepest-descent method is that
  the zeroes of the OPs condense on the Jordan arcs of the branch-cuts $\mathcal B$  in the complex plane;
these arcs may form an arbitrarily complicated {\bf forest of  loop-free trivalent trees}\footnote{Here ``forest'' is in the sense of graph-theory.}.

The implementation of the steepest descent method requires a modification of the usual Airy parametrix near the turning points where three edges of a tree are incident (see Sec. \ref{se:parametrix}).

\subsection{Boutroux and admissibility}
The Boutroux condition is a transcendental constraint on the coefficients of the polynomial defining the (nodal) hyperelliptic curve. At first sight it may not be clear that such Boutroux curves do exist and even less clear how constrained is the topology of $\mathfrak H_0$.

This problem is addressed in Section \ref{se:reconstr}; once more the approach is completely ``backwards''.  We start with an relatively brief summary of Strebel theory of quadratic differentials \cite{strebel} adapted to our simple situation. This allows us to introduce Strebel's coordinates in which the Boutroux condition is just a linear constraint, hence trivially satisfiable.
 The importance of quadratic differentials for the study of $2\times 2$ Riemann--Hilbert problems has been pointed out in other papers (e.g. \cite{kuijlaars, kenarno}) but we are unaware of a systematic application to the steepest descent method as extensive as in the present paper.

These Strebel coordinates parametrize different ``cells'' of different topology of $\mathfrak H_0$ but of the same dimension; in particular we can --given a topological forest graph satisfying some simple requirements-- {\bf reconstruct} by cutting-and-pasting a Riemann surface with the desired properties.  This process is known in the literature of quadratic differentials as {\bf welding} \cite{strebel}: in our simple application the welding merges some half-planes and infinite strips into a Riemann--sphere.

The condition of admissibility is also easily imposed in this approach since it simply singles out cells of specific topology. 
In a way in that section we are solving a more general problem of characterizing arbitrary polynomials  by the topology of the graph of their  critical horizontal trajectories (see Sec. \ref{se:reconstr} for the details).
\subsection{Spinor bundles and commuting difference operators}
In this first part we deal with formal and heuristic aspect of the asymptotic analysis. The formul\ae\  that we obtain will be shown {\em a posteriori} to represent the strong asymptotics of the orthogonal polynomials using the nonlinear steepest descent method. Of course the arguments used should be regarded as ``temporary'', pending a rigorous proof (that is contained in the second part of the paper).

The main idea is the following: 
it is well known that any sequence of OPs satisfies a three-term recurrence relation which can be written in semi--infinite  matrix  form with a tridiagonal symmetric matrix which we denote by $\X$. For generalized OPs w.r.t. complex moment functionals \cite {BEH1} the same is true but the matrix $\X$ has complex coefficients.

If the potential is ``semiclassical'' (i.e. if $V'(x)$ is a rational function and the integrations supported on arbitrary arcs, see \cite{BEH1} for details)  
 the quasi-polynomials $\psi_n:= p_n {\rm e}^{-N\frac V 2}$ satisfy a differential recurrence relation as well, with coefficients that form another matrix $\Y$ which has some closer relationship with $\X$. In the simplest case of polynomial potentials $V(x)$ then
\be
-\frac 1 N \pa_x [\psi_0,\psi_1,\dots]^t = \Y [\psi_0,\psi_1,\dots]^t\label{diffrec}
\ee
with $\Y$ antisymmetric and finite--band
\be
\Y = \frac 1 2 \le(V'(\X)_u - V'(\X)_\ell\ri)
\ee
the subscripts denoting the ``upper'' and ``lower'' part of the matrices. The commutation relations $[\frac 1N \pa_x, x] = \frac 1 N$ imply the so--called {\bf string equation}
\be
[\X, \Y] = \frac 1 N \1\ .\label{stringeq}
\ee
An heuristic approach to the asymptotics is that of sending $N\to\infty$ and also ``look'' at these matrices down the diagonal; if we shift the in dices of the matrices by $N$ and send $N$ to infinity then these matrices become doubly infinite. Moreover the string equation turns into a commutativity equation
$[\Y ,\X]=0$, while retaining their band-structure and symmetry.

In this regime these matrices can be regarded as {\bf commuting finite difference operators}, a class of objects studied in several places in the literature. Indeed there exists a fairly general classification of commuting difference operators \cite{kricnov} which is the starting point of this part of the paper.

In the case at hand the main data are:
\begin{itemize}
\item a smooth  hyperelliptic curve $\mathcal L$ of genus $g$ with $X$ being an invariant meromorphic function of degree $2$ with (symmetric) poles at two points;
\item an arbitrary meromorphic function $Y$ with prescribed pole structure\footnote{This is actually a generalization of the setting of \cite{kricnov}.}
\item a suitably generic effective  divisor $\Gamma$ of degree $g$.
\end{itemize}
These data allow us to reconstruct the matrices $\X$ and $\Y$ as
the matrices expressing the multiplication operators  by $X$ and
$Y$ on a suitable basis of meromorphic functions.

We extend this approach by introducing the Serre dual basis (of meromorphic forms) and a pairing (implementing Serra's duality) which takes the simple form of a residue.

At this level the connection to OPs is still rather distant: in order to close the gap we need to consider a ``self-dual'' picture in which the meromorphic functions of \cite{kricnov} and our Serre--dual counterparts are replaced by sequences of {\bf spinors} which are {\bf self-dual} with respect to Serre--duality. This forces conditions on the divisor $\Gamma$ and determines it up to the choice of a half--integer characteristic.

The actual application to the asymptotics of OP comes after tensoring this spinor bundle by a suitable {\bf flat  line bundle}; such bundle is provided automatically by the Riemann--Hilbert analysis and by the Boutroux condition.

Indeed the Boutroux condition  in this setting is  $\oint Y\rd X \in i\R $ for all closed loops; this means that exponentiation of $\int Y \rd X$ provides naturally transition functions for a {\bf unitary} line bundle with characters given by the periods of the Boutroux differential $Y\rd X$.
This last step provides the backbone of the asymptotic analysis of Sec. \ref{se:relat}
 
 It should be added that this algebro--geometric part of the paper is more general and should yield the main ingredients for the asymptotic analysis of the most general semiclassical OPs, i.e. with potentials $V$ whose derivative is rational and with integrals restricted to arbitrary arcs ({\bf hard--edges}).
 However the technical details of the implementation of Deift--Zhou steepest descent method (Sec. \ref{se:relat}) become rather more involved and we prefer to stay in the more standard setting of polynomial potentials for that part of the paper.

 \paragraph{Acknowledgements}
 The authors would like to thank  Peter Zograf for explaining the theory of Strebel differentials and 
 Andrew McIntyre for several discussion on topology of critical trajectories, Kenneth T.R. McLaughlin for several suggestions about relevant literature and Peter Miller for explaining how to obtain the density of zeroes of ordinary OP in the multicut case.
 Additionally we would like to thank the referee for suggesting several improvements to the exposition and historical background.

\section{Notation and main tools}
\label{se:notation}
For a given smooth genus-$g$ curve $\mathcal L$ with a fixed choice of symplectic homology basis  of $a$ and $b$-cycles,  we denote by
$\omega_j$ the normalized basis of holomorphic differentials
\be
\oint_{a_j} \omega_\ell  = \delta_{j\ell}\ ,\qquad \oint_{b_j} \omega_{\ell} = \tau_{j\ell} = \tau_{\ell j}\ .
\ee
We will denote by
$\Theta$ the theta function
\be
\Theta (\z):= \sum_{\vec n\in \Z^g} {\rm e}^{i\pi \vec n\cdot \tau \vec n - 2i\pi \z\cdot \vec n}
\ee
The Abel map (with  base-point $p_0$) is defined by
\bea
\mathfrak u: \mathcal L\to \C^g\\
\mathfrak u(p) = \le[\int_{p_0}^p \omega_1,\dots, \int_{p_0}^p \omega_g\ri]^t
\eea
and is defined up to the {\em period lattice} $\Z + \tau \cdot \Z$.
For brevity we will omit any symbolic reference to the Abel map when it appears as argument of a Theta function: namely if $p\in \mathcal L$ is a point and it appears as argument of a Theta-function, the Abel map (with a certain basepoint) will be understood.\\
We denote by $\K$ the vector of Riemann constants (also depending on the choice of the basepoint)
\be
\K_j = -\sum_{\ell=1}^{g} \le[ \oint_{a_\ell} \omega_\ell(p) \int_{p_0}^p \omega_j(q) - \delta_{j\ell} \frac
{\tau_{jj}}2   \ri]
\ee
where in this expression the cycles $a_j$ are realized as loops with basepoint $p_0$ and the inner integration is done along a path lying in the canonical dissection of the surface along the chosen representatives of the basis in the homology of the curve.

The Riemann constants have the crucial property that for a nonspecial divisor $\Gamma$ of degree $g$, $\Gamma  = \sum_{j=1}^g \gamma_j$, then the ``function''
\be
f(p) = \Theta(p - \Gamma - \K)
\ee
has zeroes precisely and only at $p=\gamma_j$, $j=1\dots g$.

We will also have to use Theta functions with (complex) characteristics; for any two complex vectors $\vec \epsilon, \vec \delta$ the theta function with these (half) characteristics is defined via
\bea
\Theta\le[{\vec \epsilon \atop \vec \delta}\ri] (z) := \exp\le(2i\pi\le( \frac {\epsilon \cdot \tau \cdot \epsilon }8 + \frac 1 2 \epsilon \cdot z + \frac 1 4 \epsilon \cdot \delta\ri) \ri) \Theta\le(z + \frac {\vec \delta} 2 + \tau \frac {\vec \epsilon }2 \ri)
\eea
Here the (half) characteristics of a point are defined by
\be
2\z = \vec \delta + \tau \vec \epsilon
\ee
where the factor of $2$ is purely conventional so that half integer characteristics have integer (half)-characteristics. In the sequel we will always use these half-characteristics.
This modified Theta function  has the following periodicity properties, for $\lambda, \mu\in \Z^g$
\bea
\Theta\le[{\vec \epsilon \atop \vec \delta}\ri] (z+\lambda  + \tau\mu)  = \exp\le[ i\pi (\vec \epsilon \cdot \lambda - \vec \delta \cdot \mu) -i\pi \mu\cdot \tau \cdot \mu  - 2i\pi z\cdot \mu \ri]
\Theta\le[{\vec \epsilon \atop \vec \delta}\ri] (z)
\eea
\bd
The prime form $E(\zeta,\zeta')$ is the $(-1/2,-1/2)$ bi-differential on
$\mathcal L\times \mathcal L$
\bea
E(\zeta,\zeta') = \frac {\Theta_\Delta
  (\zeta-\zeta')}
{h_\Delta (\zeta) h_{\Delta}  (\zeta') }
\\
h_{\Delta} (\zeta)^2 := \sum_{k=1}^{g}
\pa_{\mathfrak u_k}\ln\Theta_\Delta\bigg|_{\mathfrak
  u=0} \omega_k(\zeta)\ ,
\eea
where $\omega_k$ are the normalized Abelian holomorphic differentials,
$\mathfrak u$ is the corresponding Abel map and $\Delta = \le[\alpha\atop
  \beta\ri]$ is a half--integer odd characteristic (the prime form does
not depend on which one).
\ed
The prime form $E(\zeta,\zeta')$ is antisymmetric in the argument and
it is a section of an appropriate line bundle, i.e. it is multiplicatively multivalued on $\mathcal L\times \mathcal L$; indeed we have the multiplicative multivaluedness
\bea
&&
E(\zeta + a_j,\zeta') = E(\zeta,\zeta')\\
&& E(\zeta + b_j, \zeta') = E(\zeta,\zeta') \exp{ \le(-\frac {\tau_{jj}}2  - \int_\zeta^{\zeta'} \omega_j\ri)}
\eea
In our notation for the (half)-characteristics, the vectors $\alpha,\beta$ appearing in the definition of the prime form are actually integer valued.
We also note for future reference that the half order differential $h_\Delta$ is in fact also multivalued according to
\bea
h_\Delta(p+a_j) = {\rm e}^{i\pi \alpha_j} h_\Delta(p)\\
h_\Delta(p +b_j) ={\rm e}^{-i\pi \beta_j} h_\Delta(p).
\label{basespinor}\ .
 \eea 
Following the common usage in the literature, for a meromorphic function $F$ (or section of some line--bundle) we will use the notation $(F)$ for its divisor of zeroes/poles \cite{FarkasKra}. The writing
\be
(F)\geq -k p + m q
\ee
means that $F$ has at most a pole of order $k$ at $p$ and  a zero of multiplicity at least $m$ at $q$.

For $\mathcal L$ a hyperelliptic algebraic curve of genus $g$ realized as a double cover of the plane,  we will denote with $p^\star$ the image of the point $p\in \mathcal L$ under the hyperelliptic involution that interchanges the two sheets.

To conclude this section we recall some expressions which are well known \cite{fay} and will be used later. Suppose that the hyperelliptic surface is given by the equation $w^2 = \prod_{j=1}^{2g+2} (x-\a_j)$ and let us denote with $\infty_\pm$ the two points above $x=\infty$ where $w\sim \pm x^{g+1}$. 
If we write the normalized first--kind differentials $\omega_j$ as
\be
\omega_j = \sum_{k=1}^{g} \sigma_{jk} \frac {x^{k-1} \rd x} w
\ee
then it is a direct check to verify that 
\be
\Theta_\Delta(p-\infty_+)  =  \frac {C_\Delta} x  (1+\mathcal O(x^{-1}))  \ ,\qquad
C_\Delta :=- \sum _{j=1}^g\sigma_{j,g}\pa_j\Theta_\Delta(0)\label{Cdelta}
\ee
\section{Part I: Jacobi matrices and difference operators}
\label{se:Jacobi}

We follow \cite{kricnov} and  consider the following setting:
\begin{enumerate}
\item An hyperelliptic Riemann surface $\mathcal L$ of genus $g$, with $X$
  invariant under the holomorphic involution and with divisor
  $(X)\geq -\infty_+ -\infty_-$.
\item A meromorphic function $\wh Y$ with divisor
\bea
(\wh Y)\geq -\sum k_\mu \infty_\mu - \sum \xi_j - k_-\infty_-  + \infty_+
\eea
where the $\xi_j$'s are chosen in some subset of the Weierstrass
points $\rd X(\xi_j)=0$\footnote{
Note that there is no loss of generality in assuming that
$X(\infty_\mu)\neq X(\infty_\nu)$ (i.e. the poles of $\wh Y$ have
distinct $X$ projection) because we may add to $\wh Y$ a suitable
rational function of $X$ to reduce ourselves to this situation.
This is why we did not put any pole at $\infty_+$; moreover we can
always add a constant to $Y$ so that it has a zero at $\infty_+$.}.
For later use we also point out that we could equivalently
consider a meromorphic function such that $Y(p^\star) = -Y(p)$;
indeed we may ``antisymmetrize'' the above function by adding a
rational function of $X$. This symmetry will be of use later on so
we will introduce the special notation 
\bea
 Y(p) := \frac 1 2\le(\wh Y(p^\star) -\wh Y(p) \ri)
\eea
\item  A nonspecial divisor $\Gamma$ of degree $g$, $\Gamma =
  \sum_{j=1}^g \gamma_j$.
\end{enumerate}

The genericity assumption on $\Gamma$ is that all the divisors $\Gamma
+ r(\infty_+-\infty_-)$ are also non-special
\be
{\bf i} (\Gamma + r(\infty_+-\infty_-))=0
\label{gener1}
\ee
where ${\bf i}(\mathfrak D)$ denotes the dimension of the space of
differentials with divisor exceeding $\mathfrak D$ \cite{FarkasKra}.

The genericity condition (\ref{gener1}) implies also  
\be
{\bf i} (  \Gamma + (r-1)\infty_+ - (r+1)\infty_-)) =1\ .
\label{gener2}
\ee
Indeed, if we had ${\bf i} (  \Gamma + (r-1)\infty_+ - (r+1)\infty_-)) \geq 2$ then, by Riemann--Roch's theorem, there would exist a meromorphic function with $(f)\geq -\Gamma -r(\infty_+-\infty_-) + \infty_+ + \infty_-$. But then we would have that both $f$ and $Xf$ (clearly linearly independent) would have divisor $\geq -\Gamma  -r(\infty_+-\infty_-) $ so that $ {\bf r} (-\Gamma  -r(\infty_+-\infty_-))\geq 2$ and  --again by Riemann--Roch's thm.-- we would have ${\bf i}( \Gamma +r(\infty_+-\infty_-)) = {\bf r} (-\Gamma  -r(\infty_+-\infty_-)) -1 \geq 1$, a contradiction with the assumption (\ref{gener1}).

By Riemann--Roch's theorem and condition (\ref{gener1}) it follows
that for each $r\in \Z$ there is a unique (up to multiplicative
constant)  meromorphic function with divisor \be (P_r) \geq
-\Gamma -r\infty_+ +r\infty_- \ee

The second condition  (\ref{gener2}) states that for each
$r\in \Z$ there is a unique (up to multiplicative constant)
meromorphic differential $F_n$ with divisor satisfying
\be
(F_r) \geq \Gamma  + (r-1)\infty_+ - (r+1)\infty_-
\ee
From these properties it follows that the product $F_r P_s$ is
a differential with at most two simple poles when $r=s$ and otherwise
it has only one pole. Therefore
\be
\res{\infty_-} P_rF_s \propto \delta_{rs} \label{Serre}
\ee
This condition is just a manifestation of Serre duality.

Since --for $r=s$-- the product $P_rF_s$ is a third kind differential with simple poles,
its residues cannot vanish and hence we can normalize the two dual
sequence so that their  their Serre pairing (\ref{Serre}) is actually
$\delta_{rs}$.

Expressions for the functions $P_r$ and differentials $F_s$ in terms of Theta functions can be obtained following standard references, for example \cite{kricnov}. Since $P_r,F_s$ play only a temporary role in our discussion, we delay explicit formul\ae\ until we arrive at the final objects of interest.

The two sequences can be normalized in such a way that 
\be
\res{\infty_-} F_s P_r = \delta_{rs}\ .
\ee
Given the pole structure of $X(p)$
we  see that $XP_r$ is a linear combination of $P_{r+1},
P_r, P_{r-1}$: indeed
\be
(XP_r) \geq -\Gamma -(r+1)\infty_+ + (r-1)\infty_-
\ee
and the dimension (again generically) of the space of meromorphic
functions with divisor exceeding the above one
is ${\bf r} = 3$ and  it is spanned by the above three
meromorphic functions \cite{kricnov}.

We can express the coefficients of this three-term
recurrence relation
\be
XP_r = \gamma_{r+1}P_{r+1} + \beta_rP_r +\wt\gamma_{r} P_{r-1}
\ee
 in two distinct ways:
first and foremost, using the duality (\ref{Serre}) 
\be
\gamma_{r+1} = \res{\infty_-} F_{r+1}XP_r\ ,\ \beta_r =
\res{\infty_-} F_r XP_r\ ,\qquad \wt \gamma_{r}  = \res{\infty_-}
F_{r-1} XP_r\ . 
\ee
 This provides an explicit expression in terms
of Theta functions if we express $X$ as well in terms of them 
\be
X(p) = X_0\frac {\Theta_\Delta (p-z_0)
  \Theta_\Delta(p-z_1)}{\Theta_\Delta(p-\infty_+)\Theta_\Delta(p-\infty_-)}\label{xtheta}
\ee
where $z_0,z_1$ are the two zeroes of $X$ and $X_0$ is a constant.
A second independent way is obtained  as follows
\bea
X P_r(p) &\& = \frac {c_r}{\det(P_{r+j}(z_k))}_{j,k=1,2} \det\le[
\begin{array}{ccc}
P_{r+1} (p) & P_{r}(p) & P_{r-1}(p)\\
P_{r+1} (z_0) & P_{r}(z_0) & P_{r-1}(z_0)\\
P_{r+1} (z_1) & P_{r}(z_1) & P_{r-1}(z_1)\\
\end{array}
\ri] 
\eea
 where $z_0,z_1$ are the two zeroes of $X$ (interchanged
by the hyperelliptic involution). The constant $c_n$ is
expressed in terms of Theta functions by matching the behaviors of
both sides at one of the two infinities $\infty_\pm$ and using
(\ref{xtheta}).
The dual sequence $F_n$ satisfies --by duality (\ref{Serre})-- the
transposed  recurrence relation
\be
XF_r = \wt \gamma_{r+1} F_{r+1} + \beta_r F_r + \gamma_{r} F_{r-1}\ .
\ee
%


%
\subsection{Kernel and ``Christoffel--Darboux'' pairing}
We want to define the sequence of  ``projectors'' $K_r$ $r\in \Z$
formally as the expressions 
\bea 
&& K_r(p,\xi) =
\sum_{j=-\infty}^{r-1} P_j(p) F_j(\xi) 
\eea 
so that
$\res{\infty_-} K_r(p,\xi)P_s(\xi) = P_s(p)$ if $s\leq r-1$ and
zero otherwise. Clearly these expressions make little sense as
they stand since they involve infinite series whose convergence should then be proved:  what we want to
have is a kernel $K_r(p,\xi)$ which is a differential in $\xi$ and
a function in $p$ that satisfies the following properties:
\begin{enumerate}
\item As a differential in $\xi$ it has
\begin{enumerate}
\item a zero of order $r-1$ at $\infty_+$ and a pole of order $r$ at
  $\infty_-$
\item zeroes at $\Gamma$
\item simple pole at $\xi=p$ (the diagonal) with residue $+1$.
\end{enumerate}
\item As a function of $p$ it has
\begin{enumerate}
\item A pole of order $r-1$ at $\infty_+$ and a zero of order $r$ at
  $\infty_-$
\item Poles at $\Gamma$
\item A simple pole at $p=\xi$.
\end{enumerate}
\end{enumerate}
These properties define it uniquely as (see \cite{fay} pag 27. for
similar kernels) and an explicit formula can also be written in terms of Theta--functions. Once more we will only write formulas for the final objects.


%
The kernel $K_n(p,\xi)$  enjoys the {\bf Christoffel--Darboux} property (an exercise using
  the recursion relations or the divisor properties)
\bea
(X(p)-X(\xi)) K_r(p,\xi) = [F_{r-1}(\xi),F_{r}(\xi)] \le[\matrix{0
    & \gamma_r\cr -\wt \gamma_r& 0}\ri]
\le[\matrix{P_{r-1}(p)\cr P_{r}(p)}\ri] =\\
=\gamma_r P_{r}(p)
F_{r-1}(\xi)- \wt \gamma_rP_{r-1}(p) F_{r}(\xi)
\eea
Therefore we have the
\bt[``Christoffel--Darboux'' theorem]
\label{CDI}
\bea
K_r(p,\xi) = \frac {\gamma_r P_{r}(p)
F_{r-1}(\xi)- \wt \gamma_rP_{r-1}(p) F_{r}(\xi)} {X(p)-X(\xi)}
\eea
\et
Taking the residue
\be
1 = \res{\xi=p} K_r(p,\xi) = \frac {\gamma_r P_{r}(p)
F_{r-1}(p)- \wt \gamma_rP_{r-1}(p) F_{r}(p)} {dX(p)}
\ee
we have a representation of the differential $dX$ as
\be
dX  = \gamma_r P_{r}
F_{r-1} - \wt\gamma_r  P_{r-1}  F_{r}
\ee
 \subsection{Flat line bundles}
\label{se:spinors}
In the asymptotic analysis of orthogonal polynomials using the
Riemann--Hilbert method that follows,
 we will need to tensor the line bundle whose sections correspond to the meromorphic functions $P_n$ and the Serre-dual line bundle (the $F_r$'s) by a  suitable flat line bundle.

 This line bundle can be described in much more general terms as associated to an arbitrary second--kind differential\footnote{We could in fact use arbitrary meromorphic differentials but this would introduce some slight additional complication in the formul\ae, and we leave this to another publication.}.
 
 More precisely let $\eta$ be a meromorphic differential such that  all residues are zero (second kind differential) or integers. Let $\int\eta$ be its Abelian integral; it is defined on the universal covering of the curve $\mathcal L$ less the poles of $\eta$.
 \bd
 Near a pole $c$ of $\eta$  we define the exponential singular part $E_{\eta,c}$ of $\int \eta$ as
 \be
E_{\eta,c} (p) = \exp\le(\oint \eta(\xi)
\ln(z(\xi)-z(p))\ri)
 \ee
 where $z(\xi)$ is a local coordinate $z(c)=0$ and the integral is along loop surrounding $\xi=c$ such that $p$ is outside the loop. The function is independent of the choice of local parameter up to multiplication of a holomorphic function with nonzero value at $c$.
 \ed
 Note that, since the residue of $\eta$ at $c$ is at most an integer, this ``quantization'' makes irrelevant which branch of the logarithm is used.
 
 \paragraph{The twisted line bundle $\mathfrak L_\eta$.}
 Associated to $\eta$ there is a line--bundle $\mathfrak L_\eta$ with transition functions $E_{\eta,c} $ at the poles of $\eta$
%

Tensoring by $\mathfrak L_\eta$ the line bundle described by $\Gamma$ means that the sections of the line bundle will be "functions"  $\varphi$ with the following properties
\begin{enumerate}
\item Poles at $\Gamma$
\item Near a pole  $c$ of $\eta$: \ \ $\ds  \varphi_n(p) E_{\eta,c}(p)  = \mathcal O(1)$.
\end{enumerate}

The formul\ae\ for the dual sequences of wave functions/forms  require minimal modifications and the expressions in terms of Theta function is an exercise that we delay to the next section.


\subsection{Spinors and the symmetric picture}
\label{sec:symmspinors}
In the applications stemming from the (formal)  asymptotics of orthogonal polynomials the sequence of OP
should be orthogonal to itself: i.e. we should put in some way the two
dual sequences on a symmetric footing; moreover the hyperelliptic
involution should yield the dual sequence directly. This ``symmetry''
requirement fixes the divisor $\Gamma$. 
A similar construction in the theory of the algebro-geometric solutions of soliton equations goes back to the works \cite{cher1,cher2}.
Indeed, let us look for a spinor $\s$ with the properties
\begin{enumerate}
\item it has a  simple pole at $\infty_+$
\item it has simple zeroes at $\Gamma$
\item no other poles or zeroes.
\end{enumerate}
Such a spinor exists {\em provided} that $\Gamma$ satisfies some condition to be specified below (note that the degree of the above divisor is the
correct one, $g-1$): the square of a spinor must be a differential
(with divisor of degree $2g-2$) that   has
divisor
\be
(\mathfrak s^2) = 2\Gamma -2\infty_+
\ee
Since the image in the Jacobian of the canonical class is $-2\K$ we
must have
\bea
2\Gamma - 2\infty_+ = -2\K\ ,\qquad  \Rightarrow
\Gamma + \K = \infty_+ + \nu\\
2\nu = 0
\eea
[these equations are written understanding the Abel map and modulo the lattice of periods].
This determines  the divisor $\Gamma$ --up to the choice of a half-period $\nu$-- from Jacobi's inversion theorem. 
We then define
\bea
\pi_r:\propto P_r\s\ ,\qquad \pi_r^\star :\propto \frac {F_r}\s\\
\fK_r (p,\xi) = {\s(p)} K_r(p,\xi)\frac 1{\s(\xi)}
\eea

Exploiting the holomorphic equivalence between flat bundles and unitary flat bundles we can  express $\mathfrak s$ as section of the tensor product with the unitary line bundle characterized by
\be
\chi(a_j) = {\rm e}^{i\pi \mathcal A_j}\ ,\qquad \chi(b_j) = {\rm e}^{i\pi \mathcal B_j}\ ,\label{unitarycharacter}
\ee
where $\mathcal A, \mathcal B$ are the half-characteristics of $\nu$
\bea
2\nu = \mathcal A + \tau \mathcal B.
\eea
Since $\nu$ is a half period, then $\mathcal A, \mathcal B \in \Z^g$.

\subsubsection {$\Theta$--functional expressions}
In this paragraph we provide the explicit expressions of the spinors $\pi_n$, since they will be the final object of interest in our application.

Both $\pi_r,\pi_r^\star$ are spinors (half integer differentials) belonging to a square-root of the canonical bundle of the curve tensored by the line--bundle $\mathfrak L_\eta$ with transition functions $E_{\eta,c}$. and with divisors 
\bea
(\pi_r)\geq -(r+1)\infty_+ + r \infty_-\ ,\qquad
(\pi_r^\star )\geq r\infty_+ - (r+1) \infty_-\\
\pi_{r}(p+\gamma) = \chi(\gamma ) \pi_r(p)\ ,\ \  
\pi_{r}^\star (p+\gamma) = \chi^{-1}(\gamma ) \pi_r^\star(p)\ ,\ \forall\gamma \in \pi_1(\mathcal L).
\eea
with $\chi(\gamma)$ defined (in a basis) by eq. \ref{unitarycharacter}.

Their explicit expressions are obtained using standard arguments and are given below
\bea
\pi_r&\&:= \frac 1{\sqrt{h_r}}
  \frac {\Theta_\Delta^r (
  p-\infty_-)}{\Theta_\Delta^{r+1}(p-\infty_+)}\Theta\le[{\mathcal A + \vec \epsilon\atop \mathcal B + \vec \delta }\ri]
  (p+r\infty_- -(r+1) \infty_+  ) h _\Delta(p) {\rm e}^{- \int_{\alpha_1}^p \eta} \cr
\pi_r^\star &\& := \frac 1{\sqrt{h_r}}
 \frac {\Theta_\Delta^{r}(p-\infty_+)}{\Theta_\Delta^{r+1} (p-\infty_-)}
\Theta\le[{-\mathcal A - \vec \epsilon\atop-\mathcal B - \vec \delta }\ri] (p -(r+1)\infty_- + r\infty_+)
h_\Delta(p) {\rm e}^{\int_{\alpha_1}^p \eta}
\label{spinorchar}\\
&& \epsilon_i:= \frac 1 2 \oint_{a_j} \eta\ ,\qquad 
\delta_i:= -\frac 1 2\oint_{b_j} \eta\ .\nonumber
\eea
The constants $h_n$ are just the suitable normalizations so that
\be
\res{\infty_+}\pi_r \pi^\star_s =\delta_{rs}\label{SerreSpinDuality}
\ee
They are
\be
h_r = \frac {\Theta_{\Delta}(\infty_+-\infty_-)} { \Theta\le[{\mathcal A + \vec \epsilon\atop \mathcal B + \vec \delta }\ri](r(\infty_--\infty_+) )\Theta\le[{-\mathcal A-\vec \epsilon\atop -\mathcal B-\vec \delta }\ri]((r+1)(\infty_+-\infty_-))}
\label{normconst}
\ee
and the spinorial kernel of the Christoffel--Darboux projector now reads
\bea
\fK_r(p,\xi):=&\&  
\le[\frac {\Theta_\Delta (\xi-\infty_+) \Theta_\Delta( p- \infty_-)}
{\Theta_\Delta (\xi - \infty_-)\Theta_\Delta(p-\infty_+)} \ri]^{r} 
\frac
{
   \Theta\le[
        {\mathcal A +\vec \epsilon\atop \mathcal B  +  \vec \delta }
        \ri]
    (r(\infty_--\infty_+) + p  - \xi)
}
{
     E(\xi,p) \Theta\le[
          {\mathcal A+\vec \epsilon\atop \mathcal B + \vec \delta }
          \ri]
       (r(\infty_- - \infty_+))
}
  {\rm e}^{- \int_\xi^p \eta}
\eea
In the following applications $\eta$ will be an antisymmetric differential (under the hyperelliptic involution); in this case we have the  symmetry
\be \pi_r(p^\star) =-
\pi_r^\star(p)
 \ee 
 
 which explains the notation; indeed this
follows from the fact that the Abel map (based at a Weierstrass
point) of $p^\star$ is the opposite of that of $p$ and from the
symmetries of the Theta functions with characteristics
\be
\label{eq:sym} \Theta\le[{A\atop B}\ri] (z) =
\Theta\le[{-A\atop -B}\ri] (-z) 
\ee
together with the antisymmetry of $\eta(p) = -\eta(p^\star)$.
With these normalizations the recurrence relations become
automatically symmetric because
\bea
&& \gamma_r = \res{\infty_+}\pi_r X \pi_{r+1}^\star = -\res{\infty_+\star} \pi_r^\star X
\pi_{r+1} =  -\res{\infty_-} \pi_r^\star X
\pi_{r+1} =  \res{\infty_+} \pi_r^\star X
\pi_{r+1}= \wt \gamma_r
\eea
and hence
\be
 X\pi_r = \gamma_{r} \pi_{r+1} + \beta_r \pi_r + \gamma_{r-1}\pi_{r-1}\label{spinrecrel}
\ee

Moreover it follows from the previous Christoffel--Darboux identities (Thm. \ref{CDI})
that
\bea
\fK_r(p,\xi) &\&= \gamma_r \frac {\pi_{r-1}(p)\pi_r^\star(\xi) -
  \pi_{r}(p)\pi_{r-1}^\star(\xi)}{X(p)-X(\xi)}\\
dX(p) &\& = \gamma_r \bigg( \pi_{r-1}(p)\pi_{r}^\star(p) -
\pi_{r}(p)\pi_{r-1}^\star(p)\bigg) \label{dx}
\eea
\subsection{Ladder matrices and Lax matrix}
We preliminary point out that any spinorial Baker--Akhiezer function $\pi_k$ can be written as a linear combination in terms of any other two consecutive wave-functions (forms) with polynomial coefficients in $X$: indeed the recurrence relations (\ref{spinrecrel}) can be rewritten in matrix form as
\be
{\boldsymbol {\Pi}}_{r+1}:= \le[\matrix{ \pi_{r+1} \cr \pi_{r}}\ri] = \le[
\begin{array}{cc}
 \frac {X-\beta_r}{\gamma_r}
&  -\frac {\gamma _{r-1}}{\gamma_r} \\
1&0
\end{array}\ri]
 {\boldsymbol \Pi}_r =: \A_r(X) {\boldsymbol  \Pi}_r\label{ladder}
\ee

The ladder matrix $\A_r(X)$ is invertible and the inverse is linear as well in $X$.
Therefore
\bea
&& {\boldsymbol \Pi}_{r+k+1}  = \A_{r+k}\cdots \A_r {\boldsymbol \Pi}_{r}=: \A_{r}^{r+k}{\boldsymbol \Pi}_r\ ,\qquad k\geq 0\\
&& {\boldsymbol \Pi}_{r-k-1} = {\A_{r-k-1}}^{-1} \cdots {\A_{r-1}}^{-1} {\boldsymbol \Pi}_r \ ,\qquad k\geq 0\ .
\eea
Even more directly, denoting  by $\Amat_r = \pmatrix{0&\gamma_r\cr - \gamma_r & 0}$, the kernels
$K_r$ are written as
\be
K_r(p,\xi) = \frac {{\boldsymbol \Pi}_r^\star(\xi) ^t\Amat_r {\boldsymbol \Pi}_r(p)}{X(p)-X(\xi)}
\ee
Then we have the identity
\be
{\boldsymbol \Pi}_{j}(p) = \res{\xi=p} {\boldsymbol \Pi}_j(\xi) \fK_r(p,\xi) = -\sum_{a=\pm} \res{\infty_a} {\boldsymbol \Pi}_j(\xi) \fK_r(p,\xi) =
\sum_{a=\pm} \res{\infty_a} \frac {{\boldsymbol \Pi}_j(\xi)
  {\boldsymbol \Pi}_r^{\star,t}(\xi)\Amat_r}{X(p)-X(\xi)} {\boldsymbol \Pi}_r(p)\ .
\ee
which follows from the fact that --in any local parameter-- $\fK_{r}(p,\xi)= \frac {\sqrt{\rd z} \sqrt{\rd z'}}{z-z'}(1+ \mathcal O(z-z')$, with $z= z(p), z' = z(\xi)$.
It is easily seen that the matrix
\be
\A_r^j(x):=
\sum_{a=\pm} \res{\infty_a} \frac {{\boldsymbol \Pi}_j(\xi)
  {\boldsymbol \Pi}_r^{\star,t}(\xi)\Amat_r}{x-X(\xi)}
\ee
is a {\bf polynomial} in $x$ of degree $|r-j|$ and computes directly
the product of the ladder matrices above.

\subsubsection{Lax matrix}
Although $Y{\P}_r$ in general does not have the same pole structure as the sequence of the  $\pi_r$'s, nevertheless we can express it in terms of the same vector ${\P}_r$.

Let us define the rational function $V'_\alpha$ of $x$,
$\alpha=\mu,-$ such that \bea
&& \wh Y(p)- V'_\mu(X(p)) = \mathcal O(1)\ \hbox{ near }\infty_\mu \\
&& \wh Y(p)- V'_-(X(p)) = \mathcal O(X^{-1})\hbox { near }\infty_-\ .
\eea
The function $V'_-$ is a polynomial of degree $k_-$ whereas $V'_{\mu}$ are polynomials in $\frac 1{x-X(\infty_\mu)}$ of degree $k_\mu$ and without constant coefficient.
Define then the total potential
\be
V'(x): = \sum_{\alpha=\mu,-} V'_\alpha(x)\ .
\ee
This rational function of $x$ has the property that $\wh Y(p)-V'(X(p))$ is analytic near the points $\infty_\mu$ and $\infty_-$.
Then  we find (using that $\wh Y$ has a zero at $\infty_+$)\footnote{Recall that $\xi_j$ represent those Weierstrass points (branchpoints for $X$) which coincide with a simple pole of $Y$ (or $\hat Y$).}
\bea
\wh Y\P_r (p) =&\& \res{\xi=p} \wh Y(\xi)\P_r(\xi) \fK_r(p,\xi)= -\res{\mathfrak U, \infty_-} \wh Y\P_r(\xi) \fK_r(p,\xi)  =  \\
=&\& -\sum_{\alpha= \mu,-} \res{\xi = \infty_\alpha} \wh
Y(\xi)\P_r(\xi) \fK_r(p,\xi) - \sum_{j} \res{\xi = \xi_j}  \wh Y(\xi) \P_r(\xi) \fK_r(p,\xi) =
\cr
=&\&  -\sum_{\alpha= \mu,-} \res{\xi = \infty_\alpha}
V'(X(\xi))\P_r(\xi) \fK_r(p,\xi) - \sum_{j} \res{\xi = \xi_j}  \wh
Y(\xi)\P_r(\xi) \fK_r(p,\xi) =
\cr
=&\& -\sum_{\alpha= \mu,-} \res{\xi = \infty_\alpha} V'(X(\xi))\P_r(\xi)  \frac {{\P}_r^{\star,t}(\xi) \Amat {\P}_r(p)}{X(p)-X(\xi)} - \sum_{j} \res{\xi = \xi_j}  \wh Y(\xi)\P_r(\xi) \fK_r(p,\xi) =
\cr
=&\& \pmatrix{ 0&0\cr 0& V'(X(p))}\P_r(p) -\sum_{\alpha= \mu,-}
\res{\xi = \infty_\alpha} \frac{V'(X(\xi)) - V'(X(p))}{X(p)-X(\xi)} \P_r^{\star,t}(\xi) \Amat {\P}_r(p)+\cr
&\&  - \sum_{j} \res{\xi = \xi_j}  \wh Y(\xi)\P_r(\xi) \fK_r(p,\xi) =\\
&\& \pmatrix{ 0&0\cr 0& V'(X(p))}\P_r(p) -\sum_{\alpha= \mu,-} \res{\xi = \infty_\alpha} \frac{V'(X(\xi)) - V'(X(p))}{X(p)-X(\xi)} \P_r(\xi)  {\P}_r^{\star,t} (\xi) \Amat {\P}_r(p)+\cr
&\&  - \sum_{j} \res{\xi = \xi_j}  \wh Y(\xi)\P_r(\xi) \frac{ {\P}_r^{\star,t}(\xi) \Amat {\P}_r(p)}{X(p)-X(\xi)}
\eea
Summarizing we have --in matrix form--
\bea
 \wh  Y {\P}_r &\&= \wh D_r^{(\eta)}(X) {\P}_r\cr
\wh D_r^{(\eta)} (x) &\&= \le[\begin{array}{cc}
0&0\\
0 & V'(x)
\end{array}\ri] - \sum_{\alpha=\mu,+} \res{\infty_\alpha} \frac {V'(X)-V'(x)}{X-x} {\P}_r(\xi) {\P}_r^{\star,t}(\xi) \Amat_r +\cr
&&- \sum_{j} \res{\xi = \xi_j} \frac { \wh Y(\xi){\P}_r(\xi) {\P_r^{\star,t}}(\xi)}{X-x} \Amat_r
\label{Laxmatrix0}
\eea
Quite clearly, the matrix $\wh D_r^{(\eta)}(x)$ is a  rational matrix with the same pole structure as the derivative of the ``potential'' $V'(x)$: at the other poles ($\xi_r$)  of $\wh Y$, it has simple poles with a nilpotent residue.

Similarly we could repeat the computation for the antisymmetric function $ Y$; however it is immediate to see that\footnote{Indeed $\wh Y(p^\star) + \wh Y(p)$ is invariant under the hyperelliptic involution and hence it is a rational function of $X$, namely $V'(X)$ in our notation.}
\be
 Y = \frac 1 2 V'(X) - \wh Y
\ee
so that the matrix $\wh D_r^{(\eta)}(X)$ representing the multiplication by $\wh Y$ differs from the matrix $D_r^{(\eta)}(X)$ only by the multiple of the identity $\frac 1 2 V'(X) \1$
\be
D_r^{(\eta)}(x) = \frac {V'(x)}2 \1  - \wh D_r^{(\eta)}(x)\ ,\qquad
 [\P_r,\P_r^\star]Y\sigma_3 =   D_r^{(\eta)}(X) [\P_r, \P_r^\star]\ .\label{Laxmatrix}
\ee
where we have used that $\P_r^\star$ solves the eigenvector  equation with eigenvalue $-Y$ since it corresponds to evaluation on the other sheet of the $X$--projection and $Y$ is antisymmetric.
\paragraph{The WKB Ansatz.} 
We now spend some words on the heuristics of the whole construction and on its logical and historical underpinnings.
As pointed out in the introduction (at least for polynomial potentials) the OPs` satisfy a three term-recurrence relation of the same form as (\ref{spinrecrel}) and also a {\em differential} recurrence relation (\ref{diffrec}). The compatibility relation between the two is contained in eq. (\ref{stringeq}), often referred to as the ``string equation'' for the recurrence coefficients \cite{moore2}.  It can be rewritten in matrix form as the compatibility relation between the following differential/difference equations
\bea
 - \frac 1N  \pa_x\Psi_{n} =  D_{n}(x)  \Psi_{n} \ ,
\Psi_{_{n+1}}\!\!\!= \le[
\begin{array}{cc}
 \frac {x-\beta_n}{\gamma_n}
&-\frac {\gamma_n}{\gamma_{n-1}} \\
1 & 0
\end{array}
\ri]\Psi_n\, ,\quad
\Psi_n(x):=  \le[\begin{array}{cc}
p_{_n}(x)&  \phi_{_n}\\
p_{_{n-1}}(x) & \phi_{_{n-1}}
\end{array}\ri] {\rm e}^{-\frac N{2`} V(x)\sigma_3}
\label{compat}
\eea 
where $n=N+r$ (with $r\in \Z$ bounded as $N\to\infty$), $\phi_n$ are the Cauchy transforms of ${\rm e}^{-NV}p_n$'s   and $\gamma_n, \beta_n$ denote the recurrence
coefficients for the orthogonal polynomials (see for example
\cite{BEH2} for explicit formul\ae\ for the matrix $D_n$ in a very
general setting of semiclassical orthogonal polynomials). The heuristic approach is then to {\em assume} the
validity of Wentzel-Kramers-Brillouin (WKB) ansatz for $\Psi_n = \Psi_{N+r}$ in
the form 
\be
 \Psi_{N+r} \sim \Xi_r {\rm e}^{-N\sigma_3 
\int^x Y(\xi)\rd \xi}
\label{WKB} 
\ee 
where $\Xi_r$ (see eq. (\ref{Ximatrix})) should be expanded (as well as
$ Y$) in formal series of $N^{-1}$. Plugging the Ansatz
(\ref{WKB}) into the differential equation (\ref{compat}) yields, to
leading order 
\be 
\Xi_{r}Y\sigma_3 =  D_r^{(\eta)} (x) \Xi_r 
\ee 
which is nothing
more than an {\em eigenvalue/eigenvector} equation precisely of the
type appearing in (\ref{Laxmatrix}) (with $\eta = N\,Y\rd X$). In short the WKB approach
amounts to substituting the ODE in (\ref{compat}) by its {\em
symbol}, thus turning it into a spectral problem.

This line of approach was developed in the works  \cite{moore1, moore2}, based in turn on the isomonodromic techniques in the theory of Painlev\'e\ equations \cite{itsnovokshenov}. In the context of OPs it was applied to the ``one--cut'' case in \cite{BlIt} and to multi--cut settings in \cite{pastur}.

The self--consistency of the WKB method calls for the periods of the exponent $\int^x \wh Y(x) \rd x$ to have no real part, for otherwise the asymptotic expansion in inverse powers of $N$ for $\Xi_N$ would not be valid under analytical continuation; such condition ({\bf Boutroux condition}) will be the main requirement in the second part of the paper.  

In particular, as anticipated in the introduction,  the {\em asymptotic spectral curve} (\ref{1-2}) with the Boutroux condition (\ref{1-3}) should be thought of as the ``limiting spectral curve'' of the matrices $D_n(x)$ appearing in the differential equation of eqs. (\ref{compat}). 

The matrix $\Xi_{r}$ consists of genuine Baker--Akhiezer functions. The relation with spinors and the Serre duality (\ref{SerreSpinDuality})  should be understood as a large--$N$ relic of the orthogonality relations: indeed, at a very formal/suggestive level, 
the orthogonality reads $\int p_n p_m{\rm e}^{-NV(x)} \rd x= \delta_{nm}$ which suggests that a symmetric way of considering the large-$N$ limit should be by formally splitting $\rd x= \sqrt{\rd x}\sqrt{\rd x}$ so that $p_{N+r}{\rm e}^{-\frac N2 V(x)}\sqrt{\rd x} \sim \pi_{r}$. The $N$--dependence will be in the asymptotic line bundle determined by the differential $\eta = N \, Y \rd x$ on the hyperelliptic spectral curve, while the finite perturbation $r$ will give the lattice behavior (of Toda type).

Of course, the reader should regard these considerations  as a mere back-of-the-mind motivation towards an  educated guess, the full justification of which comes only {\em a posteriori} via a Deift--Zhou nonlinear steepest-descent analysis (contained in the second part of the paper) adapted to particular hyperelliptic curves satisfying the Boutroux condition. The existence/construction  of these is the content of the third part of the paper.

Once the strict connection between the $\pi_r$'s and the uniform
asymptotics over compact sets of the orthogonal polynomials are made
rigorous, we could conclude that indeed $\det (y \1 - D_n(x))$
admits a {\em bona-fide} limit and this limit is a Boutroux curve.

Indeed it will be shown in Section \ref{se:relat} that  the solution $\Psi_{N+r}$ of the Riemann--Hilbert problem for the orthogonal polynomials (eq. (\ref{compat}) is indeed approximated by the matrix $\Xi_r(x){\rm e}^{N\int Y \rd x\sigma_3}$  (with $\Xi_r$ given in (\ref{Ximatrix}))  {\em uniformly over compact sets not intersecting the branchcuts}.

Thus we have that the matrix of the differential equation $D_{N+r}(x)=\frac 1 N \Psi'_{N+r}(x)\Psi_{N+r}^{-1}(x)$ (which is a polynomial in $x$ of constant degree equal to $\deg V'(x)$ for polynomial potentials) {\em also  tends uniformly} to $D_{r}^{\eta}(x) = \Xi(x)^{-1} Y(x)\sigma_3 \Xi(x)$, which is our eigenvalue-eigenvector equation (\ref{Laxmatrix}) for the line-bundle $\eta = N Y\rd X$. Therefore {\em a fortiori} also the spectral curve must tend to the asymptotic Boutroux hyperelliptic curve. Since the convergence is uniform over  compact sets and the objects $D_{N+r}$ and its characteristic polynomials are all polynomials in $x$ of bounded degree, the convergence is established in any bounded set of the plane.

 \subsection{The spinor $h_ \Delta$}
Following  (\cite{fay}, page. 13) we can explicitly write the half order differential $h_\Delta$ (up to a multiplicative constant which is irrelevent in our application
 since it would then be reabsorbed in the normalizations). Let us denote by $\a_j  = X(\xi_j),\ j=1,\dots 2g+2$ the critical values of $X$  (at the Weierstrass points) and
\be
W^2 := \prod_{j=1}^{2g+2} (X-\a_j)
\ee
There are $4^g$ half periods in the Jacobian: they are all in one-to-one correspondence with the $4^g$ partitions $\{i_1,\dots,i_{g+1-2m}\}\cup\{ j_1,\dots, j_{g+1+2m}\}$ of the set $\{1,2,\dots, 2g+2\}$ for $m\geq 0$ as follows:
\begin{enumerate}
\item All even {\em non-singular} half-periods are given by the choice $m=0$ and are the image in the Jacobian of
\be
\sum_{k=1}^{g+1}\a_{i_k} - \infty_+-\infty_-- \K
\ee
\item all odd {\em non-singular} half-periods are given by the choice $m=1$ and are the image in the Jacobian of
\be \sum_{j=1}^{g-1} \a_{i_j} -\K \ee
\item All other half-periods are singular and they are even or odd according to the parity of $m$.
\end{enumerate}

Any choice of odd, nonsingular half-integer characteristics
corresponds to a choice of $g-1$ points amongst the Weierstrass
points, $\xi_{i_1},\dots, \xi_{i_{g-1}}$, therefore \be h_\Delta^2
\propto \prod_{k=1}^{g-1} (X-\a_{i_k}) \frac {dX}{W} \ee In other
words, with proper understanding of the analytic continuation on
the (double cover of the) hyperelliptic curve, the spinor $h_\Delta$
can be expressed as (up to overall constant) \be h_\Delta  = \frac
{\prod_{k=1}^{g-1} (X-\a_{i_k})^{\frac 1 4}}{\prod_{k=1}^{g+3}
(X-\a_{j_k})^{\frac 1 4}} \sqrt{dX}\label{hdelta} \ee

\subsection{Riemann Hilbert problem}
The spinors (\ref{spinorchar}) exhibit a multiplicative behavior which depends on the half period $\nu =\frac 1 2  \vec{\mathcal A} + \frac 1 2 \tau \cdot \vec{\mathcal B}$ ($\vec {\mathcal A}, \vec {\mathcal B}\in \Z^g$)
\bea
&& \pi_r(p+a_j) = {\rm e}^{i\pi \mathcal A_j} \pi_r(p)\ ,\qquad
\pi_r(p + b_j) = {\rm e}^{i\pi  \mathcal B_j} \pi_r(p)\\
&& \pi_r^\star(p+a_j) = {\rm e}^{-i\pi  \mathcal A_j} \pi_r^\star(p)\ ,\qquad
\pi_r^\star(p + b_j) = {\rm e}^{-i\pi \mathcal B_j} \pi_r^\star(p)\\
\eea

These phases define a {\bf character}
\be
\chi_\nu:\pi_1(\mathcal L) \to \Z_2\ ,
\ee
and in general so does any half--period (or half--integer characteristics).

The matrix
\be
\Xi_r:= \frac 1{\sqrt{\rd X}} \le[\matrix{\pi_r & i \pi_r^\star \cr \pi_{r-1} &i  \pi_{r-1}^\star}\ri]
\label{Ximatrix}
\ee
satisfies
\be
Y \Xi_r\sigma_3 = D_r^{(\eta)}(X)\Xi_r
\ee
for $D_r^{(\eta)}(X)$ given by the formula (\ref{Laxmatrix}).
Moreover, from (\ref{dx}) it follows
\be
\det\Xi_r = \frac i {\gamma_r}
\ee
The matrix $\Xi_r$ is a {\em bona-fide} multivalued matrix-valued function on the (desingularization of the) spectral curve
\be
0=H(X,Y)=\det(Y\1 - D_r^{(\eta)}(X))\ .
\ee
It can be thought of as a multivalued function of $x=X(p)$ with branchpoint singularities at the branchpoints $\alpha_j = X(\xi_j)$.

On the sheet of the map $x= X$ containing the point $\infty_+$ it solves a certain Riemann--Hilbert problem which we describe below; the main complication 
arises from the proper understanding of the square--root of the differential $\rd X$. We note that the entries of $\Xi_n$ are proportional to (on each sheet)
\be
Q_\Delta (x):= \frac {\prod_{k=1}^{g-1} (x-\a_{i_k})^{\frac 1 4}}{\prod_{k=1}^{g+3} (x-\a_{j_k})^{\frac 1 4}} = \frac {h_\Delta}{\sqrt{\rd X}}\ .\label{Qfunction}
\ee
Clearly this ``function'' makes sense only after a suitable surgery on the plane, which is what we describe in the paragraph below; moreover it depends on the choice
 of the characteristic $\Delta$ under the identification between half--periods and partitions of the Weierstrass point described earlier.

Each entry of $\Xi$ is of the form   (refer to eq.
\ref{spinorchar}) 
\bea 
\frac {\pi_{r}}{\sqrt{\rd X}} &\&= \frac 1{\sqrt{h_r}}
  \frac {\Theta_\Delta^r (
  p-\infty_-)}{\Theta_\Delta^{r+1}(p-\infty_+)}\Theta\le[{\mathcal A + \vec \epsilon\atop \mathcal B + \vec \delta }\ri]
  (p+r\infty_- -(r+1) \infty_+  ) {\rm e}^{- \int_{\alpha_1}^p \eta} Q_\Delta(X(p))  \cr
  &\& =: F_r(p)Q_\Delta(X(p))\cr
\frac{\pi_r^\star}{\sqrt{\rd X}} &\& := \frac 1{\sqrt{h_r}}
 \frac {\Theta_\Delta^{r}(p-\infty_+)}{\Theta_\Delta^{r+1} (p-\infty_-)}
\Theta\le[{-\mathcal A - \vec \epsilon\atop-\mathcal B - \vec \delta }\ri] (p -(r+1)\infty_- + r\infty_+)
{\rm e}^{\int_{\alpha_1}^p \eta} Q_\Delta(X(p)) = \cr
&\&=: F_r^\star(p) Q_\Delta(X(p))
\label{entries}
\eea
where the functions  $F_r$ and $F_r^\star$ have the monodromy
\bea
&\& F_r(p+\gamma) = \chi_{\nu+\Delta}(\gamma) F_r(p)\cr
&\& F_r^\star (p+\gamma) =\chi_{\nu+\Delta}(\gamma)  F_r^\star(p)\ \ \qquad \forall \gamma \in \pi_1(\mathcal L)\ ,
\eea
where we have used $\chi_\nu\chi_\Delta = \chi_{\nu+\Delta}$ and $\chi_{-\Delta} = \chi_{\Delta}$ (this last valid for half--periods only).
It follows also from the above formul\ae\ that each entry of $\Xi_r$ has a singularity at the branchpoints $\alpha_j$ of type $(x-\alpha_j)^{-1/4}$. 
In fact $\Theta_\Delta(p-\infty_-)$  (and $\Theta_\Delta(p-\infty_+)$) has simple zeroes at the Weierstrass points $\alpha_{i_k} = X(\xi_{i_k})\ ,k=1,\dots g-1$ appearing 
in eq. (\ref{Qfunction}); since the local coordinate is $\sqrt{x-\alpha_{i_k}}$ we see in eqs. \ref{entries} that at {\bf all} branchpoints we have the advocated behavior.

 When thinking of $F_r, F_r^\star$ as functions on a simply connected domain of $\C$ (one sheet of the $X$--projection), they define functions with singularities only at the branch-points and essential singularities at the $X$--projections of the poles of the twisting differential $\eta$.

If $\eta$ is an antisymmetric differential (w.r.t. the hyperelliptic involution)\footnote{ There is not much loss in generality in assuming that it is antisymmetric, because it can be always antisymmetrized by an exact differential which does not change its characteristics.} then there is the further symmetry (which follows directly from the explicit formula defining them)
\be
F_r(x) = - F_r^\star(x^\star)\ .
\ee
\paragraph{(Nonstandard) surgery.} Let  $\mathcal B$ be a set of branchcuts for the projection $X$: specifically, if $\a_i$ are the critical values of $X$ then the smooth hyperelliptic curve is written as
\be
W^2 = \prod_{i=1}^{2g+2} (X-\a_i)\ .
\ee
Then $\mathcal B$ is a collection of mutually non-intersecting oriented arcs $\Sigma_j$, joining two points $\a_{j_1},\a_{j_2}$ in such
 a  way that one branch of  $W$ can be defined as a single valued function on $\C\setminus \mathcal B$. The ``standard'' way of performing
 these cuts is to join $\a_{2i}$ to $\a_{2i+1}$ (whatever numbering has been chosen). We point out that there are ``nonstandard'' ways of performing
 an equally satisfying surgery; the only condition is that at each point $\a_i$ originates  an {\bf odd number of cuts} (we will need this generality in the following).
  We also require (which will be enough for our later application) that $\C\setminus \mathcal B$ is {\bf connected} and that each connected component of $\mathcal B$
 has an {\bf even} number of vertices. We add some oriented arcs (called {\bf gaps}) joining each connected component of $\mathcal B$ to the next, and the last one
 to $\infty$. We denote by $\Sigma$ the collection of all oriented cuts $\mathcal B$ and oriented gaps (see for example Fig. \ref{fig:sigma}).
On the resulting simply connected domain $\C\setminus \Sigma $ we have the following Riemann--Hilbert problem.

Near the poles $c$ we have
\bea
\Xi_r(p) \sim E_{\eta,c}(p) ^{\sigma_3}\ \ \ \ \hbox{near } c \label{RHinf}\\
\sigma_3:= {\rm diag} (-1,1), \ \ X(p)=x.\nonumber
 \eea
 The jumps
discontinuities of this Riemann-Hilbert problem are given in the
following paragraph.

\paragraph{Jumps on $\Sigma$.}
For $x\in \Sigma\setminus\{\a_i\}_{i=1\dots 2g}$ let  $\wt\gamma(x)$ be a closed positively oriented loop in $\C\setminus \Sigma$ intersecting sigma only at $x$ (the simple connectivity of $\C\setminus \Sigma$ implies that the homotopy class is unique).

 Define $\sharp_\Delta(x)$ as  the difference between the number of Weierstrass points (in the interior region cut by  $\wt \gamma(x)$) entering in the numerator of $Q_\Delta$ (which {\em define} $\Delta$) and the number of the ones in the denominator. It has the following {\bf properties} which are easily proved:
\begin{itemize}
\item it is odd on the cuts;
\item it is even on the gaps;
\item if $x$ is on the last gap extending to infinity, then $\sharp_\Delta(x) = -4 = (g-1)-(g+3)$.
\end{itemize}
The jumps of $Q_{\Delta}(x)$ on the cuts and the gaps are given by
$e^{2\pi i(1\pm\frac{1}{4}\sharp_\Delta(x))}$ depending on the
orientations of the corresponding cut/gap. In particular the
function $Q_\Delta(x)$ is continuous across the last gap.

Any $\Z_2$ (which we think of as the multiplicative group consists
of $1$ and $-1$) character on $\pi_1(\mathcal L)$ induces an assignment of
signs on the cuts and gaps of $\Sigma$ as follows. Define a loop
$\gamma(x)$ such that:
\begin{itemize}
\item $\gamma(x)$ is a closed loop on $\mathcal L$  based at $\xi_1$
($X(\xi_1)=\a_1$) the Weierstrass point chosen as base--point for
the Abel map; \item the $X$--projection of $\gamma(x)$ is a {\bf
positively oriented loop}  intersecting $\Sigma$ {\bf only} at
$\a_1 = X(\xi_1)$ and at $x$.
\end{itemize}
The above recipe defines $\gamma(x)$ on $\mathcal L$ {\bf up to orientation}; however this ambiguity is inessential for us because we will be evaluating $\Z_2$ characters only (and $(-1)^{-1} = -1$).

Given an arbitrary $\aleph:\pi_1(\mathcal L) \to \Z_2$
we assign to each {\em oriented} cut $\Sigma_j$ and each {\em oriented} gap $\wt \Sigma_\ell$ the sign
\bea
&\& (-)^{\Sigma_j} := \aleph (\gamma(x)) \langle X(\gamma(x))\cdot \Sigma_j\rangle\ ,\qquad \hbox{ for } x\in \Sigma_j\cr
&\& (-)^{\wt \Sigma_j} := \aleph (\gamma(x))\qquad \hbox{ for } x\in \wt\Sigma_\ell \label{aleph}
\eea
where $\langle X(\gamma(x))\cdot \Sigma_j\rangle$ denotes the intersection number for oriented curves in $\C$.
Some simple properties are worth pointing out:
\begin{itemize}
\item If we denote by $\Sigma_1$ the cut attached to $\alpha_1$, then for $x\in \Sigma_1$  the loops $\gamma(x)$ are homotopically trivial (in $\mathcal L$), hence the character $\aleph(\gamma(x))=1$;
\item there are $2g+1$ amongst cuts and finite gaps (i.e. excluding the gap that extends to infinity);
\item as $x$ moves along the links of $\Sigma$ then $\gamma(x)$ spans $2g$ homologically independent loops (defined up to orientation) in $\mathcal L$;
\item any $\Z_2$ character on $\pi_1(\mathcal L)$ corresponds (in one-to-one fashion) to a half-integer characteristic on the Jacobian of  $\mathcal L$.
\end{itemize}
These facts imply that we can arbitrarily assign signs to all finite links of $\Sigma$ (except $\Sigma_1$) and then {\bf define} a $\Z_2$ character on $\pi_1(\mathcal L)$ by
 using eqs. (\ref{aleph}). This is consistent since the rank of $\pi_1(\mathcal L)$ is $2g$ and there are (excluding $\Sigma_1$) the same number of finite links in $\Sigma$.

With this preparatory material we can formulate the jump relations of $\Xi$ on $\Sigma$
\bl
The matrix $\Xi_r(x)$ on $\C\setminus \Sigma$ satisfies the following jump relations:
on the {\bf cuts}
\be
\Xi_r(x)_+ = (-1)^{\frac {\sharp_\Delta(x)-1}2} \chi_{\Delta+\nu} \langle X(\gamma(x))\cdot \Sigma_j  \rangle
\Xi_r(x)_- \le(\begin{array}{cc}
0&1\\
-1&0
\end{array}\ri)
\ee
while on the {\bf gaps}
\be
\Xi_r(x)_+ = (-1)^{\frac{\sharp_\Delta(x)}2}\chi_{\nu+\Delta}(\gamma(x)) \Xi_r(x)_-
\ee
\el
The proof is a simple inspection of the properties of $F_r(x), F_r^\star(x)$ and $Q_\Delta$ (using the argument principle). The main point that we make here is that on the gaps the jump is at most a
 sign and on the cuts it is given by $\pmatrix{0&1\cr -1&0}$ up to a sign (which is computed in detail by the formula).

We have remarked above that we can assign arbitrary signs to all finite links of $\Sigma$ except $\Sigma_1$ and lift this assignment to a $\Z_2$ character of $\pi_1(\mathcal L)$. Therefore we have proved
\bp
\label{RHnorm}
If we choose  the orientation of $\Sigma_1$ such that $(-1)^{\frac {\sharp_\Delta(x)-1}2} \langle X (\gamma(x))\cdot \Sigma_1\rangle = 1$ then,
for any choice of $\Delta$ and of orientations of the other cuts/gaps, there is a corresponding (unique) choice of the half-period $\nu$ such that the matrix $\Xi_n$ solves the {\bf sign--normalized} Riemann--Hilbert problem
\bea
\label{RHcut}
&\& \Xi_r(x)_+ =  \Xi_r(x)_- \le(\begin{array}{cc}
0&1\\
-1&0
\end{array}\ri)\qquad \hbox {on the {\bf cuts}} \\
&\& \Xi_r(x)_+ =  \Xi_r(x)_- \label{RHgap}\qquad \hbox{ on the {\bf gaps}}\\
&\& \Xi_r(x) = \mathcal O\le( \frac 1{(x-\alpha_j)^{\frac 1 4}} \ri) ,\ \ x\to\alpha_j\label{RHgrowth}\ ,
\eea
namely it has no jumps on the gaps.
\ep

\section{Part II: asymptotics of orthogonal polynomials}
\label{se:relat}
The construction in section \ref{sec:symmspinors} suggests that the
spinors could be interpreted as the large $N$ asymptotics of some
orthogonal polynomials. In this section we will show that this is
indeed the case. We will make use of the steepest decent method
\cite{D}, \cite{DKV}, \cite{DKV2}, \cite{V}, \cite{its} to approximate
the Riemann-Hilbert problem satisfied by orthogonal polynomials
with semi-classical potentials \cite{BEH1} for large $N$. The
solutions of these  Riemann-Hilbert problems then
represent the strong asymptotics of the orthogonal polynomials in
the large $N$ limit. The main result of this section is that, when
the meromorphic functions $Y$ and $C$ and the Riemann surface
satisfy a certain condition  (Definition \ref{de:adm}), the
spinorial matrix constructed in Sec. \ref{se:spinors} provides the solution to this
deformed Riemann-Hilbert problem and hence gives the strong
asymptotics of the corresponding orthogonal polynomials away from
the branch points of $Y$ where Airy asymptotics must be used instead.

The setting is not as general as in the first part of the paper: the potential will be just polynomial, namely the divisor of poles of $y = Y(p)$ will coincide with that of $x = X(p)$.
Additional complications would arise in presence of other singularities and hard-edges but very interesting and somewhat surprising features are already present for this simplest class of potentials: the general situation is addressed in a future publication.\par\vskip 6pt

Let $(\mathcal L,x,y)$ be a triple such that  $\mathcal L$ is a smooth hyperelliptic surface of genus $g$, $x=X(p)$ is an involution-invariant meromorphic function with two simple  poles and  $y$ is a meromorphic
functions on the Riemann surface $\mathcal L$ represented
as a (nodal)  hyperelliptic relation between  $x$ and $y$
\begin{eqnarray*}
y^2=\Pi_{i=1}^{2g+2}(x-\alpha_i)M^2(x)
\end{eqnarray*}
where the $\alpha_i$ are the (distinct)  $x$-values of  the Weierstrass points and $M(x)$ is a polynomial with roots not coinciding with any of the $\alpha_j$'s.

Let $V(x)=\sum_{1}^{d+1}{{u_i}\over{i}}x^i$ be a polynomial of
order $d+1$ (without constant term) such that
\be
Y(p)\sim {1\over 2}V^{\prime}(X(p))+O(1),\quad
p\rightarrow\infty_+ \label{defpotential}
\ee
on $\mathcal L$. The degrees of $V, M$ and the genus are related by
\be
g+1+\deg (M)  = \deg(V')\ .\label{degrees}
\ee
We define the Stokes' ray for this triple $(\mathcal L,x,y)$ as
follows
\begin{eqnarray}\label{eq:Stokes}
\mathcal R_k&=&\left\{x\in\mathbb{C},\arg(x)\in\vartheta+{{(2k-1)\pi}\over{2(d+1)}}\right\}\nonumber\\
\vartheta :&=&{{\arg(u_{d+1})}\over{d+1}},\quad k=0,\ldots,2d+1
\end{eqnarray}

The situation which will be relevant for our discussion is the following, which we formalize in a definition
\bd
\label{de:Boutroux}
The triple $(\mathcal L,x,y)$ is said to satisfy the {\bf Boutroux condition} if all contour integrals $\oint_\gamma  y\rd x \in i\R$ are purely imaginary.
\ed
Since $y$ is anti-symmetric under the hyperelliptic involution  and has only poles above $x=\infty$, there are $2g+1$ homologically independent classes to consider.
It will be proved later in Section \ref{se:reconstr} that such condition can be fulfilled. In fact these ``Boutroux'' triples will be constructed.

Let $\alpha_1$ be one of the branchpoints: we can define the following function on the curve $\mathcal L$
\be
h(p) = \Re\le(\int_{\alpha_1}^p y(s)\rd s\ri)\ .
\ee
The choice of $\alpha_1$ or any other branchpoint does not affect the definition of $h$ (see second bulleted item after Def. \ref{de:adm}). 
Here we think of $h(p)$ as a function on the hyperelliptic curve itself: the contour of integration is immaterial because of the Boutroux condition (the only additive monodromy of the integral is purely imaginary). Therefore $h(p)$ is a well--defined harmonic function on $\mathcal L\setminus \{\infty_\pm\}$.
If we consider it as a function on the $x$-plane, then it has two branches which differ only by a sign: in particular the harmonic continuation of one branch on the punctured $x$--plane $\C\setminus\{\alpha_j\}_{j=1,\dots, 2g+2}$ has  the exact same multiplicative monodromy of the analytic continuation of $y$ (again because of  the Boutroux condition).
\bl
Under the Boutroux--condition, the function $h(x) = \Re \le(\int_{\alpha_1}^x y(s)\rd s\ri) $ has only multiplicative monodromy with values $\pm 1$ and is otherwise independent of the choice of contour of integration where $y_1(x)$ is the branch of
$y$ that behaves like ${1\over 2}V^{\prime}(x)$ near $x=\infty$. Its zero level set is well defined.
\el
The multivaluedness of $h(x)$ is the same as the multivaluedness of $y(x)$: therefore we can  make appropriate cuts on the $x$-plane for which $y(x)$ becomes single valued on the resulting domain. On the same domain then $h$ will be harmonic, possibly with jump-discontinuities across those cuts.  If $x$ belongs to one of these cuts and $h(x)_\pm$ denote the boundary values on the two sides, we have
\be
h(x)_+ = -h(x)_-\ .
\ee
If we can choose the cuts within the zero-level set of $h$, then $h$ will be {\bf continuous} on the whole plane and {\bf harmonic} away from the cuts. For the time being we formalize this into the following definition.

\begin{definition}\label{de:adm}
Let $(\mathcal L,x,y)$ be a triple such that $\mathcal L$ is represented as a nodal
hyperelliptic curve by
\be
y^2=M^2(x)\Pi_{i=1}^{2g+2}(x-\alpha_i)
\ee
where $\alpha_i$ are distinct and that $M(x)$ is a polynomial with roots not coinciding with any of the $\alpha_j$'s. This triple is called {\bf admissible} (and {\em noncritical}) if
\begin{enumerate}
\item $\oint_{\gamma}ydx$ is imaginary for any closed curve
$\gamma\in\mathcal L$ and $\res{p=\infty_+}y\rd x=1$ ({\bf normalized Boutroux condition}\footnote{The adjective ``normalized'' refers to the normalization of the residue at $\infty_+$; clearly, given a non-normalized Boutroux triple one can get a normalized one by rescaling $y$ by a real constant.}).

\item It is possible to define branch cuts of $y$ in such a way that
\begin{enumerate}
\item all the cuts are {\bf finite} Jordan arcs denoted by $\Sigma_i$
joining two branchpoints; 
\item  $h(x)$ is continuous on the whole
plane and harmonic away from the cuts; 
\item for all cuts $h(x)$ (or $-h$)  is negative on both sides of each cut.
\end{enumerate}
\item  For $\beta_i$ any root of $M(x)$ then $h(\beta_i)\neq 0$ ({\em noncriticality}).
\end{enumerate}
\end{definition}
Since the zero levelset  $$\mathfrak H_0:= \{x|\ h(x)=0\}$$ is independent of the harmonic continuation it makes sense to study its topological properties.

Before proceeding, let us make a few observations of the set
$\mathfrak H_0$ in the presence of the Boutroux condition.

\begin{itemize}
\item The set $\mathfrak H_0$ consists of a finite union of Jordan arcs.
\item All branch-points $\alpha_j$ belong to $\mathfrak H_0$: indeed $\int_{\alpha_i}^{\alpha_j} y dx$ is half of a closed loop on the Riemann--surface $\mathcal L$ and --by the Boutroux condition-- it is thus purely imaginary. Therefore $h(\alpha_j) = \Re\int_{\alpha_1}^{\alpha_j} y \rd x = 0$.
\item Since we assume all the branch points to be simple, there are $3$ arcs originating from each branch point (as a simple computation in a local coordinate shows).
 They can either lead to  another branch point or  towards $\infty$. 
 We call a branch that ends at another
branch point a {\bf closed arc} (or branch)  and one that leads to $\infty$ an {\bf open
arc}.  Moreover, the only self-intersection points of  the set
$\mathfrak H_0$ are the branch points.
\item The set $\mathfrak H_0$ cannot contain a closed finite loop $\gamma$:
if this happened, then there would necessarily be another closed loop  within the region bounded by $\gamma$ and   without any
branch point or singularity inside it. Since then (one branch of) $h$ would be harmonic on this simply connected domain,  continuous on its closure and with zero boundary value, $h(x)$ would vanish identically by the maximum modulus theorem, a contradiction.
\item The function $h(x)$ is continuous across a cut $\Sigma_i$ if and only if
$\Sigma_i\subset \mathfrak H_0$.
\item There can only be an odd number of branch cuts coming out of a
branch point for otherwise the continuation of  $y(x)$ in a neighborhood of the
branch point would have no multivaluedness and the point could not be a branch--point. In our situation we can run either one or three
  cuts at each branch--points while remaining in $\mathfrak H_0$.  If there is only one branch-cut then we cannot choose the branch 
of $h$ so as to have a  definite sign in a neighborhood of the branch-point. If there are $3$ branch-cuts, then we can choose a branch
 of $h$ such that $h$ is {\bf continuous} (not harmonic) with a semi-definite sign in a neighborhood of the branch-point.
\end{itemize}
These simple observations imply  that $\mathfrak H_0$ is a {\bf trivalent} graph
with no closed loop with trivalent vertices at the branchpoints. It may contain possibly some open Jordan arcs not containing any branch-point 
(i.e. extending from $\infty$ to $\infty$). We will dwell at length on the topology of such graphs in Sec. \ref{se:reconstr}.

 The second condition in the definition of admissibility (Def \ref{de:adm}) implies that all
branch cuts belong to $\mathfrak H_0$ (because otherwise $h$ would not be continuous across the cut): therefore none of the
branch points between different connected  components of $\mathfrak H_0$ can be
connected by a branch cut. We then have 
\begin{lemma}\label{le:number}
If $(\mathcal L,x,y)$ is admissible, then each connected component of
$\mathfrak H_0$ contains an even number of branch points.
\end{lemma}

{\bf Proof.} We can form another graph ${\cal B}$ using the branch
points as vertices and the branch cuts as edges. Let the connected
components of this graph be ${\cal B}_i$. We will show that each
${\cal B}_i$ contains an even number of vertices.

 First note that by the last bulleted item in the above list of facts,  each
branch point $\alpha_i\in {\cal B}_i$ can only be connected to an
odd number (specifically $1$ or $3$)  of branch points through branch cuts coming out of it.
Let $\beta$ be a branch point connected to $\alpha_i$
through a branch cut: then $\beta$ can only be connected to an odd
number of branch points. Apart from $\alpha_i$, all of these
points can only be connected to $\alpha_i$ through this branch
point $\beta$, or a closed loop would form in the graph. Therefore
the branch points that are connected to $\beta$ will add an even
number of new branch points to the connected component of ${\cal
B}_i$ of the graph.

By repeating this argument, we see that the total number of branch
points in this component must be even. {\bf Q.E.D.}

\bd
We will call a branch
point with $0,1,2$  open arcs incident to it,  a branch point (vertex) of {\bf type I,
II or  III} respectively.
\ed

In the set $\mathfrak H_0$ there are open arcs which do not contain any branch-points: removing those arcs yields a trivalent graph $\mathfrak X_0$  (the {\bf critical graph}) with some {\bf open edges} (i.e. edges attached to only one vertex).
We can now define the branch cut structure of such  a trivalent tree-like (or forest-like) graph.
\begin{definition}\label{de:cut}
Let $\mathfrak X_0$ be a connected trivalent graph with no closed loop and open end-edges. Then the {\bf branch cut structure} of
the graph $\mathfrak X_0$ is  a subgraph ${\cal B}$ of $\mathfrak X_0$ containing all its vertices and such that
\begin{enumerate}
\item  Each vertex has either $1$ or $3$ edges incident to it,
\item  $\mathcal B$ has no open-edges.
\item Each connected component of ${\mathcal B}$ has an even number of vertices.
\end{enumerate}
The edges of ${\cal B}$ will be called {\bf branch cuts}.
The definition applies also to graphs $\mathfrak X_0$ with several connected components provided each component has  the aforementioned properties.
\end{definition}

The following lemma shows that a branch cut structure for a graph
with an even number of vertices exists and is unique.
\begin{lemma}\label{le:even}
Let $\mathfrak X_0$ be a  graph with an even number of vertices
$\alpha_1,\ldots,\alpha_{2k}$ as  in
definition \ref{de:cut}. Then the branch cut structure ${\cal B}$ of $\mathfrak X_0$ exists and is unique.
\begin{figure}
\resizebox{15cm}{!}{ 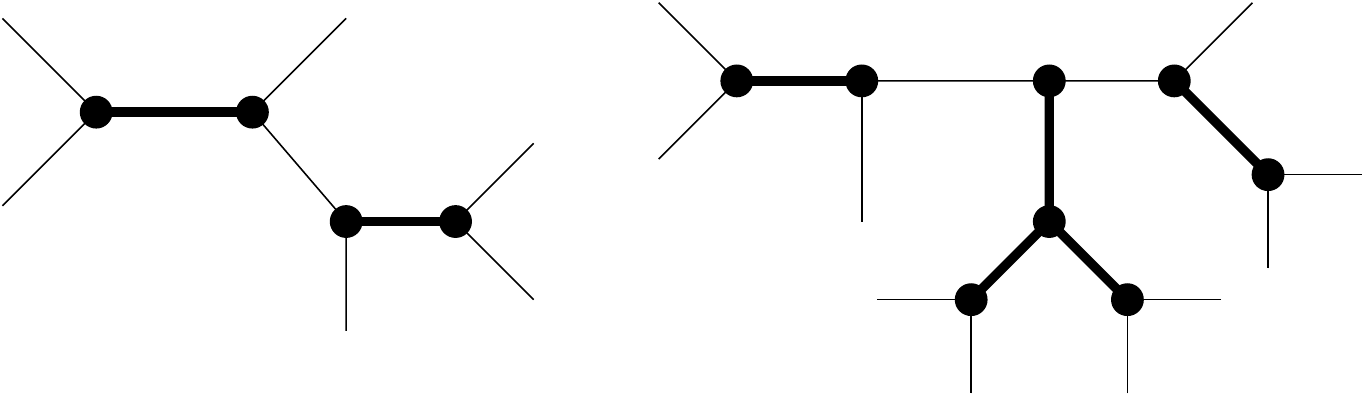}
\caption{A trivalent open graph with the properties of Def. \ref{de:cut} and its branchcut structure (thick lines).}
\end{figure}
\end{lemma}
{\bf Proof.}
We can reason independently on each connected component of $\mathfrak X_0$, so there is no loss of generality in assuming $\mathfrak X_0$ to be connected.

First remove all open edges and call $\dot \mathfrak X_0$ the result of this first pruning: it is still connected and  with only trivalent, univalent  and bivalent vertices.

If the resulting graph $\dot \mathfrak X_0$ does not contain any
bivalent vertices then  ${\cal B}=\dot
\mathfrak X_0$. 

Viceversa, if $\dot \mathfrak X_0$ contains some bivalent vertices, then 
each of these belongs to a maximal chain of bivalent vertices; the chain connects two subgraphs of $\dot \mathfrak X_0$ which we denote by  $L$ and $R$ (also with only uni/bi/trivalent vertices). 
They connect to the chain by an oddvalent vertex (if the vertex of --say-- $L$ is univalent, this means that $L$ itself is a single vertex).

Now, if the chain has an even number of vertices, then either both  $L$ and $R$ have an even number of vertices or they both have an
odd number of vertices; we need to remove every second link in the chain so that each bivalent vertex is turned into an univalent 
one, and there is only one way of doing so  such that all the
resulting graphs contain an even number of vertices.

If the chain has an odd number of vertices, then either $L$ or (exclusively) $R$ have also an odd number of vertices; once more there is only one way of removing every second link in the chain so as to leave each subgraph with an even number of vertices.

The result of this second  pruning is a collection of elementary segments (two univalent vertices joined by a link, which is what remains of the bivalent chain)
  and the two subgraphs $L$ and $R$ (possibly with one extra vertex on one of the two if the chain was odd).
 By repeating this procedure we could reduce the graph into a
collection of subgraphs $\{\dot \mathfrak X_i\}$ which do not
contain any bivalent vertex. The branch cut structure ${\cal B}$
of $\mathfrak X_0$ is then given by ${\cal B}=\cup_i\dot \mathfrak
X_i$.
 {\bf Q.E.D.}
\par \vskip 5pt

In order to apply this construction to an admissible triple $(\mathcal L,x,y)$ we adopt the following strategy:
\begin{itemize}
\item  Define $h(x) = \Re\int^x y_1\rd s$, where $y_1$ is the
branch of $y$ that behaves like ${1\over 2} V^{\prime}(x)$ near
$x=\infty$ with some arbitrarily chosen branch cuts. We can then
define $\mathfrak H_0 = \{h(x)=0\}$; 
\item remove the arcs of
$\mathfrak H_0$ that contain no branch-point of $y$ and call
$\mathfrak X_0$ the resulting trivalent tree-like graph.
\item The branch-cut structure $\mathcal B$ of $\mathfrak X_0$ can be used
to {\em redefine} the branch-cuts of $y_1$ (whence the name of
``branch-cut structure''). These branch-cuts may be
``nonstandard''. Indeed it may happen that a branchpoint has {\bf
three} branch-cuts connecting it to as many other branch-points.
Although unusual this is consistent with the multi-valuedness of
$y$.
\end{itemize}
The admissibility requirements then is the condition that the
locally harmonic function $h$ could be chosen such that
\begin{enumerate}
\item $h$ is continuous on the $x$ plane and harmonic away from the branch-cuts;
\item the sign on {\bf both sides} of {\bf each} branch-cut is negative.
\end{enumerate}
At this point the reader may wonder how strong a condition is this and if there are any examples; in Sec. \ref{se:reconstr} we will show how to reconstruct an arbitrary admissible triple from the topological structure of its (admissible) graph $\mathfrak X_0$. For the time being we present only two figures exemplifying two curves of the same genus satisfying the Boutroux condition but one being admissible and the other non-admissible.

\begin{figure}
\resizebox{15cm}{!}{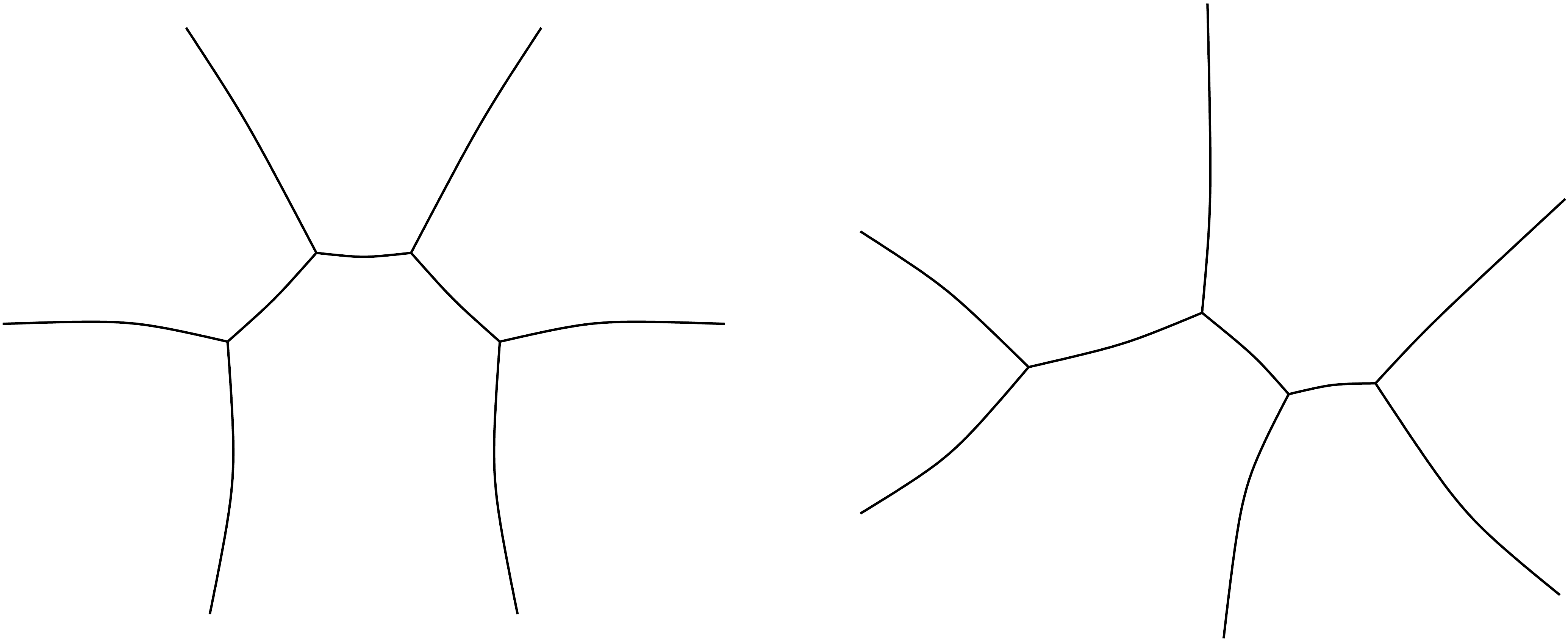}
\caption{
The critical graph $\mathfrak H_0 = \mathfrak X_0$ of two Boutroux curves: the one on the left is admissible (marked are the signs of $h$), the right is not admissible.}
\label{fig:admiss}
\end{figure}

\bx
The first example is an admissible triple (left in Fig. \ref{fig:admiss}, with
\be
y^2 = (x+1+r)(x-1-r) (x-{\rm e}^{i\pi/3}) (x-{\rm e}^{2i\pi /3})\ ,\qquad, r \simeq 0.4144...
\ee
The second curve is not admissible, although it satisfies the Boutroux condition (rather approximately, on the left of Fig. \ref{fig:admiss})
\be
y^2 \simeq (x-2-0.15 i)(x+2-0.6 i)(x-1)(x-0.88i)
\ee
\ex

For an admissible triple, we will
choose branch cuts of $y$ that satisfy condition 2 in Def. \ref{de:adm},
which are the ones defined by the branch cut structure ${\cal B}$
of the graph $\mathfrak X_0$.
\subsection{The $G$-function}
\label{Gfunction}
\begin{figure}[htbp]
\begin{center}
\resizebox{15cm}{!}{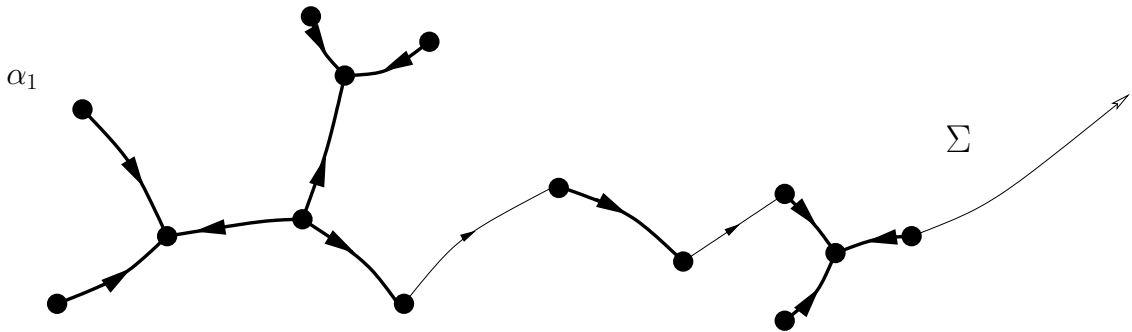}
\caption{The set $\Sigma$ connecting the different components of the branchcuts $\mathcal B$.
}
\label{fig:sigma}
\end{center}
\end{figure}

The main character of the construction is (as in \cite{its, DKV}) the so-called $G$-function. In this context it is simply the Abelian integral $\int y\rd x$; of course attention must be paid to its multivaluedness (which is only additive on the Riemann surface $\mathcal L$).

The construction of the previous section has yielded a forest  graph $\mathcal B = \sqcup_i \mathcal B_i$ made of several connected components each of which is  a loop-free treelike closed graph with only odd-valent (1 or 3) vertices.

Let us order these components in some arbitrary fixed way; using some arcs not intersecting $\mathcal B$ we then connect a vertex of $\mathcal B_1$ with a  vertex of $\mathcal B_2$, then one of $\mathcal B_2$ to one of $\mathcal B_3$ and so on and so forth, adding an open arc that extends to $\infty$ from the last connected component of $\mathcal B$ : we call $\Sigma$ the resulting set (see Fig. \ref{fig:sigma}).
We will call these arcs joining the consecutive $\mathcal B_i$'s the {\bf complementary arcs} or {\bf gaps} and denote them by $\wt \Sigma_j$, whereas we denote by $\Sigma_i$ the edges of $\mathcal B$.

 The set $\dot \C:= \C\setminus \Sigma$ is simply connected and hence we can define unambiguously
\be
G^{(i)}(x) = \int_{\alpha_1}^x y_i(s)\rd s\ ,\ \ \ i=1,2,
\ee
where $y_1 = -y_2$ are the two branches of $y$ and $\alpha_1$ is  a univalent vertex of $\mathcal B_1$ and $y_1\sim \frac 12 V'(x)$
 at $x=\infty$.
 The contour of integration is  taken to lie within the simply connected domain $\dot \C$.

We also assume that the edges of $\mathcal B$ have been oriented, so as to be able to distinguish the left from the right side. We then formulate

\begin{lemma}
\label{le:monG}
The function $G(x):= G^1(x,\Sigma)$ has jump discontinuities on $\Sigma$
as follows.
\begin{eqnarray}\label{eq:monG}
G (x)_+&=&G (x)_-+i\wt \sigma_j,\quad
x\in \wt \Sigma_j\nonumber \\
G (x)_+&=&-G (x)_-+i\sigma_j,\quad x\in \Sigma_j
\end{eqnarray}
where $\sigma_i$ and $\wt \sigma_i$ are real.
\end{lemma}
{\bf Proof}. 
Let us first consider a gap $\wt \Sigma_j$ and $x\in \wt\Sigma_j$: the two paths from $\alpha_1$ to $x$ on the left/right of the gap lift to a closed  loop $\gamma$ on the curve $\mathcal L$ because it encircles some number of connected components $\mathcal B_1, \dots, \mathcal B_j$, each of which contains an even number of branchpoints. Therefore
\be
G (x)_+ = G (x)_ - +  \int_\gamma y\rd x\ ,
\ee 
which proves the first identity, with $\wt \sigma_j = -i \int_\gamma y\rd x$.

In order to prove the second identity we first note that 
\bea
&& G^1 (x)_\pm = -G^2(x)_\pm\ \ \hbox { on a branchcut}.
\eea
Let $x\in \Sigma_j$ (an edge of  $\mathcal B$); a closed loop on the Riemann surface of $y$ consists of a contour joining $\alpha_1$ to $x$ on the 
left of the cut in the first copy of $\dot C$ and a contour joining $\alpha_1$ to $x$ on the right of the cut on the other copy of $\dot C$. Therefore 
\be
G^1 (x)_+ = G^2(x)_- + \int_\gamma y\rd x\ .
\ee  
From this the first assertion follows immediately with $\sigma  = -i \int_\gamma y \rd x$. {\bf Q.E.D.}\par \vskip 5pt

Note that $h(x) = \Re G^1(x)$ and has no discontinuities (as we know already) on the gaps $\wt \Sigma_j$, where it is actually harmonic, and is continuous on the cuts $\Sigma_j$'s where --however-- is not differentiable (but it admits normal derivative on both sides).

From the definition of  Stokes' ray (\ref{eq:Stokes}), we can
define sectors ${\cal S}_k$ in
$U=\mathbb{C}/(\cup_{1}^{g+1}\Sigma_i)$ such that

1. As $x\rightarrow\infty$, the sector ${\cal S}_k$ is bounded by
$\mathcal R_{2k}$ and $\mathcal R_{2k+1}$ (\ref{eq:Stokes})

2. Let the ${\cal Y}_+$ be the set
\begin{eqnarray*}
{\cal Y}_+=\left\{x\in U, h(x)> 0\right\}
\end{eqnarray*}
then each sector is asymptotically contained in ${\cal Y}_+$.

We will need later the following counting of edges and vertices
\bl
\label{edgecount}
Let $\mathcal B = \sqcup_{i=1}^K \mathcal B_i$ be the decomposition of the branchcut structure into connected components. Let $b_i$ be the number (which is even) of vertices and $e_i$ the number of edges in $\mathcal B_i$.
Then
\be
e_i = b_i-1
\ee
Therefore the total number of edges $E$ is
\be
E = 2g+2 - K\ .
\ee
\el
{\bf Proof}. The first formula follows immediately by induction on the number of vertices; the second follows by considering that the sum $\sum_{i=1}^K b_i = 2g+2$.  {\bf Q.E.D.}\par \vskip 5pt

\subsection{The Stokes--Kirchoff normalized differential of second kind}
A crucial r\^ole will be played by a suitable  normalized differential of the second kind which we baptize ``Stokes'' because of its relation (to be shown) with the Stokes matrices.
\bd
\label{de:stokesdiff}
A {\bf Stokes differential} is a second-kind differential $\rd w$  with the following properties
\begin{enumerate}
\item it is antisymmetric w.r.t. the hyperelliptic involution;
\item it has  poles (without residues) of degree at most $g+1$ at $\infty_\pm$;
\item all periods around the connected components of the branchcut structure $\mathcal B$ are zero (or integer multiples of $2i\pi$).
\end{enumerate}
\ed
Some {\bf properties} follow immediately:
\begin{itemize}
\item The Stokes differential belongs to a $\Z^{K-1}$ lattice of affine spaces of dimension $2g+1 - K$. Indeed there are a span of $2g$ antisymmetric second-kind differentials with  poles of that order at $\infty_\pm$ and without residue; imposing that the contour-integrals around each of the connected components is a multiple of $2i\pi$ (the residue at $\infty_\pm$ is zero so there are only as many loops as the finite  gaps)  imposes $K-1$ affine constraints.
\item Since the periods are zero (mod $2i\pi \Z$) around $\mathcal B_i$'s, its Abelian integral $w(x) = \int_{\alpha_1}^x \rd w$ defines a {\bf single valued} (mod $2i\pi \Z$)  function on $\C\setminus \mathcal B$  with a pole of order at most $g+1$ at infinity.
\item By using similar arguments used for the $G$-function (together with the antisymmetry), we can show that it solves a certain scalar Riemann--Hilbert problem
\bea
w(x)_+ &\& = -w(x)_- + i \mu_j \ ,\qquad x\in \Sigma_j\label{RHPw} \\
\mu_1 &\& = 0\nonumber
\eea
on the edges of $\mathcal B$. Here $\Sigma_1$ is the edge attached to $\alpha_1$ (the Weierstrass basepoint of the integration defining $w$); $\mu_1$ is zero because       $\alpha_1$ and $x\in \Sigma_1$ can be joined by a {\em contractible} loop whose projection on the $x$-plane intersects $\Sigma$ only at $x$ and $\alpha_1$.
\end{itemize}

The other important condition that we need to impose is specified in the next definition; note first
that since there are no closed loops in the graph $\mathcal B$ we can always choose orientations of the edges of $\mathcal B$ in such a way that all the orientations of edges incident to any vertex are all incoming or outgoing.
 We assume one such  orientation in the following definition.
\bd
\label{de:Kirchoff}
The Stokes differential $\rd w$ is said to satisfy the {\bf (complex) Kirchoff's law} if for each trivalent edge of $\mathcal B$, denoted $\Sigma_{i_1},\Sigma_{i_2}, \Sigma_{i_3}$ the three incident edges (with their chosen orientation), then
\be
{\rm e}^{i \mu_{i_1}} + {\rm e}^{  i \mu_{i_2}} + {\rm e}^{ i \mu_{i_3}} = 0
\ee
where all edges are oriented so that at each trivalent vertex they are all incoming or outgoing.
\ed

It should be clear that such Kirchoff-Stokes differentials do exist, since the Kirchoff's constraint poses only some number of nonlinear constraints. We need the exact count of these constraints.

If $b_i$ is the number of vertices in $\mathcal B_i$ then there are
$\frac {b_i}2 +1$ univalent vertices  (of type I) and
hence $\frac {b_i}2 -1$ trivalent vertices  (recall that $b_i$ is
even). Summing up over all connected components of $\mathcal B$ we
find the total number $T$ of trivalent vertices \be T  =
\sum_{i=1}^K \le(\frac {b_i}2 -1\ri) =  g+1-K. \ee Therefore \bp
\label{Kirkdim} The {\bf Kirchoff--Stokes} differentials  on a
Boutroux curve form a $\Z^{2g+1-K}$ lattice of manifolds of
dimension $g=genus(\mathcal L)$. \ep The lattice aspect is due to
the obvious fact that we can arbitrarily add integer multiples of
$2\pi$ to each $\mu_j$.

\subsection{Asymptotics}
The following lemma shows that the spinors constructed in section
\ref{se:spinors} satisfy a Riemann-Hilbert problem closely
related to the one satisfied by $G$. This will allow us to express
the large $N$ asymptotics of certain orthogonal polynomials in
terms of these spinors.

The relation requires that we specify the choice of the flat line bundle $\mathfrak L_\eta$ of Sec. \ref{se:spinors} associated to a Stokes--Kirchoff differential: we will set 
\be
\eta =\rd w +  N y\rd x \label{Linf}\ .
\ee
Note that this $\eta$ has residue $\pm N\in \Z$ at $\infty_\pm$ (because of the normalized Boutroux condition).  
We will call the line--bundle associated with this choice of Stokes--Kirchoff differential the {\bf asymptotic line bundle} and denote it with $\mathfrak L_\infty$.

We note that it is the tensor product of a line bundle $\mathfrak L_{\rd w}$ with the line--bundle $\mathfrak L_{Ny\rd x}$; this last one, because of the Boutroux condition and since $N\in \N$, is a {\bf unitary} line-bundle.
\begin{lemma}
\label{le:monspinor}
Let $\Sigma$ be defined as in Sec. \ref{Gfunction} and the  half integer
characteristics chosen as specified in Prop. \ref{RHnorm} for the given $\Delta$ and choice of orientations of
 edges of $\mathcal B$\footnote{The orientation of the edge $\Sigma_1$ attached to the basepoint of integration forces the orientation of 
the edges of the connected component of $\mathcal B$ to which it belongs (all edges should be either incoming or outgoing from the trivalent vertices).}. Then  the matrix
\begin{eqnarray}
\Psi:=\Psi_{N,r}  =\le[
\begin{array}{cc}
\frac 1{c_r} & 0 \\
0 & -i c_{r-1}
\end{array}
\ri]\Xi_r(x) {\rm e}^{(N G + w(x))\sigma_3}
\label{eq:monspinor}
\end{eqnarray}
(where the constants $c_r$ are specified in the proof: see eq. (\ref{Ximatrix}) for the definition of $\Xi_r(x)$)
satisfies the following Riemann-Hilbert problem
\begin{eqnarray}
\Psi_+(x)&=&\Psi_-(x)\pmatrix{0&e^{-iN\sigma_j -i\mu_j}\cr
         -e^{iN\sigma_j + i\mu_j}&0\cr}, \quad
x\in\Sigma_j \cr
\Psi_+(x)&=&\Psi_-(x)\pmatrix{e^{-iN\wt\sigma_j}&0\cr
0&e^{iN\wt \sigma_j}\cr},\quad x\in \wt \Sigma_j \cr
\Psi(x) &=& \mathcal O \le(\frac 1{(x-\alpha_j)^{\frac 14}}\ri)\ ,\qquad x\to\alpha_j\cr
\Psi(x)&=&
(\1 + \mathcal O(x^{-1}) x^{r \sigma_3} ,\quad x\rightarrow\infty
\label{eq:jumpcond}
\end{eqnarray}
and $\Psi(x)$ is holomorphic on $\dot \C = \C \setminus \Sigma$.
\end{lemma}
\br
An essentially identical RH problem appears in the work \cite{DIKZ}, where it was similarly reduced to the problem of Prop. \ref{RHnorm}. The idea perhaps was first used in \cite{DZ}.
\er
{\bf Proof of Lemma \ref{le:monspinor}.} 
The growth condition near the branchpoints follows from (\ref{RHgrowth}).

From  the general definition in eq. (\ref{spinorchar})
specialized to the differential (\ref{Linf}) we see that the matrix $\Psi$ is analytic in a punctured neighborhood of $\infty$. Indeed the essential singularity of the spinors $\pi_r, \pi_r^\star$ at $\infty$ is removed by the multiplication by the exponential $\exp \int \eta\sigma_3$. This leaves us with a power growth at $\infty$; specifically
\be
\frac {\pi_r(p(x)){\rm e}^{\int_{\alpha_1}^{p(x)} Ny\rd x + \rd w}} {\sqrt{\rd x}}\sim  c_r x^r\ ,\qquad x\to \infty \label{polp}
\ee
has a pole of degree $r$ at $\infty_+$ (which is on the sheet chosen for $p = X^{-1}(x)$) and a zero of order $r$ at $\infty_-$. This is so because (by their definition) the spinors $\pi_r {\rm e}^{\int \eta}$ have a pole of order $r+1$ at $\infty_+$ and a zero of order $r-1$ at $\infty_-$; since $\rd X$ has a double pole at both $\infty_\pm$ and appears in the square--root, (\ref{polp}) follows. Similar considerations show that
\be
\frac {\pi_r^\star(p(x)){\rm e}^{ - \int_{\alpha_1}^{p(x)} Ny\rd x + \rd w}}{\sqrt{\rd x}} \sim  \frac 1 {c_r} x^{-r-1}\ ,\qquad x\to \infty \label{polpstar}\ .
\ee
Formulas (\ref{polp}, \ref{polpstar}) {\em define} the constants $c_r$; the reason why $c_r$ appears in the denominator in (\ref{polpstar}) is simply that --by construction of Serre duality--
\be
1 = \res{\infty_+} \pi_r\pi_{r}^\star  = \res{\infty}\bigg( c_r x^{r} + \mathcal O(x^{r-1})\bigg) \bigg( \frac 1{c_r} x^{-r-1} + \mathcal O(x^{-r-2})\bigg) \rd x
\ee

If we remove the exponential part from $\Psi$ then it satisfies the Riemann--Hilbert problem (\ref{RHcut}, \ref{RHgap}); with the exponential part the character specified by
\be
\begin{array}{rcl}
\exp\oint_\bullet \eta:\pi_1(\mathcal L)&\longrightarrow& \C^\times\\
\gamma & \mapsto & {\rm e}^{\oint_\gamma Ny\rd x+\rd w}\ .
\end{array}
\ee
appears as the jump relations of the Abelian integral $\int Ny\rd x + \rd w$, thus yielding the proof. {\bf Q.E.D.}\par \vskip 5pt

For completeness we report the expressions for the entries of the matrix $\Psi_{N,r}$. 
The two vector complex characteristics ${\epsilon}_{_N}$ and $ (- {\delta}_{_N})$ are half of   the $a$ and $b$-periods (on the chosen dissection of the curve in terms of a homology basis) of the differential $Ny\rd x + \rd w$  following the notation of eq. (\ref{spinorchar})\footnote{We only add a subscript $N$ to emphasize the dependence on the large parameter $N$.}:
\bea
\le[{\epsilon}_{_N}\ri]_j = \frac 1 2 \oint_{a_j} \big(Ny\rd x + \rd w\big)\cr
\le[{\delta}_{_N}\ri]_j = -\frac 1 2 \oint_{b_j} \big(Ny\rd x + \rd w\big)
\label{characters}
\eea

After straightforward simplifications,  the  entries of $\Psi_{N,r}$  read (using (\ref{entries}) specified to the case of line bundle associated to $\eta = Ny\rd x + \rd w$)
\bea
\le(\Psi_{N,r}\ri)_{11} &\& =  \wt c_r
  \frac {\Theta_\Delta^r (
  p-\infty_-)}{\Theta_\Delta^{r+1}(p-\infty_+)}
  \frac{\Theta\le[{\mathcal A+ {\epsilon}_{_N} \atop \mathcal B+{\delta}_{_N} }\ri]
  (p+r\infty_- -(r+1) \infty_+  )}{ \Theta\le[{ \mathcal A +{\epsilon}_{_N} \atop \mathcal B+{\delta}_{_N} }\ri]
  (r(\infty_- - \infty_+)  )}Q_\Delta(X(p))\ ,\cr
  \le(\Psi_{N,r}\ri)_{12} &\& = i \wt c_r
  \frac {\Theta_\Delta^r (
  p-\infty_+)}{\Theta_\Delta^{r+1}(p-\infty_-)}
  \frac{\Theta\le[{- \mathcal A - {\epsilon}_{_N} \atop -\mathcal B - {\delta}_{_N} }\ri]
  (p+r\infty_+ -(r+1) \infty_-  )}{ \Theta\le[{-\mathcal A -{\epsilon}_{_N} \atop -\mathcal B- {\delta}_{_N} }\ri]
  (r(\infty_+ - \infty_-)  )}Q_\Delta(X(p))\cr
  \le(\Psi_{N,r}\ri)_{21} &\& = \frac i{\wt c_{r-1}}
  \frac {\Theta_\Delta^{r-1} (
  p-\infty_-)}{\Theta_\Delta^{r}(p-\infty_+)}
  \frac{\Theta\le[{\mathcal A+ {\epsilon}_{_N} \atop \mathcal B+{\delta}_{_N} }\ri]
  (p+(r-1)\infty_- -r \infty_+  )}{ \Theta\le[{ \mathcal A +{\epsilon}_{_N} \atop \mathcal B+{\delta}_{_N} }\ri]
  (r(\infty_- - \infty_+)  )}Q_\Delta(X(p))\ ,
  \cr
  \le(\Psi_{N,r}\ri)_{22} &\& =\frac 1{\wt c_{r-1}}  
  \frac {\Theta_\Delta^{r-1} (
  p-\infty_+)}{\Theta_\Delta^{r}(p-\infty_-)}
  \frac{\Theta\le[{- \mathcal A - {\epsilon}_{_N} \atop -\mathcal B - {\delta}_{_N} }\ri]
  (p+(r-1)\infty_+ -r \infty_-  )}{ \Theta\le[{-\mathcal A -{\epsilon}_{_N} \atop -\mathcal B- {\delta}_{_N} }\ri]
  (r(\infty_+ - \infty_-)  )}Q_\Delta(X(p))\ ,
  \label{asymport}\\
&&  \wt c_r:= \frac {{C_\Delta}^{r+1}
}{\Theta_\Delta^r(\infty_+-\infty_-)}\ , 
\eea
 where $C_\Delta$ was
introduced in (\ref{Cdelta}). The constants $\{\wt c_r\}_{r\in \Z} $
that appear in the formul\ae\ above have been determined by the
normalization that $\Psi_{1,1}  = x^r(1+\mathcal O(1/x))$ and
$\Psi_{2,2} = x^{-r} (1+\mathcal O(1/x))$.

\subsection{Semiclassical generalized orthogonal polynomials}

The main theorem of this section is that --broadly speaking-- when $(\mathcal L,x,y)$ is an admissible triple
 the spinors $\pi_n$ associated with it give the large
$N$ asymptotics of some orthogonal polynomials of the type considered in \cite{BEH1}.

These polynomials are defined and related to our Boutroux-admissible curve as follows:
let $(\mathcal L,x,y)$ be an admissible triple as in Def. \ref{de:adm} and let us choose a Kirchoff-Stokes differential $\rd w$ (and corresponding Stokes function $w$) as in Def. \ref{de:stokesdiff} and with the same notations as in the remarks that follow that definition.

Each component $\mathcal B_i$ with $b_i$ vertices has  $b_i/2 + 1$ type-III  (univalent) vertices on the boundary of as many distinct components $\mathcal Y_j$ of $\mathcal Y_+$, each of which contains (at least) one Stokes sector; we can choose $b_i/2 +1  $ distinct oriented contours $\Gamma_{i,\ell}, \ell=1,\dots, \frac {b_i}2 $  connecting one Stokes sector to the remaining ones and passing through the edges of $\mathcal B_i$.
It is easily seen that each edge belongs to at least one such contour.

\begin{figure}
 \resizebox{6cm}{!}{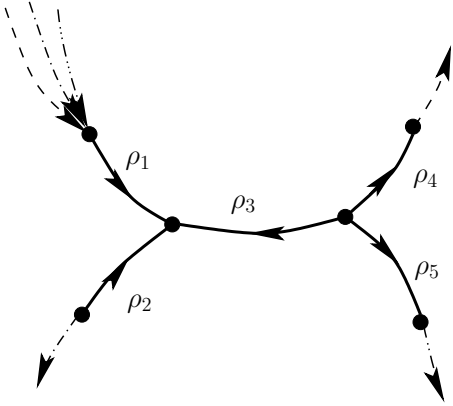}
\caption{A connected component $\mathcal B_i$ with $b_i=6$ vertices, $b_i - 1=5$ edges and $b_i/2-1 = 2$ trivalent vertices. Shown are the net currents $\rho_1, \dots, \rho_5$ through the edges, satisfying the $2$ Kirchoff constraints; we can find always appropriate complex currents to the contours (in textured linestyle) so as to give the desired net currents through the edges.}
\label{Kirchoff}
 \end{figure}

We associate a complex weight $\varkappa_{i,\ell}$ to each oriented contour $\Gamma_{i,\ell}$ as follows;
the periods of $\rd w$ which enter its RHP (\ref{RHPw}) define ``complex currents''
\be
\rho_j := {\rm e}^{-i\mu_j}\ ,
\ee
on each edge $\Sigma_j$ of $\mathcal B$. Because of the Kirchoff's condition
in Def. \ref{de:Kirchoff} we can always find other complex currents $\varkappa_{i,\ell}$ to associate to the contours $\Gamma_{i,\ell}$ in such a way that the ``net current'' through each link is precisely $\rho_j$. Note that $\rho_1 = 1$ always\footnote{This will mean that one of the Stokes matrices is normalized, which we can always accomplish by a conjugation.}
(see for example Fig. \ref{Kirchoff}): note that the weights $\varkappa_{i,\ell}$ do not depend on the $2\pi\Z$ arbitrariness entering the definition of the $\mu_j$'s.

\begin{figure}[!h]
\resizebox{11cm}{!} {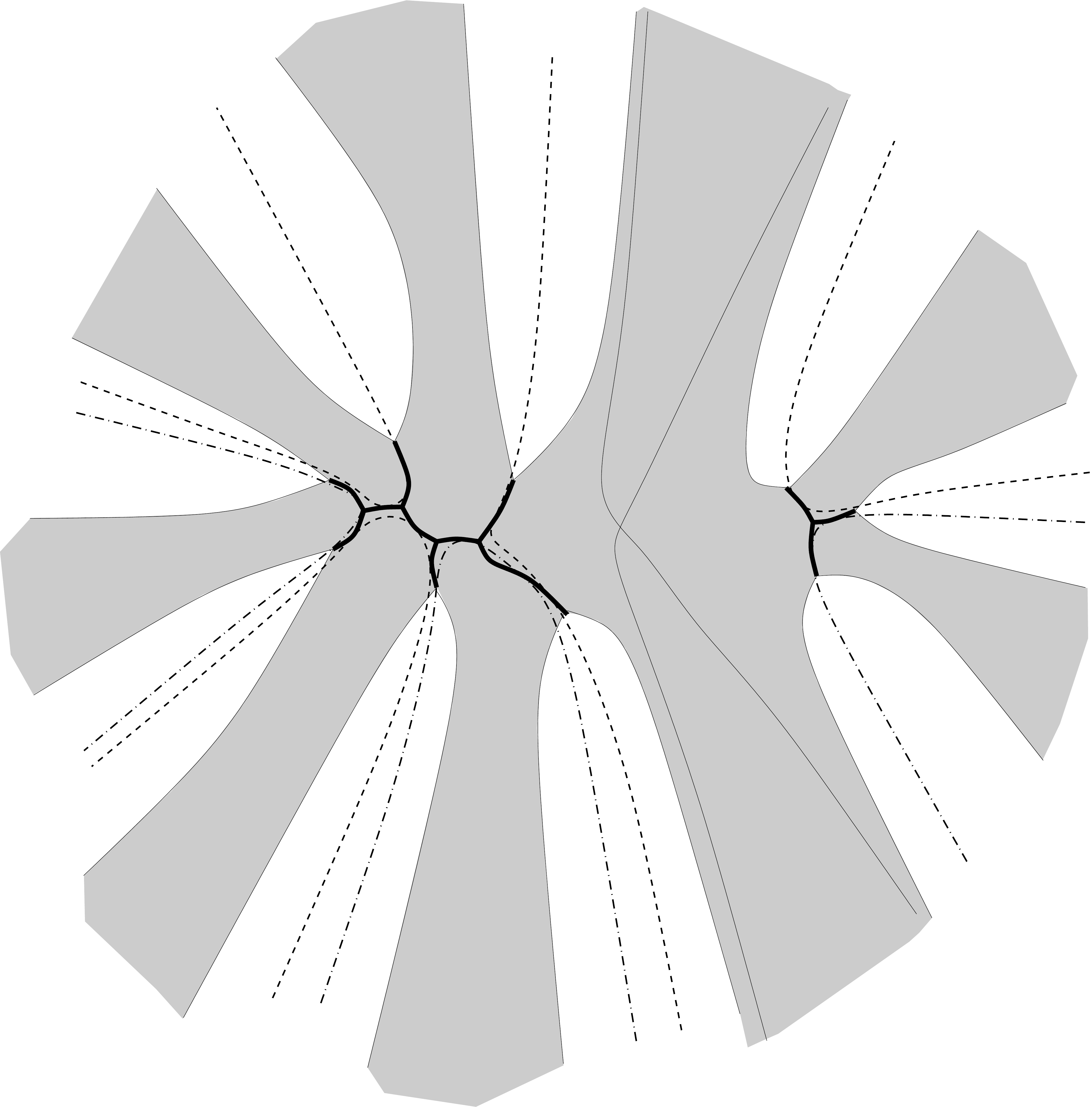}
\caption{In this figure (which is only suggestive and not a numerical output) the grey area is where $h<0$ and the black thick lines are where $h=0$ and constitute the branchcut $\mathcal B$; here the function $h$ is continuous but not harmonic. The thin line in the grey area denotes the levelset of $h(x)=c<0$ passing through a saddle-point. 
Since there are $9$ sectors within as many white areas the picture corresponds to some potential $V(x)$ of degree $9$.
 One should conveniently think of the grey area as the ``sea'' and the white areas as the emerged ``continents''. The paths $\Gamma_{k,j}$ should connect the infinity within a continent to the infinity within another continent never going on the sea; for this purpose they can pass through the branchcuts as ``causeways''. In each (oriented) path runs a complex current $\varkappa_{k,j}$ and the net currents within the edges of $\mathcal B$ automatically satisfy Kirchoff's law. The only genericity requirement is that none of the  net currents ought to vanish.
   In this picture there are only two components $\mathcal B_1$ and $\mathcal B_2$ (left and right respectively) with $10$ and $4$ branchpoints; $\mathcal B_1$ connects $6$ continents, each of which contains a sector where $\Re(V)\to +\infty$, whereas $\mathcal B_2$ connects $3$ such continents.  
 The paths $\Gamma_{k,j}$ (dashed and dot-dashed) once defined can be deformed arbitrarily respecting the connectivity they provide between Stokes' sectors without affecting the values  of 
   (\ref{eq:ortho}) 
   due to Cauchy's theorem (they are already depicted after the deformation but the reader should think of them at first as going exactly through the thick links).
   The existence of a situation with all the topological features of this example (and many others) will be proved in the last part of the paper.
   }
\label{Fig5bis}
\end{figure}

We repeat this procedure for all connected components $\mathcal B_1,\dots \mathcal B_K$ of the branchcut structure  $\mathcal B$.
A visual example is contained in Fig. \ref{Fig5bis} and its detailed caption.

Define now the integral operator
\be
\int_\varkappa := \sum_{i=1}^K \sum_{\ell = 1}^{b_i/2} \varkappa_{i,\ell} \int_{\Gamma_{i,\ell}}\ .
\ee
Recall that the Boutroux curve defines a ``potential''  $V(x)$ via (\ref{defpotential}). 
Let $p_n(x)$ be the {\bf monic} orthogonal polynomials such that
\begin{eqnarray}\label{eq:ortho}
{1\over {2\pi i}}\int_{\varkappa}\rd x p_m(x)p_n(x)e^{-N V(x)}=h_n\,\delta_{nm}\ .
\end{eqnarray}

Denote by $\phi_n(x)$ and $\tilde{\phi}_n(x)$ the following
functions
\begin{eqnarray*}
\phi_n(x)&=&p_n(x)e^{-{1\over 2}NV(x)}\\
\tilde{\phi}_n(x)&=&\frac{e^{{1\over 2}N V(x)}}{2i\pi}\int_{\varkappa}{{e^{-{1\over
2}N V(z)}\phi_n(z)}\over{z-x}}dz
\end{eqnarray*}

With these preparatory notations and remarks we are ready to formulate the main theorem of the section.
\begin{theorem}[Strong asymptotics]
We have the following asymptotic estimates
\label{stron}
\begin{itemize}
\item
For any compact set not contained in $\C \setminus \mathcal B $,  the following matrix-valued function 
\begin{eqnarray*}
\Phi_{N,r}(x):=\pmatrix{\phi_{N+r}(x)&\tilde{\phi}_{N+r}(x)\cr
\frac {2i\pi}{h_{N+r-1}}\phi_{N+r-1}(x)&\frac {2i\pi}{h_{N+r-1}}\tilde{\phi}_{N+r-1}(x)\cr}
\end{eqnarray*}
has the asymptotic behavior as $N\to \infty$ and $r\in \Z$ fixed
\begin{eqnarray}
\label{eq:away}
\Phi_{N,r}\sim\Psi_{N,r}(x){\rm e}^{-NG (x)\sigma_3}\ ,\quad
         x\in\mathbb{C}/(\cup_{i=1}^{g+1}\Sigma_i)
\end{eqnarray}
away from the branch cuts. The matrix  $\Psi_{N,r}$ was introduced in (\ref{eq:monspinor})) and the entries given by (\ref{asymport}).

\item On each edge  $\Sigma_i$ of $\mathcal B$  the asymptotic behavior as  $N\to \infty$ is given by
\begin{eqnarray}\label{eq:on}
\Phi_{N,r}\sim\le(\Psi_{N,r} (x)\pmatrix{e^{-NG (x)}&0\cr \pm{\rho_i}^{-1}e^{NG (x)}&e^{NG (x)}}\ri)_\pm,
         \quad x\in \Sigma_i
         \end{eqnarray}
\end{itemize}
\end{theorem}
%

{\bf Proof of Thm. \ref{stron}}. From
\cite{BEH1}, the matrix $\Phi_{N,r}(x)$ satisfies the following
Riemann-Hilbert problem
\begin{eqnarray*}
(\Phi_{N,r})_+(x)&=&(\Phi_{N,r})_-(x)\pmatrix{1&{\kappa}(x)\cr
                                    0&1\cr},\quad x\in\Gamma:= \bigcup_{i,\ell} \Gamma_{i,\ell}\\
\Phi_{N,r}(x)&=&\exp\pmatrix{-{1\over 2}N V(x)+(N+r)\log x&0\cr
                       0&{1\over 2}N V(x)-(N+r)\log x\cr},\quad
                       x\rightarrow\infty
\end{eqnarray*}
where the $+$ and $-$ indices denote the values of the function on
the left and right hand sides of the contour respectively. The
piecewise constant function ${\kappa}(x)$
is given by
\be
 \kappa(x)= \le\{\begin{array}{cc}
\varkappa_{i,\ell} & x\in \Gamma_{i,\ell}\setminus \mathcal B\\
\rho_j & x \in \Sigma_j \subset \mathcal B
\end{array}\ri.
\ee

We can now transform the above Riemann-Hilbert problem by
multiplying $e^{NG (x)\sigma_3}$ on the right of
$\Phi_N(x)$, where $\sigma_3$ is the Pauli matrix $\sigma_3={\rm
diag}(1,-1)$. Let
\be
\Phi^1(x):= \Phi_N(x) {\rm e}^{{NG }(x)\sigma_3}\ .
\ee

This new matrix  has jump discontinuities on $\Sigma$ and $\Gamma$
(see Figure \ref{fig:sigma} for $\Sigma$). The Riemann--Hilbert
problem is  transformed into the following for $\Phi^1(x)$:
\begin{eqnarray*}
\Phi_+^1(x)&=&\Phi_-^1(x)\pmatrix{e^{N(G _+-G _-)}&\wt\kappa(x)
e^{-N(G _-+G _+)}\cr
                                    0&e^{N(G _--G _+)}}, \quad
                                    x\in\Gamma \supset \mathcal B\\
\Phi_+^1(x)&=&\Phi_-^1(x)\pmatrix{e^{N(G _+-G _-)}&0\cr
                                    0&e^{N(G _--G _+)} }, \quad
                                    x\in\Sigma/(\Sigma\cap\Gamma)\\
\Phi^1(x)&=&\le(I+O\left({1\over x}\right)\ri) x^{r\sigma_3},\quad x\rightarrow\infty
\end{eqnarray*}
Note that $\Sigma\setminus (\Sigma \cap \Gamma)$ is the union of all gaps and there is actually no jump on the last gap that extends to $\infty$ (because the residue of $\eta$ at infinities is an integer).
\paragraph{Single cuts.}
Suppose ${\cal B}_i$ is a connected component of
${\cal B}$ with only one branch cut $\Sigma_i$ (and hence $\mathcal B_i = \Sigma_i$)

\begin{figure}[htbp]
\begin{center}
\resizebox{7cm}{!}{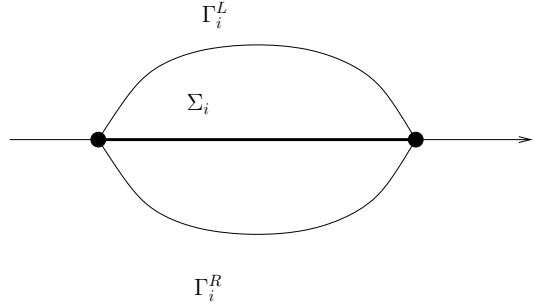}\caption{Lens--opening near an isolated branchcut.}
\label{fig:normallens}
\end{center}
\end{figure}

The jump matrix of $\Phi^1(x)$ on the branch cut $\Sigma_i$ is
then
\begin{eqnarray*}
\Phi_+^1(x)&=&\Phi_-^1(x)\pmatrix{e^{N(G _+-G _-)}&\rho_ie^{-N(G _-+G _+)}\cr
                                    0&e^{N(G _--G _+)}\cr}, \quad
                                    x\in\Sigma_i
\end{eqnarray*}
As in \cite{DKV} and \cite{V}, we will make use of the
factorization
\begin{eqnarray*}
\pmatrix{a&-b\cr
         0&a^{-1}\cr}=\pmatrix{1&0\cr
         -a^{-1}b^{-1}&1\cr}\pmatrix{0&-b\cr
         b^{-1}&0\cr}\pmatrix{1&0\cr -ab^{-1}&1\cr}
\end{eqnarray*}
and write the jump matrix as
\begin{eqnarray}
&&\pmatrix{e^{N(G _+-G _-)}&\rho_ie^{-N(G _-+G _+)}\cr
                                    0&e^{N(G _--G _+)}\cr}\\&=&\pmatrix{1&0\cr
         \rho_i^{-1}e^{2NG _-}&1\cr}\pmatrix{0&\rho_ie^{-N(G _-+G _+)}\cr
         -\rho_i^{-1}e^{N(G _-+G _+)}&0\cr}\pmatrix{1&0\cr
         \rho_i^{-1}e^{2NG _+}&1\cr}\label{jphi1}
\end{eqnarray}
We can now follow the same technique in \cite{DKV}, \cite{V} and
deform the contour into a `lens' as follows. Let $\Gamma_i^L$ and
$\Gamma_i^R$ be contours joining $\alpha_{2i-1}$ and $\alpha_{2i}$
on the left and right hand sides of $\Sigma_i$ such that $h(x)$ is
negative on both $\Gamma_i^L$ and $\Gamma_i^R$ (See figure
\ref{fig:normallens}). This is possible because of admissibility.

We can then deform the Riemann Hilbert problem near these branch
cuts as follows. Let
\begin{eqnarray}\label{eq:singlecut}
\Phi^2(x)&=&\Phi^1(x), \quad \textrm{ for $x$ outside the
lens-shaped
regions,}\nonumber\\
\Phi^2(x)&=&\Phi^1(x)\pmatrix{1&0\cr
         -\rho_i^{-1}e^{2NG (x)}&1\cr}, \quad \textrm{for $x$ in the left hand side lens
         regions}\\
\Phi^2(x)&=&\Phi^1(x)\pmatrix{1&0\cr
         \rho_i^{-1}e^{2NG (x)}&1\cr}, \quad \textrm{for $x$ in the right hand side lens
         regions}\nonumber
\end{eqnarray}

\begin{figure}[htbp]
\begin{center}
\resizebox{10cm}{!}{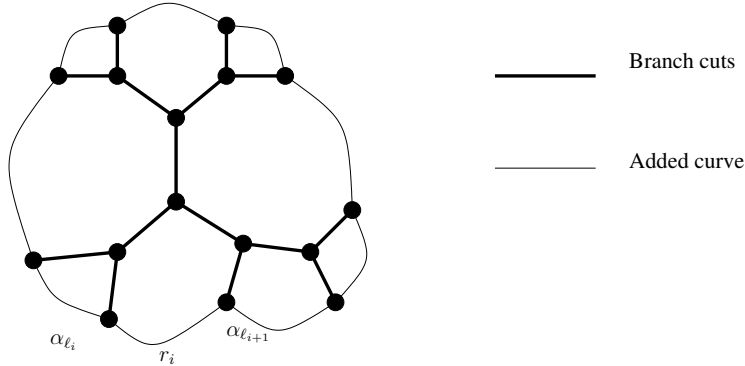}
\caption{Lens--opening near a tree of $\mathcal B$.}
\label{fig:biglens0}
\end{center}
\end{figure}

\paragraph{Multiple cuts.}
Now suppose that ${\cal B}_j$ is a component of ${\cal B}$ that
contains more than one branch cut. Then apart from the boundary
points of ${\cal B}_j$, all the branch points are of type I. Let
the boundary points be $\alpha_{\ell_1},\ldots,\alpha_{\ell_n}$.
Since this connected component of $\mathcal B$ is a loop--free
trivalent tree (in a finite region of the plane) and a
neighborhood of it (less $\mathcal B_j$) lies in the set $h(x)<0$
by definition of admissibility, we can then join these points by
curves that lies within $\{h(x)<0\}$ as in figure
\ref{fig:biglens0}.

\begin{figure}[htbp]
\begin{center}
\resizebox{10cm}{!}{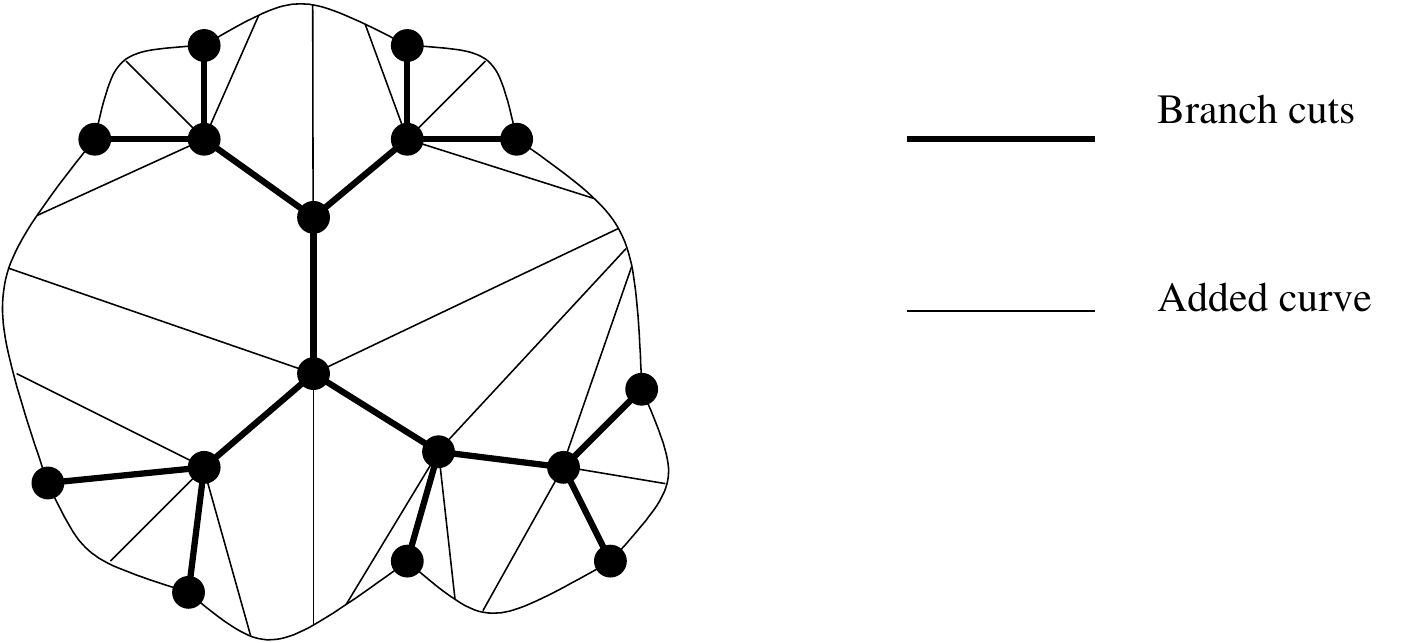}
\caption{Lens--opening near a tree of $\mathcal B$ and added jumps.}
\label{fig:biglens}
\end{center}
\end{figure}

Let the curve between $\alpha_{\ell_i}$ and $\alpha_{\ell_{i+1}}$ be $r_i$.  From each trivalent vertex we run three contours  joining the vertex to the closest  amongst the arcs $r_i$'s without intersecting the cuts already made (see Fig. \ref{fig:biglens} for a self-explanatory exemplification).
On each side of each edge $\Sigma_m$ of $\mathcal B_j$ there are precisely two regions bounded by $\Sigma_m$ a contour amongst the $r_j$'s and the added contours from the trivalent vertices; we denote these two regions $D_{m,L}$ and $D_{m,R}$;
we then deform the Riemann-Hilbert problem near ${\cal B}_j$ to an
equivalent one for a $2\times 2$ matrix-valued function
$\Phi^2(x)$ defined as follows in the various regions
 \begin{eqnarray}\label{eq:multicut}
\Phi^2(x)&=&\Phi^1(x), \quad \textrm{ for $x$ outside the
lens-shaped
regions,}\nonumber\\
\Phi^2(x)&=&\Phi^1(x)\pmatrix{1&0\cr
         -\rho_m^{-1}e^{2NG (x)}&1\cr}, \quad x\in D_{m,L}\nonumber\\
\Phi^2(x)&=&\Phi^1(x)\pmatrix{1&0\cr
         \rho_m^{-1}e^{2NG (x)}&1\cr}, \quad x\in D_{m,R}\nonumber 
\end{eqnarray}
for each branch cut $\Sigma_m \in{\cal B}_j$. 

\begin{figure}[htbp]
\begin{center}
\resizebox{8cm}{!}{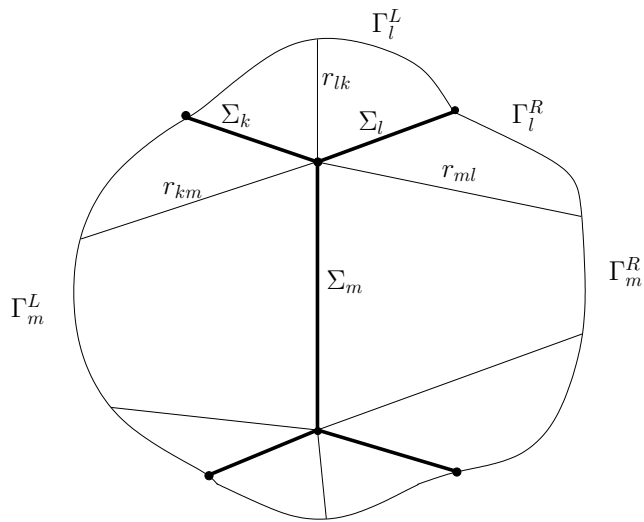}
\caption{Another tree and lens-opening with added jumps.}
\label{fig:biglenscut1}
\end{center}
\end{figure}

The matrix $\Phi^2(x)$ then has jumps on $\Gamma \cup \Sigma$ and the contours shown in Fig. \ref{fig:biglenscut1}. On $(\Gamma \cup \Sigma)\setminus \mathcal B$ it has the same jumps as $\Phi^1(x)$ (eq. \ref{jphi1}). On the contours in Fig. \ref{fig:biglenscut1} its jumps are the following:
\begin{eqnarray}\label{eq:jump2}
\nu^2(x)&=&\pmatrix{1&0\cr
         \rho_m^{-1}e^{2NG (x)}&1\cr},\quad x\in \Gamma_m^L\nonumber\\
\nu^2(x)&=&\pmatrix{1&0\cr
         \rho_m^{-1}e^{2NG (x)}&1\cr},\quad x\in \Gamma_m^R\\
\nu^2(x)&=&\pmatrix{0&\rho_me^{-N(G _-+G _+)}\cr
         -\rho_m^{-1}e^{N(G _-+G _+)}&0\cr},\quad x\in\Sigma_m\nonumber\\
\nu^2(x)&=&\pmatrix{1&0\cr
         f_{ml}(x)e^{2NG (x)}&1\cr},\quad x\in r_{ml}
\end{eqnarray}
where $f_{ml}(x)$ is either of the following, depending on the
orientation of the contours
\begin{eqnarray*}
f_{ml}(x)=\pm\rho_m^{-1}\pm \rho_l^{-1}
\end{eqnarray*}
in either of the above case, $f_{jk}(x)e^{2NG (x)}\rightarrow 0$ as
$N\rightarrow\infty$ uniformly away from the branch points.

From the jump conditions (\ref{eq:jump2}) and (\ref{eq:monG}), we
see that, away from the branch points, the Riemann-Hilbert problem
for $\Phi^2(x)$ approaches the following uniformly as
$N\rightarrow\infty$.
\begin{eqnarray*}
\Phi^2(x)\rightarrow\Phi^{\infty}(x)
\end{eqnarray*}
where $\Phi^{\infty}(x)$ satisfies the following jump conditions
\begin{eqnarray}\label{eq:jumpinfty}
\nu^2(x)&=&\pmatrix{0&\rho_je^{-iN\sigma_j}\cr
         -\rho_j^{-1}e^{iN\sigma_j}&0\cr},\quad x\in\Sigma_j\nonumber\\
\nu^2(x)&=&\pmatrix{e^{iN\wt \sigma_j}&0\cr
         0&e^{-iN\wt \sigma_j}\cr},\quad x\in\tilde{\Sigma}_j
\end{eqnarray}
and $\Phi^{\infty}(x)\rightarrow I$ as $x\rightarrow\infty$.

By lemma \ref{le:monspinor}, the matrix $\Psi_{N,r}$
satisfies such the Riemann-Hilbert problem (\ref{eq:jumpcond}) and
has the same behavior as $x\rightarrow\infty$. Hence
$\Phi^{\infty}(x)=\Psi_{N,r}(x)$. The asymptotic formula (\ref{eq:away})
and (\ref{eq:on}) can now be seen by reversing the sequence of
transformations. To complete the proof, we need to show that there
exists parametrices near the branch points. We will discuss this
in the next section.

\subsection{Parametrix near branch points}
\label{se:parametrix} Since the Riemann-Hilbert problem for
$\Phi^2(x)$ does not tend to the one satisfied by
$\Phi^{\infty}(x)$ near the branch points, we still need to solve
the Riemann-Hilbert problem exactly near the branch points. The
solution of the Riemann-Hilbert problem (\ref{eq:jump2}) for
$\Phi^2(x)$ near the branch points are the local parametrices and
they can be constructed from the Airy function. This problem has
been discussed in many places in the literature \cite{D},
\cite{its}, \cite{DKV} and we shall not repeat the details of the
analysis here. Instead, we would illustrate the main idea and
provide the solution to such a problem. The only relatively new
feature --in this respect-- is the presence of turning point (the
branch-points) of type I, namely where three branchcuts are
connected.

In order to complete the asymptotic analysis and obtain a uniform estimate, we need to find a solution $\Phi^p(x)$ of the
Riemann-Hilbert problem (\ref{eq:jump2}) in a neighborhood
$O_{\alpha}$ of a branch point $\alpha$ such that as
$N\rightarrow\infty$, we have $\Phi^p(x)\rightarrow\Phi^2(x)$ near
the boundary of such neighborhood. We transform the RH problem a first time by multiplying
the solution $\Phi^2(x)$ by $\exp(-{NG }(x)\sigma_3)$ on the right
\begin{eqnarray*}
Z^p(x)=\Phi^2(x)\exp(-{NG }(x)\sigma_3), \quad x\in O_{\alpha}.
\end{eqnarray*}
As a consequence, the Riemann--Hilbert problem satisfied by the  matrix $Z^p(x)$ is transformed into a problem with constant jumps.\\

\begin{figure}[htbp]
\begin{center}
\resizebox{15cm}{!}{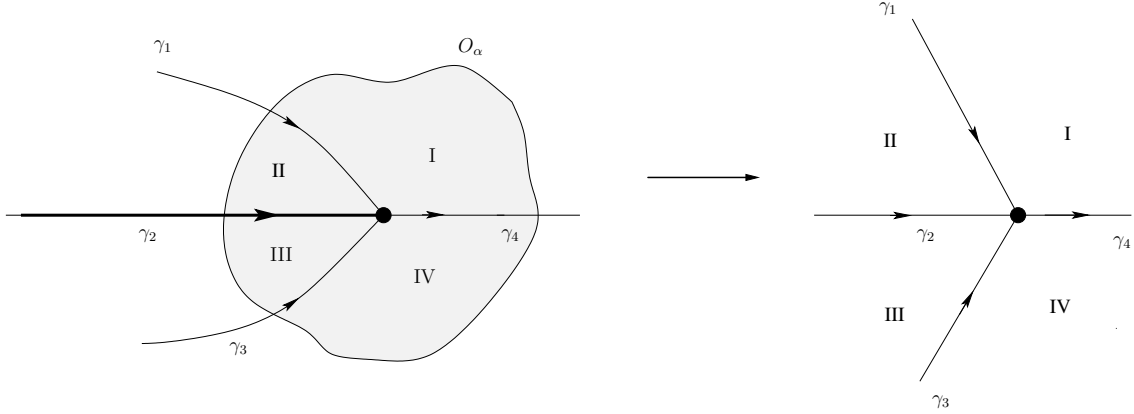}
\caption{The
function $({3\over 2}N(G(x)-G(\alpha)))^{3\over 2}$ maps the
neighborhood $O_{\alpha}$ of $\alpha$ to a neighborhood of the
origin.}
\label{fig:parametrix1}
\end{center}
\end{figure}

{\bf Univalent turning point (type III)}. If $\alpha$ is a turning point with
only one incident branch-cut, say  $\Sigma_m$,  then the
Riemann--Hilbert problem for $Z^p$ in $O_{\alpha}$ consists of the
following jumps on the contours $\gamma_i$
\begin{eqnarray}\label{eq:jumppara1}
\nu^p(x)&=&\pmatrix{1&0\cr  {\rho_m}^{-1}&1\cr}, \qquad x\in\gamma_1\nonumber\\
\nu^p(x)&=&\pmatrix{0& \rho_m\cr -{\rho_m}^{-1}&0\cr},\qquad  x\in\gamma_2\nonumber\\
\nu^p(x)&=&\pmatrix{1& 0\cr \rho_m^{-1} &1\cr},\qquad  x\in\gamma_3 \\
\nu^p(x)&=&\pmatrix{1& \rho_m\cr 0&1\cr},\qquad  x\in\gamma_4\nonumber
\end{eqnarray}
where $\rho_m$ is a constant depending on which branch cut is incident to the turning point under scrutiny. (See figure
\ref{fig:parametrix1})

{\bf Trivalent turning point (type I)}. If $\alpha$ has 3 branch cuts
attached, the function $Z^p(x)$ then
 satisfies the following Riemann-Hilbert problem
\begin{eqnarray}\label{eq:jumppara2}
\nu^p(x)&=&\pmatrix{1& 0\cr \rho_{m_1}^{-1}+\rho_{m_2}^{-1}&1\cr},\qquad  x\in\gamma_1\nonumber\\
\nu^p(x)&=&\pmatrix{0& \rho_{m_2}\cr -\rho_{m_2}^{-1}&0\cr},\qquad x\in\gamma_2 \\
\nu^p(x)&=&\pmatrix{1& 0\cr \rho_{m_2}^{-1}+\rho_{m_3}^{-1}&1\cr},\qquad  x\in\gamma_3\nonumber\\
\nu^p(x)&=&\pmatrix{0& \rho_{m_3}\cr -\rho_{m_3}^{-1}&0\cr},\qquad x\in\gamma_4\nonumber\\
\nu^p(x)&=&\pmatrix{1& 0\cr \rho_{m_3}^{-1}+\rho_{m_1}^{-1}&1\cr},\qquad  x\in\gamma_5\nonumber\\
\nu^p(x)&=&\pmatrix{0& \rho_{m_1}\cr -\rho_{m_1}^{-1}&0\cr},\qquad x\in\gamma_6\nonumber
\end{eqnarray}
where the orientations of the rays $\gamma_i$ are chosen such that
all of them are pointing away from the branch point.

A different choice of orientations will only result in a change of
sign of some of these constants. The crucial observation is that
with this choice of orientations we have
$\rho_{m_1}+\rho_{m_2}+\rho_{m_3}=0$, which implies that the
product of all the above jump-matrices is the identity  and is
precisely the Kirchoff's law for our ``currents''
$\rho_j$'s.\par\vskip 4pt

We would like to construct solutions $Z^p(x)$ to the
Riemann-Hilbert problems (\ref{eq:jumppara1}) and
(\ref{eq:jumppara2}) near the branch point $\alpha$ such that
\begin{eqnarray}
\label{eq:boundary}
Z^p(x)e^{{{NG }(x)\sigma_3}} =  \Phi^\infty(x)(\1 + \mathcal O(N^{-1})),\quad x\in
\partial O_{\alpha},\quad N\rightarrow\infty
\end{eqnarray}
The solutions near these branch points can be constructed by using
the Airy function. The Airy function $Ai(x)$ is the unique
solution of the ODE
\begin{eqnarray*}
Ai^{\prime\prime}(\xi)=\xi Ai(\xi)
\end{eqnarray*}
with the following asymptotics
\begin{eqnarray*}
Ai(\xi)&=&{1\over{2\sqrt{\pi}}}\xi^{-{1\over 4}}e^{-{2\over
3}\xi^{3\over 2}}\left(1+O\left({1\over{\xi^{3\over 2}}}\right)\right)\\
Ai^{\prime}(\xi)&=&-{1\over{2\sqrt{\pi}}}\xi^{{1\over 4}}e^{-{2\over
3}\xi^{3\over 2}}\left(1+O\left({1\over{\xi^{3\over 2}}}\right)\right)
\end{eqnarray*}
as $\xi\rightarrow\infty$ and $|\arg \xi|<\pi$. The branches of
$\xi^{1\over 4}$ and $\xi^{3\over 2}$ in the above are principal
branches with branch cut at the negative real axis.

\begin{figure}[htbp]
\begin{center}
\resizebox{15cm}{!}{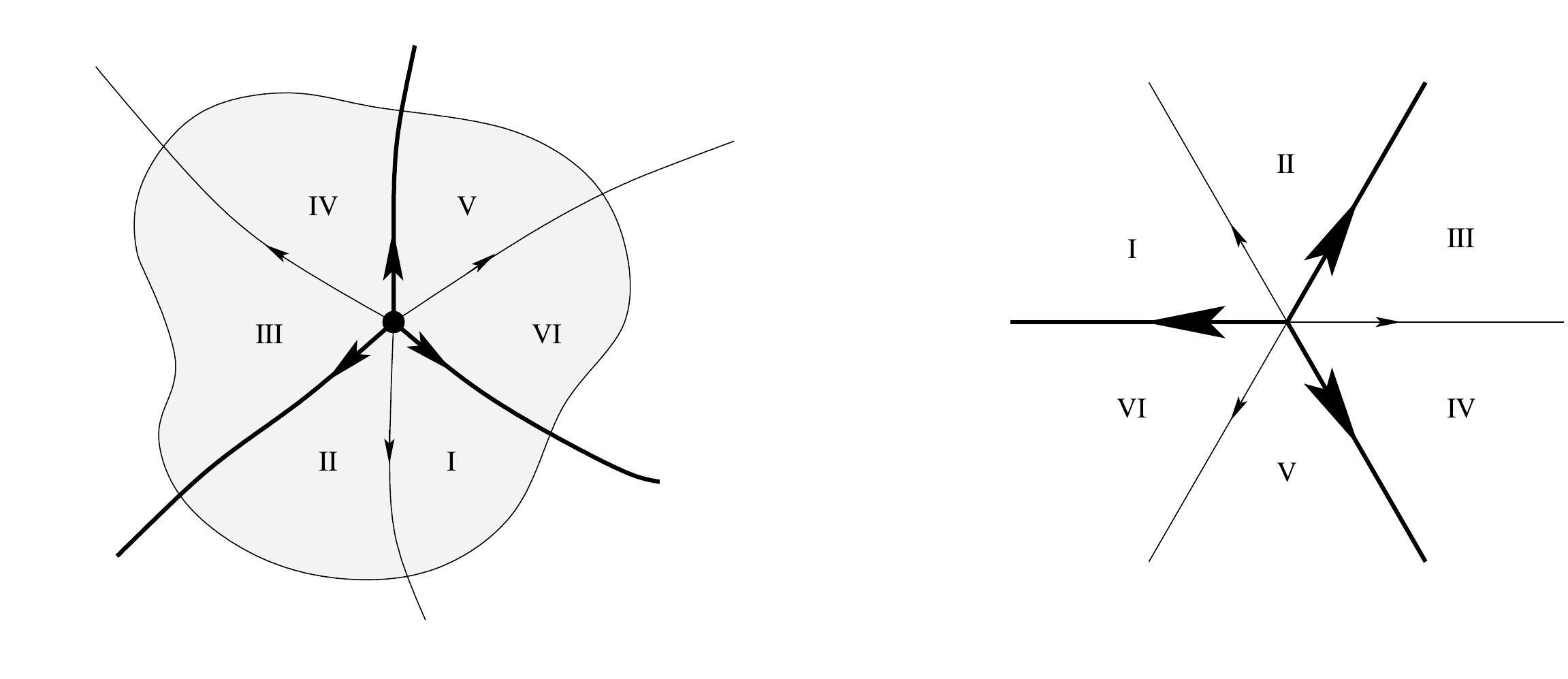}\caption{The
function $({3\over 2}N(G(x)-G(\alpha)))^{3\over 2}$ maps the
neighborhood $O_{\alpha}$ of $\alpha$ to a neighborhood of the
origin.}\label{fig:parametrix2}
\end{center}
\end{figure}

Since $G(x)-G(\alpha)$ behaves like $cx^{3\over 2}$ near a branch
point for some constant $c$, we can choose a function
$\xi=({3\over 2}N(G(x)-G(\alpha)))^{2\over 3}$ that maps the
branch cut onto the negative real axis.\footnote{This requirement
is due to the fact that in defining the Airy function $Ai(\xi)$,
we assume $|\arg \xi|<\pi$. The requirement can be dropped if we
use different branches of the Airy function. See \cite{its}} This
is possible because the function ${3\over 2}N(G(x)-G(\alpha))$
maps the branch cut onto the imaginary axis. The function $\xi$
then defines a one-to-one mapping between the neighborhood
$O_{\alpha}$ of the branch point $\alpha$ and a neighborhood of
the origin in the complex $\xi$ plane. (See Figures
\ref{fig:parametrix1}, and \ref{fig:parametrix2}.)

As in \cite{D}, a solution of the Riemann-Hilbert problem
(\ref{eq:jumppara1}) is given by the following. Let $\Psi^p(x)$ be
the matrix given by
\begin{eqnarray}\label{eq:para1}
Z^p(\xi)&=&\pmatrix{Ai(\xi)& Ai(\omega^2\xi)\cr
Ai^{\prime}(\xi)&\omega^2Ai^{\prime}(\omega^2\xi)\cr}
e^{-{{i\pi}\over 6}\sigma_3}\rho_{m}^{-{{\sigma_3}\over 2}}, x\in I\nonumber\\
Z^p(\xi)&=&\pmatrix{Ai(\xi)& Ai(\omega^2\xi)\cr
Ai^{\prime}(\xi)&\omega^2Ai^{\prime}(\omega^2\xi)\cr}
e^{-{{i\pi}\over 6}\sigma_3}\rho_m^{-{{\sigma_3}\over 2}}\pmatrix{1&0\cr -\rho_m^{-1}&1\cr}, x\in II\nonumber\\
Z^p(\xi)&=&\pmatrix{Ai(\xi)& -\omega^2Ai(\omega\xi)\cr
Ai^{\prime}(\xi)&-Ai^{\prime}(\omega\xi)\cr}
e^{-{{i\pi}\over 6}\sigma_3}\rho_m^{-{{\sigma_3}\over 2}}\pmatrix{1&0\cr \rho_m^{-1}&1\cr}, x\in III \\
Z^p(\xi)&=&\pmatrix{Ai(\xi)& -\omega^2Ai(\omega\xi)\cr
Ai^{\prime}(\xi)&-Ai^{\prime}(\omega\xi)\cr} e^{-{{i\pi}\over
6}\sigma_3}\rho_m^{-{{\sigma_3}\over 2}}, x\in IV\nonumber
\end{eqnarray}
where $\omega=e^{{2\pi i}\over 3}$. Then by using the relations
\begin{eqnarray*}
Ai(\xi)+\omega Ai(\omega\xi)+\omega^2Ai(\xi \omega^2)&=&0\\
Ai^{\prime}(\xi)+\omega^2Ai^{\prime}(\xi \omega)+\omega
Ai^{\prime}(\xi \omega^2)&=&0
\end{eqnarray*}
one can show that $Z^p(\xi)$ defined by (\ref{eq:para1}) does
satisfy the jump conditions (\ref{eq:jumppara1}).

Since the only jump discontinuity of $\xi^{3\over 2}$ is at the
branch cut where it changes sign, while $N(G(x)-G(\alpha))$ has
jumps at the gap $\tilde{\Sigma}_m$ and the branch cut $\Sigma_m$,
these two functions are not identical. However, a branch of
$\xi^{3\over 2}$ can be chosen such that the difference between
${2\over 3}\xi^{3\over 2}$ and $N(G(x)-G(\alpha))$ remains bounded
in $O_{\alpha}$.

Let $\zeta$ be the function ${NG }(x)-{2\over 3}\xi^{3\over 2}$, where
the branch of $\xi^{3\over 2}$ is chosen such that $\zeta$ remains
bounded in $O_{\alpha}$: in fact this difference is a locally constant function on $\mathcal O_\a\setminus \Sigma$.

 By considering the jump discontinuities
of ${NG }(x)$ and $\xi^{3\over 2}$, we see that $\zeta(x)$ is bounded
throughout $O_{\alpha}$ and has the following jump discontinuities
\begin{eqnarray}\label{eq:zeta1}
\zeta(x)_+&=&\zeta(x)_-+iN\tilde{\sigma}_m,\quad
x\in\tilde{\Sigma}_m\cap O_{\alpha}\nonumber\\
\zeta(x)_+&=&-\zeta(x)_-+iN\sigma_m,\quad x\in\Sigma_m\cap
O_{\alpha}
\end{eqnarray}
To fix the boundary conditions such that
$Z^p(\xi)e^{{{NG }(x)\sigma_3}}\rightarrow \Phi^\infty(x)$ as
$N\rightarrow\infty$ on the boundary of $O_{\alpha}$, first note
that $Z^p(\xi)e^{{NG }(x)\sigma_3}$ has the following asymptotic form
\begin{eqnarray*}
Z^p(\xi)e^{{NG }(x)\sigma_3} ={e^{{i\pi}\over
12}\over{2\sqrt{\pi}}}\pmatrix{\xi^{-{1\over 4}}&0\cr
0&\xi^{1\over 4}\cr}\pmatrix{1&1\cr -1&1\cr}\left(I+O\left({1\over
N}\right)\right)e^{-{{\pi i}\over
4}\sigma_3}e^{\zeta(x)\sigma_3}\rho_m^{\sigma_3\over 2},\quad
\xi\rightarrow\infty
\end{eqnarray*}
Let $Z_{as}(\xi)$ be the leading term of the above asymptotic
expansion. Then from (\ref{eq:zeta1}) and the fact that
$\xi^{1\over 4}\rightarrow i\xi^{1\over 4}$ across the branch cut
of $G(x)$, we see that $Z_{as}(\xi)$ has the following jump
discontinuities
\begin{eqnarray*}
Z_{as}(\xi)_+&=&Z_{as}(\xi)_-\pmatrix{0&\rho_me^{-iN\sigma_m}\cr
-\rho_m^{-1}e^{iN\sigma_m}&0\cr},\quad x\in\Sigma_m\cap
O_{\alpha}\\
Z_{as}(\xi)_+&=&Z_{as}(\xi)_-\pmatrix{e^{iN\tilde{\sigma}_m}&0\cr
0&e^{-iN\tilde{\sigma}_m}\cr},\quad x\in\Sigma_m\cap O_{\alpha}
\end{eqnarray*}
Therefore the matrix
\begin{eqnarray*}
E(x)=\Phi^\infty(x)Z_{as}^{-1}(x)
\end{eqnarray*}
is single-valued in $O_{\alpha}$. It is also holomorphic in
$O_{\alpha}$ because it can at worst have a square-root singularity at $\alpha$,  but since it is single-valued,
this cannot happen and hence $E(x)$ is holomorphic in
$O_{\alpha}$.

Hence the function $E(x)Z^p(x)$ is bounded in $O_{\alpha}$, tends
uniformly to $\Phi^{\infty}(x)$ near the boundary of $O_{\alpha}$
and satisfies the jump condition (\ref{eq:jumppara1}).

For a branch point with 3 branch cuts attaching to it we choose
$\xi=({3\over 2}N(G(x)-G(\alpha)))^{2\over 3}$ that maps the
branch cut $\Sigma_{m_1}$ onto the negative real axis. The
following $Z^p(\xi)$ would satisfy the jump conditions
(\ref{eq:jumppara2})
\bea
\label{eq:para2}
Z^p(\xi)&=&\pmatrix{Ai(\xi)& Ai(\omega^2\xi)\cr
Ai^{\prime}(\xi)&\omega^2Ai^{\prime}(\omega^2\xi)\cr}
e^{-{{i\pi}\over 6}\sigma_3}\pmatrix{1&0\cr -1&1\cr}\left({\rho_{m_3}}\over{\rho_{m_1}\rho_{m_2}}\right)^{{\sigma_3}\over 2}, \quad \xi\in I\nonumber\\
Z^p(\xi)&=&\pmatrix{Ai(\xi )& Ai(\omega^2\xi)\cr
Ai^{\prime}(\xi)&\omega^2 Ai^{\prime}(\omega^2\xi)\cr}
e^{-{{i\pi}\over 6}\sigma_3}\left({\rho_{m_3}}\over{\rho_{m_1}\rho_{m_2}}\right)^{{\sigma_3}\over 2}, \quad \xi\in II\nonumber\\
Z^p(\xi)&=&-\pmatrix{Ai(\xi)& Ai(\omega^2\xi)\cr
Ai^{\prime}(\xi)&\omega^2 Ai^{\prime}(\omega^2\xi)\cr}
e^{-{{i\pi}\over 6}\sigma_3}\pmatrix{0&1\cr -1&0\cr}\left({\rho_{m_1}}\over{\rho_{m_2}\rho_{m_3}}\right)^{{\sigma_3}\over 2},\quad \xi\in III\nonumber\\
Z^p(x)&=&-\pmatrix{Ai(\xi)& -\omega^2 Ai(\omega\xi)\cr
Ai^{\prime}(\xi)&-Ai^{\prime}(\omega\xi)\cr}
e^{-{{i\pi}\over 6}\sigma_3}\pmatrix{0&1\cr -1&0\cr}\left({\rho_{m_1}}\over{\rho_{m_2}\rho_{m_3}}\right)^{{\sigma_3}\over 2},\quad \xi\in IV \\
Z^p(\xi)&=&-\pmatrix{Ai(\xi)& -\omega^2Ai(\omega\xi)\cr
Ai^{\prime}(\xi)&-Ai^{\prime}(\omega\xi)\cr} e^{-{{i\pi}\over
6}\sigma_3}\left({\rho_{m_2}}\over{\rho_{m_1}\rho_{m_3}}\right)^{{\sigma_3}\over 2},\quad \xi\in V\nonumber \\
Z^p(\xi)&=&-\pmatrix{Ai(\xi)& -\omega^2Ai(\omega\xi)\cr
Ai^{\prime}(\xi)&-Ai^{\prime}(\omega\xi)\cr} e^{-{{i\pi}\over
6}\sigma_3}\pmatrix{1&0\cr
1&1\cr}\left({\rho_{m_2}}\over{\rho_{m_1}\rho_{m_3}}\right)^{{\sigma_3}\over
2},\quad \xi \in VI\nonumber
\eea
By using $\rho_{m_1}+\rho_{m_2}+\rho_{m_3}=0$, one can check that
$Z^p(\xi)$ does indeed satisfy the jump conditions
(\ref{eq:jumppara2}).

Let $\zeta(x)$ be the following function
\begin{eqnarray*}
\zeta(x)&=&{NG }(x)-\xi^{3\over 2},\quad x\in I\cup II\cup V\cup VI\\
\zeta(x)&=&{NG }(x)+\xi^{3\over 2}, \quad x\in III\cup IV
\end{eqnarray*}
where the branch of $\xi^{3\over 2}$ is chosen such that
$\zeta(x)$ is finite in $O_{\alpha}$. Since $\xi^{3\over 2}$ has
only one branch cut at $\gamma_6$ while ${NG }(x)$ has 3 branch cuts
at $\gamma_2$, $\gamma_4$ and $\gamma_6$ respectively, $\zeta(x)$
has the following jump discontinuities
\begin{eqnarray}\label{eq:zeta2}
\zeta(x)_+&=&-\zeta(x)_-+iN\sigma_{m_1},\quad
x\in\gamma_6\nonumber\\
\zeta(x)_+&=&-\zeta(x)_-+iN\sigma_{m_2},\quad x\in\gamma_2\\
\zeta(x)_+&=&-\zeta(x)_-+iN\sigma_{m_3},\quad
x\in\gamma_4\nonumber
\end{eqnarray}
To fix the boundary condition in this case, first observe that the
function $Z^p(x)$ has the following asymptotic expansions
\begin{eqnarray*}
Z^p(\xi) &\sim&{e^{{i\pi}\over 12}\over{2\sqrt{\pi}}}\xi^{-{1\over
4}\sigma_3}\pmatrix{1&1\cr -1&1\cr}\left(I+O\left({1\over
N}\right)\right)e^{-{{\pi i}\over 4}\sigma_3}e^{-{2\over
3}\xi^{3\over
2}\sigma_3}\left({\rho_{m_3}}\over{\rho_{m_1}\rho_{m_2}}\right)^{{\sigma_3}\over
2},\quad \xi\in
I\cup II\\
Z^p(\xi) &\sim&-{e^{{i\pi}\over
12}\over{2\sqrt{\pi}}}\xi^{-{1\over 4}\sigma_3}\pmatrix{1&1\cr
-1&1\cr}\left(I+O\left({1\over N}\right)\right)\pmatrix{0&1\cr
-1&0\cr}e^{({2\over 3}\xi^{3\over 2}+{{\pi i}\over
4})\sigma_3}\left({\rho_{m_1}}\over{\rho_{m_2}\rho_{m_3}}\right)^{{\sigma_3}\over
2},\nonumber\\
&&\xi\in III\cup IV \\
Z^p(\xi) &\sim&-{e^{{i\pi}\over
12}\over{2\sqrt{\pi}}}\xi^{-{1\over 4}\sigma_3}\pmatrix{1&1\cr
-1&1\cr}\left(I+O\left({1\over N}\right)\right)e^{-{{\pi i}\over
4}\sigma_3}e^{-{2\over 3}\xi^{3\over
2}\sigma_3}\left({\rho_{m_2}}\over{\rho_{m_1}\rho_{m_3}}\right)^{{\sigma_3}\over
2},\quad \xi\in V\cup VI\nonumber
\end{eqnarray*}
as $\xi\rightarrow \infty$. If we now take take $Z_{as}(\xi)$ to
be
\begin{eqnarray*}
Z_{as}(x) &=&{e^{{i\pi}\over 12}\over{2\sqrt{\pi}}}\xi^{-{1\over
4}\sigma_3}\pmatrix{1&1\cr -1&1\cr}e^{-{{\pi i}\over
4}\sigma_3}e^{\zeta(x)\sigma_3}\left({\rho_{m_3}}\over{\rho_{m_1}\rho_{m_2}}\right)^{{\sigma_3}\over
2},\quad \xi\in
I\cup II\\
Z^p(x) &=&-{e^{{i\pi}\over 12}\over{2\sqrt{\pi}}}\xi^{-{1\over
4}\sigma_3}\pmatrix{1&1\cr -1&1\cr}e^{-{{\pi i}\over
4}\sigma_3}\pmatrix{0&1\cr
-1&0\cr}e^{\zeta(x)\sigma_3}\left({\rho_{m_1}}\over{\rho_{m_2}\rho_{m_3}}\right)^{{\sigma_3}\over
2},\quad \xi\in III\cup IV \\
Z_{as}(x) &=&-{e^{{i\pi}\over 12}\over{2\sqrt{\pi}}}\xi^{-{1\over
4}\sigma_3}\pmatrix{1&1\cr -1&1\cr}e^{-{{\pi i}\over
4}\sigma_3}e^{\zeta(x)\sigma_3}\left({\rho_{m_2}}\over{\rho_{m_1}\rho_{m_3}}\right)^{{\sigma_3}\over
2},\quad \xi\in V\cup VI
\end{eqnarray*}
then from (\ref{eq:zeta2}), we see that $Z_{as}(x)$ has the same
jump discontinuities as $\Phi^2(x)$.

Therefore the matrix
\begin{eqnarray*}
E(x)=\Phi^\infty(x)Z_{as}^{-1}(x)
\end{eqnarray*}
is holomorphic in $O_{\alpha}$ and $E(x)Z^p(x)$ will be bounded in
$O_{\alpha}$ and will tend uniformly (to within $\mathcal O(1/N)$) to $\Phi^\infty(x)$ near the
boundary of $O_{\alpha}$. Also, it satisfies the jump condition
(\ref{eq:jumppara2}).

Finally, if we let $\Phi^3(x)$ be the following matrix-valued
function
\begin{eqnarray*}
\Phi^3(x)&=&\Phi^{\infty}(x),\quad
x\in\mathbb{C}/\cup_{i=1}^{2g+2}O_{\alpha_i}\\
\Phi^3(x)&=&\Phi^p(x),\quad x\in\cup_{i=1}^{2g+2}O_{\alpha_i}
\end{eqnarray*}
where $\Phi^p(x)=E(x)Z^p(x)e^{{NG }(x)\sigma_3}$. Then as in \cite{D},
\cite{DKV} and \cite{its}, one can show that the Riemann-Hilbert
problem satisfied by $\Phi^2(x)$ tends to the one satisfied by
$\Phi^3(x)$ uniformly at all points as $N\rightarrow\infty$.
Therefore the function $\Phi^3(x)$ would give the asymptotics of
the orthogonal polynomials as $N\rightarrow\infty$.

\subsection {Density of zeroes of the orthogonal polynomials}
\label{se:density}
The content of Thm. \ref{stron} is that the (monic) OPs $p_{N+r}$ behave uniformly on compact sets not intersecting the branchcut structure $\mathcal B$ as the following expressions (see \ref{asymport})
\bes
p_{N+r}(x){\rm e}^{-\frac N2 V(x)} \sim
(\Psi_{N,r})_{1,1}{\rm e}^{-N g(x)} =
\wt c_r
  \frac {\Theta_\Delta^r (
  p-\infty_-)}{\Theta_\Delta^{r+1}(p-\infty_+)}
  \frac{\Theta\le[{\mathcal A+ {\epsilon}_{_N} \atop \mathcal B+{\delta}_{_N} }\ri]
  (p+r\infty_- -(r+1) \infty_+  )}{ \Theta\le[{ \mathcal A +{\epsilon}_{_N} \atop \mathcal B+{\delta}_{_N} }\ri]
  (r(\infty_- - \infty_+)  )}Q_\Delta(X(p)){\rm e}^{-N g(x)} \ ,
  \label{pasymp}
\ees
where $\wt c_r$ is the constant (independent of $N$) computed after (\ref{asymport}),  $p$ is  point on the curve $\mathcal L$ with $X(p)=x$ and on the same sheet as $\infty_+$ and the vector of complex characteristics ${\epsilon}_{_N}, {\delta}_{_N}$ are given in eq. (\ref{characters}).

In particular (using that the Abel map of $\infty_+$ is opposite of that of $\infty_-$) it follows that
\bea
\frac 1{N} \ln \le| p_{N+r}(x) \ri|&\& \sim \frac 1 2 \Re(V(x)) - h(x) +\cr
 &\&+ \frac 1 N  \ln \le|\frac {Q_\Delta(X(p))\Theta_\Delta^r (
  p-\infty_-)}{\Theta_\Delta^{r+1}(p-\infty_+)}
\frac{\Theta\le[{ {\epsilon}_{_N} \atop {\delta}_{_N} }\ri]
  (p+r\infty_- -(r+1) \infty_+  )}{ \Theta\le[{ {\epsilon}_{_N} \atop {\delta}_{_N} }\ri]
  (r(\infty_- - \infty_+)  )}
\ri| \\&+& \frac {\ln|\wt c_r|}N, \quad h(x) := \Re g(x) \eea
uniformly over compact sets in $\C\setminus \mathcal B$.

In the RHS of (\ref{entries}) there are at most $g$ zeroes (on the first sheet) which do not belong to $\mathcal B$; therefore we have immediately
\bp
All but at most $g$ zeroes of the polynomials $p_{N+r}$ are contained in an arbitrary neighborhood of $\mathcal B$ for $N$ large enough.
\ep
In different terms this means that the almost all zeroes accumulate on $\mathcal B$.



It should be evident that the zeroes do not actually lie on $\mathcal B$ but they get closer as $N$ increases. From Thm. \ref{stron}, formula (\ref{eq:on}) it follows that in a neighborhood of $\mathcal B$ we have
\be
p_{N+r}(x){\rm e}^{-N V(x)} \sim A {\rm e}^{N G (x)} - B {\rm e}^{-N G (x)}\ ,
\ee
where $A,B$ are expressions depending in $x$ but with at most $g$ zeroes (whose count is asymptotically irrelevant).

Solutions are of the form
\be
G (x) =  \frac 1{2N} \ln \le|\frac B A \ri| + \frac i{2N} \arg(B/A) + 2i\pi \frac kN
\ee
which shows (in a slightly heuristic way) that the zeroes have asymptotically vanishing $\Re G$ (which means that they are ``close'' to the branchcuts $\mathcal B $ where $h = \Re G $ vanishes) and $\Delta \Im G  \sim \frac 1 N$.

Using Cauchy--Riemann's equations for $G $ one finds that
the tangential density along a cut is

\be
\frac 1 \pi \frac {\pa \Im G }{\pa s} =  - \frac 1 \pi \frac {\pa h}{\pa n_+}
\ee
where $\pa s$ and $\pa n$ denote the tangential and normal derivatives along the (smooth parts of the) branchcuts $\mathcal B$.

To put it differently
 the density of zeroes of $p_{N+r}$ per unit length along an arc of $\mathcal B$  as $N\to\infty$ tends to
\be
 \nu_\infty(x) := - \frac 1 {2 \pi} \le(\frac{\pa h}{\pa n_+} + \frac {\pa h}{\pa n_-} \ri) = -\frac 1 {\pi} \frac {\pa h}{\pa n_\pm} \ ,\ \ x\in \mathcal B \setminus \bigcup \{\alpha_j\} \label{density}
\ee
where  $\pa n_\pm$ denote the normal derivatives on the two sides of the cut.

Note that
 this expression is positive because the function $h$ is zero on the cut and negative (strictly) on its left/right neighborhoods (by definition of admissibility of our triple $(\mathcal L,x,y)$), hence the normal derivatives are strictly negative on the cuts (but away from the turning points).
 By elementary harmonic-function theory (or using electrostatic analogy) we then have
\be
h(z) =  \frac 1 2 \Re V(z) + \int_{\mathcal B} \ln|z-\zeta| \nu_\infty(\zeta)\rd s
\ee
(where $\rd s$ denotes the ordinary arc--length)
and hence the total mass of $\int_{ \mathcal B}\nu_\infty(x) \rd s$ is necessarily $1$ since that corresponds to the residue of $y\rd x$ (and to the constant in front of the logarithmic term of $G(x)$).
\section{Part III: Reconstruction of (admissible) Boutroux curves}
\label{se:reconstr}
In this section we prove that any admissible graph is the graph of an admissible triple.
In order to do this we take a detour in the theory of Strebel differentials, which we state here in a simple form suitable for our application; all the general statements can be found in \cite{strebel} and \cite{jenkins}.

Let $P(z) = \prod_{i=1}^k (z-a_i)^{\mu_i}$ be an arbitrary polynomial of degree $n = \sum_{i=1}^k \mu_i$. It defines a quadratic differential
\be
\phi(z) := P(z) {\rm d}z^2\ .
\ee
The {\bf metric} associated to a quadratic differential $\phi$, denoted by $|\phi|$, is (quite literally) $|P(z)| |dz|^2$; in our case this reads
\be
|\phi| = |P(z)| \bigg(dx^2 + dy^2\bigg)\ .
\ee
It is an easy exercise to see that this metric is flat, with conical singularities at the zeroes of $P(z)$ (and cuspidal singularities at the poles of order higher than $1$). The flat coordinates are given by the real/imaginary parts of
\be
w:= \int^z \sqrt{P(s)}\rd s
\ee
which is a locally defined parameter away from the zeroes (poles) of $P$.
\bd
The horizontal lines are the lines defined by $\Im(w)= $const (or equivalently $\arg(\sqrt{P(z)} dz)=0, \pi$).
\ed
Since $\sqrt{P}$ is defined up to a sign, so is $w$ (and up to translations as well), but the notion of horizontality is well--defined.
\bd
The {\bf  critical horizontal lines} are the horizontal lines which contain any of the critical points (zeroes) of $P$.\\
The {\bf critical horizontal graph} (critical graph for short) is the union of all critical horizontal lines.
\ed
It is a simple check in a local coordinate that from a zero $z=a$ of multiplicity $\mu$ there originate $\mu+2$ critical horizontal lines, with relative angles of $2\pi/(\mu+2)$ at $a$.

 Clearly the critical graph of $P$ consists of the union of a finite number of Jordan arcs.
 
 In the general theory of trajectories of quadratic differentials one may encounter trajectories that fill domains with non-empty interior, called {\bf recurrent} trajectories: by definition a trajectory $\gamma:\R\to \mathcal L$ ($\mathcal L$ an arbitrary Riemann surface) is recurrent if it belongs to its {\bf limiting set}
 \be
L_\gamma := \bigcap_{t\in\R} \bigcup_{s>t} \{\gamma(s)\}\ .
\ee
Visually these are curves filling ``ergodically'' some region.
 This is not the case in the situation at hands because of the following 
 \bp[Thm 15.2 \cite{strebel}]
 No trajectory ray of a holomorphic quadratic differential on a domain of connectivity $\leq 3$ is recurrent.
 \ep
 In our case the quadratic differential is holomorphic in $\C P^1\setminus \{\infty\} = \C$:
therefore (as follows also from \cite{jenkins}), our critical trajectories can only 
connect two critical points or one critical point to $\infty$.

We first dwell a bit on the topology of the critical graph, $\mathfrak X$.

The  statements  summarized in the following lemmas  can be found (with different notation) in \cite{jenkins} but are not difficult to show directly;  here we restrict the attention to polynomial quadratic differentials, but more general statements can be found ibidem.

\bl 
The graph $\mathfrak X$  contains no loop in the finite part
of the plane.
\el

%
Let $\C\setminus \mathfrak X = \sqcup \Gamma_j$ be the decomposition in connected components.
\bl
\label{l1}
Each connected component $\Gamma_j$ is simply connected.
\el

\bl
\label{l2}
Each (simply) connected component $\Gamma_j$ has at most two boundaries, each consisting of an infinite piecewise Jordan curve. Each of the boundary components contain at least one critical point (since it is constituted of critical lines).
\el


In each $\Gamma_j$ we can choose one boundary critical point $z_0$ as basepoint for integration and define $w_j$ to have $\Im(w_j)\geq 0$, $w_j(z_0)=0$.
\bl
\label{l3}
The function $w_j = \int_{z_0}^z \sqrt{P(s)}\rd s$ is a uniformization function from $\Gamma_j$ to an infinite horizontal strip or to the upper half-plane.
\el

Lemma \ref{l3} shows that $\C\setminus \mathfrak X$ is the union of halfplanes and strips (topologically); the complex $z$-plane itself, 
$\C_z$ is then realized as a union of closed half-$w$-planes  or horizontal $w$-strips with appropriate identification of the boundary 
points; each of these halfplanes/strips has at least one marked point on each boundary.
\bd
A marked halfplane/strip is a copy of $\Im (w)\geq 0$ ($0\leq \Im w\leq \ell$, respectively) together with a collection of marked points on boundary (at least one for each boundary) and up to horizontal translations.
\ed
From this definition it is clear that a marked halfplane with $K+1$ marked points ($K\geq 0$) has $K$ {\bf real} moduli,
 whereas a marked strip with $K+2$ points (at least one for each boundary) has $K+2$ real moduli or, better, one complex
 modulus and $K$ real moduli. These are simply the differences of two chosen critical points on different boundaries (which gives a complex parameter with nonzero imaginary part)
 and the $K$ remaining relative positions of the other marked points on the two boundaries (which are real).

Near $z=\infty$ which is a pole of order $n+4$ for the quadratic differential $\phi$, the general theory \cite{strebel}  shows  that
\begin{enumerate}
\item  any horizontal line approaches $\infty$ asymptotic to $n+2$ directions forming relative angles $\frac {2\pi}{n+2}$. 
\item \label{fact2} any noncritical horizontal line in a neighborhood of the pole is a topological circle, approaching the pole along two consecutive asymptotic rays.
\end{enumerate}

\begin{wrapfigure}{r}{0.4\textwidth}
\resizebox{0.4\textwidth}{!}{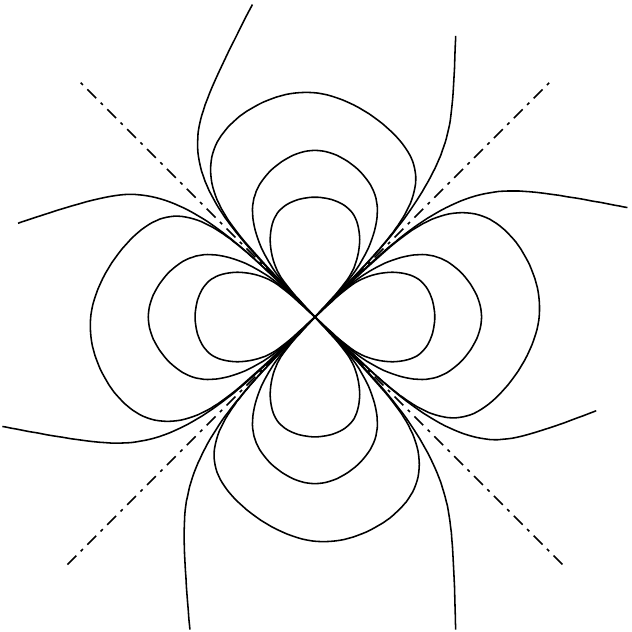}
\caption {The structure of the horizontal lines near infinity (in the local parameter $\zeta  = \frac 1 z$), here for a  polynomial of degree $2$.}
\label{Fig:infinity}
\end{wrapfigure}
This in particular forces the following 
\bl 
\label{asymstrip} The
left of the two rims of any strip approach  $\infty$
asymptotically along the same direction. Ditto for the right. 
\el
{\bf Proof.}
 Suppose that --say-- the right upper/lower rims
approach infinity along different critical directions; then, by
\ref{fact2} above, there would be a noncritical horizontal
trajectory within the strip that is topologically a circle with one point at $\infty$ and confined in $\Re(w)<<0$ or $\Re(w)>>0$. Such a
trajectory would not span the whole strip from one side to the
other. This is a contradiction, since horizontal noncritical
trajectories within a strip span the whole strip from left to
right. {\bf Q.E.D.}\par\vskip 5pt

\bl
 \label{noadjstrip} Two
strips cannot have a complete  boundary in common. 
\el
{\bf Proof.} If this were the case there would have to be at least one
critical point on the separating boundary; such point would have
only two critical lines, but then it would be necessarily regular
(the critical points are all at least trivalent). {\bf Q.E.D.}\par
\begin{figure}
\resizebox{15cm}{!}{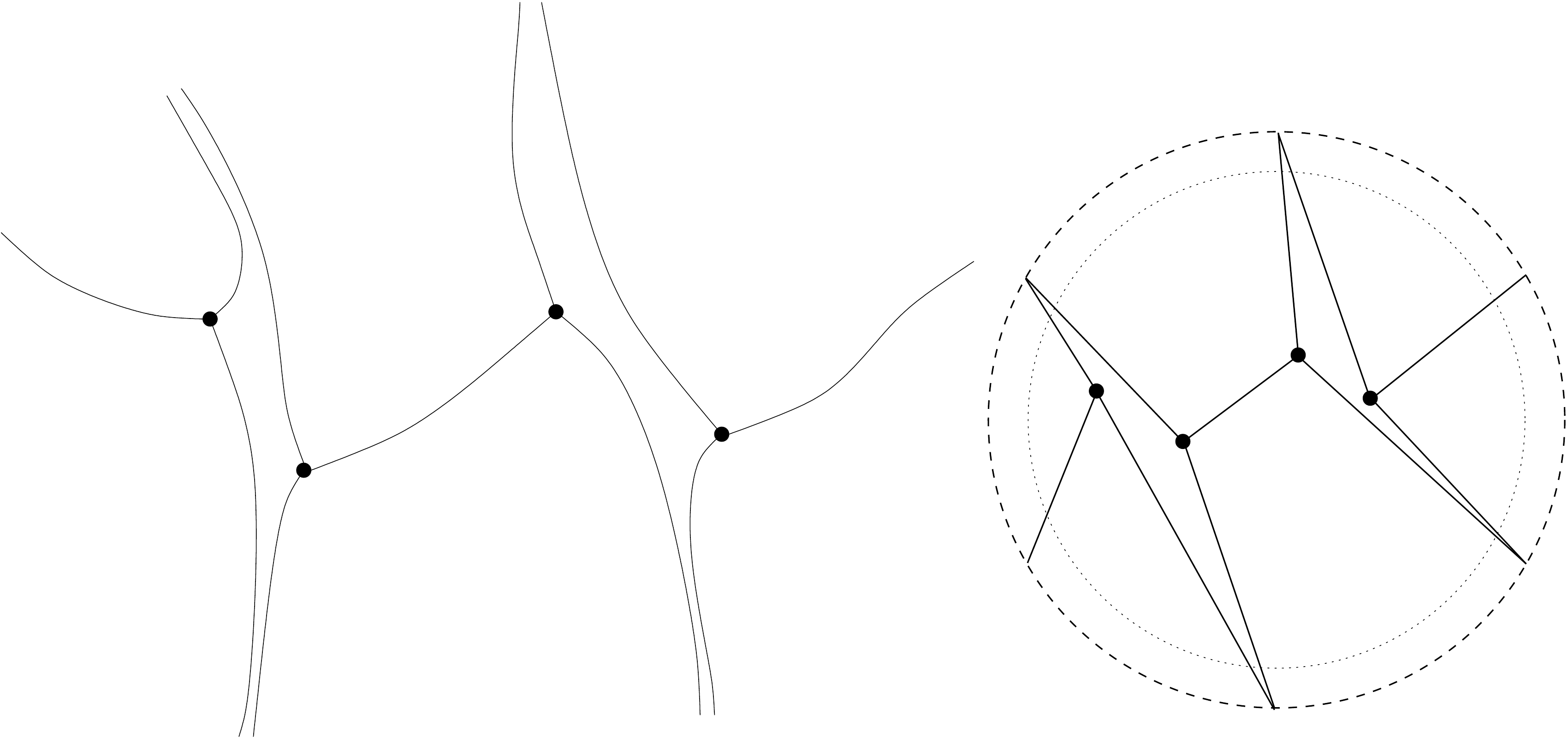}
\caption{An example of a critical graph for a polynomial of degree $4$ and its ``clock'' diagram.}
\label{clock}
\end{figure}

\vskip 5pt
We can depict the topological structure of the graph
$\mathfrak X$ as follows: we draw a disk, whose boundary
represents the asymptotic directions at infinity. On the boundary
we mark $n+2$ points representing the asymptotic directions of
approach.  Inside the disk we mark the critical points and connect
them according to the connectivity dictated by the graph
$\mathfrak X$ (see the example in Fig. \ref{clock}).

It follows from  Lemma \ref{asymstrip} that the two left (right) rims of a strip approach $\infty$ along the same direction, therefore they join at the same vertex on the clock diagram.
\paragraph {Decorations.}
Each strip has at least one critical point on each rim; we chose two such points $\alpha,\beta$  and associate to this pair the integral $\rho:= \int_\alpha^\beta \sqrt{P(x)}\rd x$,
 where the path of integration lies within said strip and the branch of the square-root is chosen so as to have a result with positive imaginary part. We call this number the {\bf modulus} of the strip (associated to the given choice of pair of points). 

If the boundary between regions has more than one critical point belonging to it then we assign to each arc between two adjacent critical points the Strebel length of that arc, 
namely $\int_\alpha^\beta \sqrt{P(x)} \rd x $, where the integral is performed along said arc and the branch is chosen so as to have a {\bf positive} result. 

The clock-diagram, together with these parameters (the complex moduli of the strips and the positive Strebel lengths of consecutive critical points on the same boundary) will be called the {\bf decorated clock diagram}.

The critical graph $\mathfrak X$ can be considered  (from a topological point of view) as a loop--free forest made of vertices of different valencies and edges connecting vertices with either other vertices or infinity (along a given direction). In the following lemma we study some elementary enumerative properties of this graph.

\bl
\label{counting}
Let $\mu_i+2$ be the valencies of the vertices, $V$ the number of vertices, $P$ the number of halfplanes and $S$ the number of strips. Let $o_i$ be the number of {\bf open}  
critical lines from the $i-$th vertex (i.e. lines that go to infinity), and $c_i$  the number of critical lines that are closed (i.e. go to some other vertex). Then
\bea
&& 2S + P = \sum o_i \label{imn}\\
&& S = V-1 - \frac 1 2\sum c_i\label{imi}
\eea
As a consequence, the number halfplanes is $n+2 =2+\sum \mu_i$.
 \el
{\bf Proof}.
First of all $o_i+c_i = \mu_i+2$ because this counts the valency of each vertex.
The sum $\sum c_i$ is even because each closed line appears in exactly two vertices (i.e. is counted twice).
Note  that each halfplane has at least one critical point (vertex) on its boundary, 
and there must be two open critical half-lines on its boundary.
On the other hand each strip has $4$ open trajectories (two for each side). So $4S+2P$ is the number of {\em sides} of open trajectories, i.e. twice the number of open trajectories $2\sum o_i$.
From this we have the first equation (\ref{imn}).

The second formula is proved as follows.
Assume first that there are no closed trajectories; in this case each vertex is on the boundary of some strip, unless there is only one vertex (in which case the formula holds trivially). This is so because the $i$-th vertex splits the plane into $v_i$ sectors, one of which must contain another vertex. Then the two rays bounding this sector are one side of a strip. Each vertex then contributes one strip and hence the number of strips is $V-1$.

Suppose now that the i-th vertex has $c_i$ closed trajectories; this means that there are $c_i$ other vertices connected to this one  (necessarily distinct, since there are no closed loops): they do not contribute {\em per se} to the number of strips. The formula follows.

Finally, by substituting (\ref{imi}) into (\ref{imn}), one has $P
= 2+\sum \mu_i = n+2$. {\bf Q.E.D.}\par\vskip 5pt

Collecting Lemmas \ref{l1},\ref{l2},\ref{l3}, \ref{counting},\ref{noadjstrip} together with the idea of the clock diagram we see that any clock-diagram must follow the rules formalized in the following definition.

\bd
\label{adm}
A (topological) clock diagram  is a regular $n+2$-gon called the {\bf clock} (whose  vertices are ``infinite'') containing 
$V$ vertices (the "finite vertices") of valencies $\mu_i+2$ such that 
$$
n =  \sum_{i=1}^V \mu_i
$$
Moreover
\begin{enumerate}
\item each edge connects two finite vertices or one finite vertex and an  infinite one;
\item there are no loops in the interior of the clock;
\item each component $\Gamma_j$ in which the interior of the clock is separated by the edges  is topologically a marked halfplane or strip.
\item each half--plane borders exactly one of the sides of the clock.
\end{enumerate}
\ed
\subsection{Inverse problem}
We want to show that for any  graph $\mathfrak X$ giving rise to a {\bf decorated} clock-diagram with the topological properties of Def. \ref{adm}, there is a polynomial whose critical graph corresponds to the given graph. This is essentially an (interesting) exercise in {\bf welding} \cite{strebel}.

In fact the problem is essentially solved and in the more general setting of rational quadratic differentials in \cite{jenkins}, so we basically rephrase the contents of \cite{jenkins} adding the details that are needed for our application. 

Suppose we have a clock-diagram; by Def. \ref{adm} the interior of the clock is partitioned in regions $\Gamma_j$ each of which  is either a halfplane or a strip (topologically). This implies that the given diagram tells us how to glue (topologically for the time being) half-planes and strips so as to have a simply connected topological space. 

We now choose an arbitrary decoration of the clock-diagram:
for each component of a boundary with $K+1$ marked points we assign (arbitrarily) $K$ real positive numbers, representing the relative distances between consecutive marked points; for each strip, we choose two marked points on the two distinct boundaries and assign an arbitrary complex parameter $\rho$, $\Im (\rho)>0$ to this pair.

By virtue of this construction we will have {\bf decorated} marked halfplanes/strips in such a way that the decorations on the boundaries (i.e. the relative distances of the marked points) match between neighboring $\Gamma_j$'s.

We think of these marked abstract halfplanes/strips $\Gamma_j$ as
realized in copies of the $w$-upper-half-plane and introduce a
{\bf flat} coordinate $w_j$ for each halfplane/strip, normalized
so that $w_j$ vanishes at one of the marked points on one
boundary.

We form the topological surface $X = \sqcup \overline{ \Gamma_j}/\sim$, where the equivalence relation is the {\bf metric}  identification of the boundaries according to the coordinates $w_j$ and the topological structure dictated by the clock diagram.

\subsubsection{Conformal structure}
The construction of the conformal structure follows \cite{jenkins}, with some minor deviations.

Our topological surface $X$ is connected and simply connected since
it is a model of the interior of the clock (and hence of the plane):
the conformal structure is defined in an interior point of
$\Gamma_j$ by the coordinate $w_j$ itself. A neighborhood $O$ of  a
critical vertex $a$ of multiplicity $\mu$ intersects $\mu + 2$
halfplanes/strips which we simply denote by
$\Gamma_{1},\dots,\Gamma_{\mu+2}$. The local coordinate $z$  in this
neighborhood $O$ is then defined by \be z(p) = {\rm e}^{\frac{2i\pi
\ell} {\mu+2}} w_{\ell}^{\frac 2{\mu+2}}, \quad p\in
O\cap\Gamma_{\ell},\qquad  w_\ell  = \int_a^z \sqrt{P(s)}\rd s,\ \
\ell = 1,\dots, \mu+2. \ee Near a smooth common boundary points
between --say-- $\Gamma_1,\Gamma_2$  of coordinates $s, \wt s$, the
coordinates are $(w_1-s)$ and $\pm (w_2-\wt s)$ on the two sides
(respectively), where the sign depends on the relative orientations
of the strips/halfplanes at that point.

It follows from these definitions of the conformal structure that $(\rd w_j)^2$ lifts to a well--defined {\bf holomorphic} quadratic differential on the (now) Riemann surface $X$; this differential has zeroes precisely at the marked points and are of multiplicities $\mu$.

In order to conclude that it defines a polynomial, we need to compactify the surface and show that the quadratic differential has a pole at the point of compactification.

\paragraph{Compactification}
We first topologically compactify $X$ using one-point compactification (Alexandrov).
 The ensuing topological space is connected, compact and simply connected.

Let $n = \sum \mu_i$ be the sum of all multiplicities of zeroes of the quadratic differential $\rd w^2$ which  --as discussed above--  is globally defined on $\overline X\setminus \{\infty\}$.

We need to define a local coordinate $\zeta$ at the compactification point $\infty$; since a neighborhood of $\infty$ intersects all regions $\Gamma_j$'s, we need to express the to-be-defined coordinate $\zeta$ in terms of the coordinates $w_j$ naturally defined in each region; moreover we must do so in such a  way that boundaries of adjacent regions are mapped to the same line in the $\zeta$-plane near $\zeta=\zeta(\infty)=0$.

\begin{figure}
\resizebox{10cm}{!}{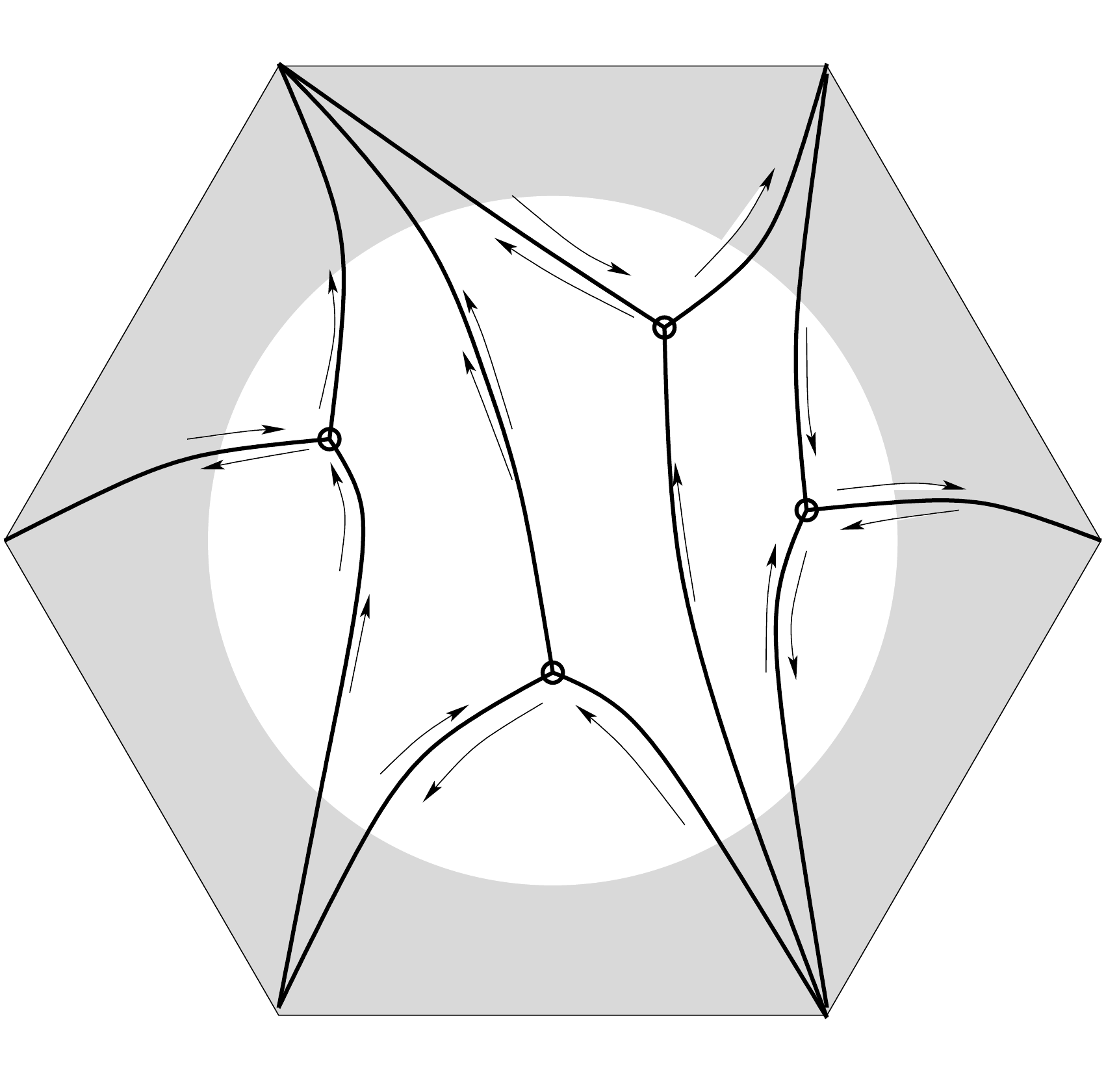}
\resizebox{5cm}{!}{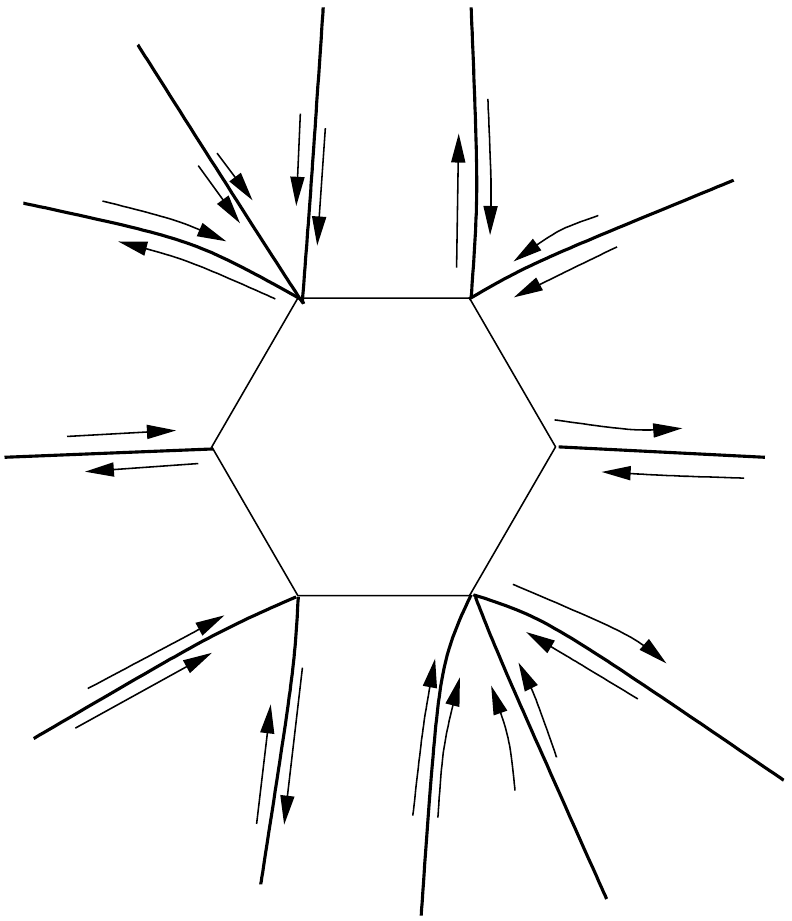 }
\caption{An example of the process of compactification and the corresponding inverted clock-diagram at the point of compactification.}
\label{fig:exampcomp}
\end{figure}

Let $U_\infty$ be a neighborhood of $\infty$ and let $V_j$ be the intersections of $U_\infty$ with all the domains $\Gamma_\ell$'s. To have a pictorial idea we can turn the clock-diagram inside-out so  that the inside of the $(n+2)$-polygon represents the point $\infty$ and the lines from the vertices the critical trajectories (see figure \ref{fig:exampcomp} for an example).

Let us proceed clockwise on such an inverted-clock diagram; because of our convention $\Im (w_j)$'s the orientation of two consecutive half-planes (the regions that border one edge of the inverted-clock diagram) must be opposite, independently on the number and orientations of the subregions of strips that may be incident on the vertex between the two planes.

Starting from region $V_1$ we can continue the differential $\rd w = \rd w_1$ to the region $V_2$, where $\rd w= \pm \rd w_2$.
If $V_2$ is one end of a strip then we may have one or the other sign, depending on  our choice of uniformizer; if $V_2$ is another half-plane then necessarily we have a minus sign according to our observation above.

Recall that there are $n+2 =2+\sum \mu_j$ half-planes (edges); therefore after going around $\infty$ once  we will have monodromy sign of $+1$ if $n$ is even, $-1$ if $n$ is odd.

In other words we have
\bl
If $n=2k$ then the differential $\rd w$ is well-defined (up to overall sign) in $U_\infty\setminus\{\infty\}$.\\
If $n=2k+1$ then the differential $\rd w$ is  well-defined on the double-cover of $U_\infty\setminus \{\infty\}$. 
\el

In the case of even $n$ (the case of interest to us) we can define a function $w$ by integrating (one branch of) $\rd w$; this function is not single-valued in general but has additive monodromy 
\be
w\mapsto w  + \oint \rd w =w+2i\pi \beta\ .
\ee
In fact it is not difficult to see that $\beta$ has a geometrical meaning of the alternating sum over all vertices of the clock of the total modular width of the strips incident to each vertex.

If $n$ is odd it is also not difficult to see that $\oint \rd w$ (where the integral winds twice around the compactification point) is always zero due to the monodromy $\rd w\mapsto -\rd w$ after one loop.

A uniformizer $\zeta = \zeta(w)$ must have the following properties
\begin{itemize}
\item $\zeta(w)$ is a locally analytic function for $|w|>>0$;
\item $\zeta(w+\beta) = \zeta(w)$;
\item $\zeta(w)$ has a  singularity of type $w^{-\frac n 2-1}$ (so as to be able to accommodate $n+2$ halfplanes in one).
\end{itemize}

We sketch the main steps of the construction 
following the ideas in  \cite{jensen, strebel}. 

Consider the intersection of  a halfplane, denoted by  $\Pi_1$, with a neighborhood of infinity; $\Pi_1$ comes equipped by definition with a coordinate $w_1$. We now proceed clockwise around infinity and consider the next region of intersection; this might be a strip or another halfplane which borders $\Pi_1$ on the boundary  $\Im(w_1)=0,\ \Re(w_1)>>1$.

Using now the coordinate of this region (possibly up to translations) we can extend the coordinate $w_1$ to this strip; if the other edge of this strip borders another strip we repeat the procedure until we reach a strip which borders with the next clockwise halfplane $\Pi_2$.

\begin{figure}
\resizebox{14cm}{!}{
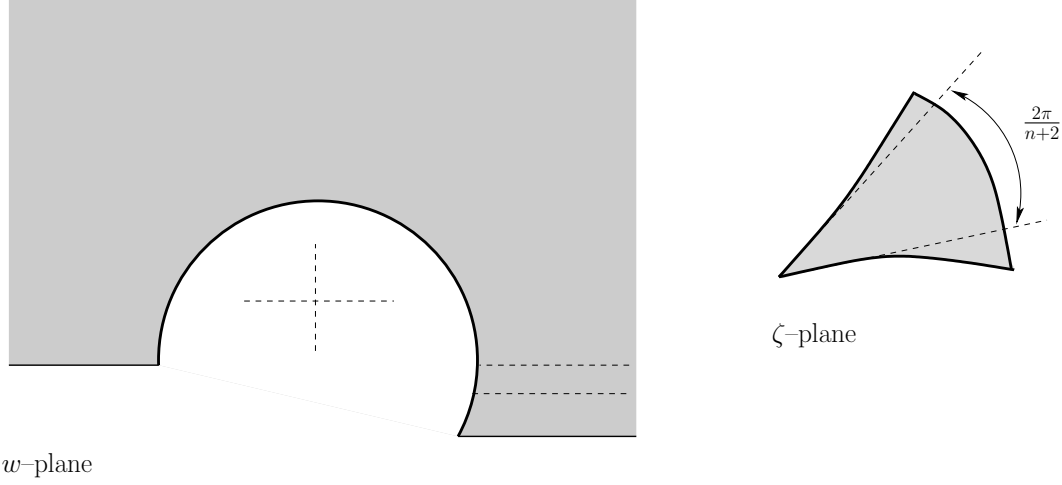}
\caption{The regions used for compactification; on the left an example of a shifted region $H_\rho + \rho'$ and on the right the image through $\zeta(w)$ of that region; the asymptotic directions have an angular separation of $2\pi/(n+2)$. Subsequent regions are mapped to the $\zeta$ plane by choosing different branches so as to fill all sectors of a neighborhood of $\zeta=0$.
Also shown are the separations between the added half-strips on the bottom right of the $w$ halfplane (in this case two strips).}
\label{compact2}
\end{figure}

The region we obtain looks like the one in Fig. \ref{compact2}

Let's denote $H_{\rho_1}$ this region, where $\rho_1$ is the sum of all complex moduli of the parts of strips that we have attached to the right of the halfplane: we repeat this construction for all $n+2$ half-planes, obtaining thus regions $H_{\rho_2}, \dots, H_{\rho_{n+2}}$.

In order to glue one such region to the next, keeping in mind that consecutive half-planes have upside-down matching, we have to glue $H_{\rho_1}$ with $-H_{\rho_2} + \rho_1$ and so on and so forth.
We obtain a chain of regions (we assume for definiteness $n$ even)
\be
H_{\rho_1} \rightarrow -H_{\rho_2} + \rho_1 \rightarrow H_{\rho_3} -\rho_1 + \rho_2 \rightarrow -H_{\rho_4}+\rho_1-\rho_2+\rho_3\cdots H_{\rho_{n+2}} + \sum_{j=1}^{n+1} (-1)^j \rho_j\ .
\ee
The resulting Riemann surface  $\mathcal V$ has  in particular two boundaries corresponding to a half-line in the negative real  $w$ axis of $H_{\rho_1}$ and and a half-line in the line $\{ w+\rho_{n+2},\ \ \Im(w)=0,\   \Re (w)>>0\} $ of the last region; we denote them by $\Sigma_L,  \Sigma_R$ respectively.

In addition on $\mathcal V$ the function $w$ is now single valued and  realizes a $n/2+1$ cover of a neighborhood of $\infty$; the two boundaries $\Sigma_L$ and $\Sigma_R$ are mapped to
\bea
&\& w(\Sigma_L) \subset \{\Im(w)=0 \ \ \Re(w) <<0\}\\
&\& w(\Sigma_R) \subset \{\Im (w) = \Im (\beta)\ \ \ \Re(w)<<0\}
\eea
where 
\be
\beta := \frac 1{2i\pi} \sum_{j}^{n+2} (-1)^j \rho_j
\ee
and the points on these lines are identified by $w \sim w+ \beta$.
%
%
%
Now consider the map $\zeta(w)$ defined implicitly (up to overall additive constant) by
\be
\zeta^{-\frac {n+2}2} + \beta \ln \zeta = w\label{zeta}
\ee
For $|w|$ large there is a univalent branch of $\zeta(w)$ which behaves as any given branch of $w^{-\frac 2{n+2}}$.

Under this map, the line $\Sigma_L$ is mapped by (one branch of)  $\zeta(w)$ to a Jordan arc $\zeta(\Sigma_L)$ approaching $\zeta=0$.
Consider a neighborhood $U$  of $\zeta =0$ with a cut along this arc, so as to have a  simply connected domain. 
The image  $w(U\setminus \zeta(\Sigma_L))$ covers a punctured neighborhood of  $w=\infty$ precisely $n/2+1$ times and realizes a  biholomorphic equivalence with the Riemann surface $\mathcal V$; because of the periodicity of $\zeta(w)$, the two sides of the cut $\zeta(\Sigma_L)$ are mapped to $\Sigma_L $ and $\Sigma_R$.

Thus the function $\zeta$ realizes the above identification of $\Sigma_L$ and $\Sigma_R$ and gives a conformal uniformization of $\mathcal X$; in addition $\zeta$ is continuous at the compactification point $\infty$ and therefore defines a conformal structure at this point\footnote{If $n$ is odd we have already shown that (on the double cover) $w$ has no monodromy and the construction works identically with the caveat that we need to use the double cover of the neighborhood of the compactification point.}.

This construction also  shows that the differential $\rd w^2$ has a pole of order $n+4$ at the compactification point $\infty$ in the local uniformizer $\zeta$.

Since $\overline X $ is holomorphically equivalent to the Riemann-sphere the differential $dw^2$ is represented in a global uniformizing coordinate $z$ as 
\be
\rd w^2 = P(z)\rd z^2\ ,
\ee
with $P(z)$ a polynomial of degree $n$: we have thus succeeded in proving the equivalence of polynomials and decorated admissible graphs.
\br
In a nutshell the above discussion boils down to the following statement: the space of  polynomials of given multiplicities of zeroes
\be
P(z) = \prod_{j=1}^K  (z-a_j)^{\mu_j}\ ,
\ee 
can  be (locally) parametrized  by decorated admissible Strebel graphs. The ``coordinates'' are (essentially)
\be
E_{j-1}:= \int_{a_1}^{a_j} \sqrt{P(z)}{\rm d}z\ ,\qquad j=2,\dots,
\ee
up to translations and dilations that leave these integrals invariant (and for some choice of the contour of integrations and branch of the square-root).

A na\"ive parameter counting confirms this fact: polynomials of given multiplicities of zeroes are parametrized by the $K$ positions of the zeroes (distinct) and an overall multiplicative constant, in total $K+1$ complex parameters. If we factor out the group of invariance of the above integrals, namely  the action of translations $z\mapsto z+c$ and dilations
\be
\wt z = \lambda z\ ,\qquad  \wt P(\wt z) = \lambda^2 P(\lambda z)
\ee 
we have the same number of parameters, $K-1$.

For our application, however, it is essential to control the global topological structure of the critical graph, something the above coordinates tell nothing about. 
One obvious reason is that the zeroes of the polynomials are on the same footing and hence --even for the same polynomial-- we could assign different coordinates. More importantly,
we could have two inequivalent clock diagrams with decorations given by the same numbers: the corresponding polynomials would then be different but with the same Strebel lengths between zeroes\footnote{To put it differently, clock diagrams with given number of finite vertices  represent {\bf cells} in the space of polynomials; the decorations give coordinates to each cell, but it is meaningless on a global level to consider only the decoration.}.
\er
\subsection{(Admissible) Boutroux curves}

In the case of relevance to our paper  we must restrict the parametrization given by the decorated clock--diagram of a polynomial to a suitable submanifold; this is the submanifold of polynomials of the form
\be
P(z) =  - M^2(z)\prod_{j=1}^{2g+2}(z-\alpha_j) = -y^2
\ee
(where the sign is just conventional for our application so that the function $h$ introduced in Def. \ref{de:adm} is the imaginary part of the Strebel flat coordinate $w$) with additional constraints on the clock--diagram and on its decoration described below.

\begin{wrapfigure}{r}{0.4\textwidth}
\resizebox{0.4\textwidth}{!}{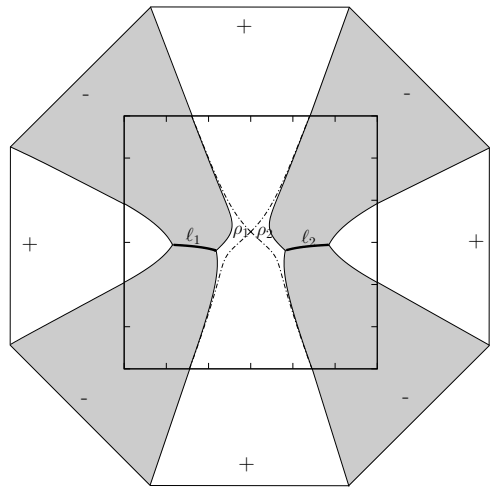}
\caption{An example of reconstruction of admissible triple. The part within the graduated square is an actual numerical output.}
\label{fig:examp1}
\end{wrapfigure}

The corresponding clock--diagram  has only trivalent vertices (the $\alpha_j$'s) and even-valent ones (the zeroes of $M$). The Boutroux condition (Def. \ref{de:Boutroux}), as discussed 
implies that the {\bf critical} $0$-level set\footnote{The {\bf critical level set} of a function is the union of all connected components of the level-set that contain at least one critical point.}
$\mathfrak X_0$ of $\Im \int_{\alpha_1}^x \sqrt{P(z)}\rd z$
is well defined independently of the choice of critical point $\alpha_j$ and independently of the choice of contour of integration.

In this case the {\bf height} function $h = \Im w$ can be defined as a continuous global function with smooth saddle points at the zeroes of $M(z)$: indeed it is easily seen that there is a consistent choice of signs for the Strebel coordinate near each even critical point (the zeroes of $M$ all appear with even multiplicity in the quadratic differential) in such a way that the height function $h = \Im (w)$ is harmonic there.

The critical horizontal graph of $P(z)$ thus naturally splits into $\mathfrak X_0\sqcup \mathfrak X_1$, where all the $\alpha_j$'s belong to $\mathfrak X_0$.
 The distances between critical  points $\alpha_j$ belonging to the same connected component of $\mathfrak X_0$ are arbitrary real numbers.
 The other moduli are the complex moduli of the strips: they must satisfy the real constraint that the heights of the strips ``between'' two connected components of $\mathfrak X_0$ must add up to zero.

The reconstruction theorems proved in the previous sections could be rephrased  by saying ``if you can draw the graph you want, then it exists (with the same topological features)''; instead of giving an abstract account of the procedure we find it more instructive and transparent to analyze some examples.

\begin{wrapfigure}{r}{0.4\textwidth}
\resizebox{0.4\textwidth}{!}{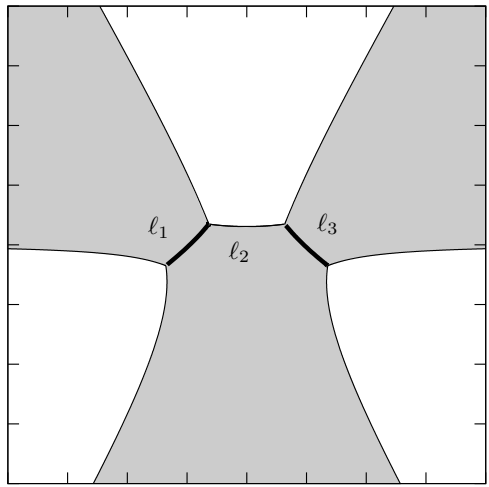}
\caption{An example of reconstruction of admissible triple of unusual topology (numerical output).}
\label{fig:examp2}
\end{wrapfigure}

\begin{figure}
\resizebox{8cm}{!}{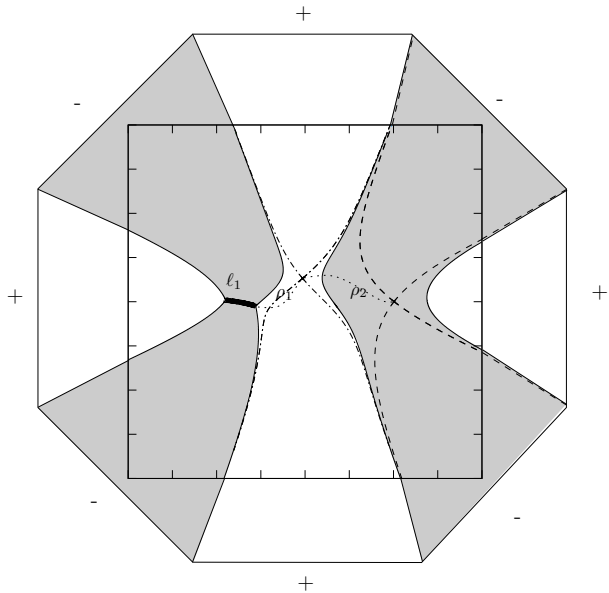}
\caption{An example of admissible triple where the Stokes sectors are not all connectible by admissible paths (the part within the graduated square is a numerical output). Here $\Im (\rho_2)> \Im (\rho_1)$.}
\label{fig:examp4}
\end{figure}

\begin{wrapfigure}{r}{0.4\textwidth}
\resizebox{0.4\textwidth}{!}{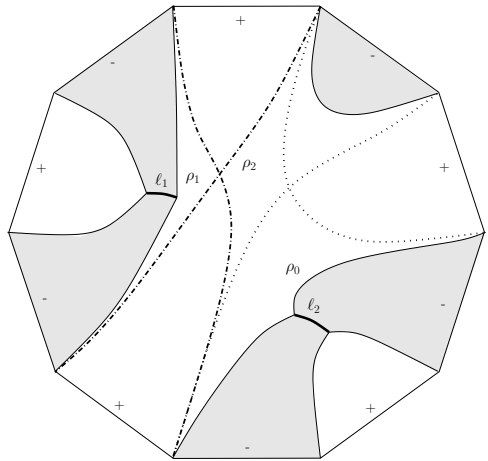}
\caption{An example of reconstruction of admissible triple of unusual topology with two double points: the constraint is that $\Im(\rho_1) = \Im(\rho_2)+\Im(\rho_0)$. The  shaded area is the where $h<0$ (the ``underwater''). The dashed curves are at positive equal height $h$ (the dash-dot one  is higher). Not a numerical output.}
\label{fig:examp3}
\end{wrapfigure}

Let us consider the examples in figures (\ref{fig:examp1}, \ref{fig:examp2}, \ref{fig:examp3}, \ref{fig:examp4}).
In Fig. \ref{fig:examp1} 
the black contours make up $\mathfrak X_0$, whereas the complete Strebel critical graph is the collection of solid and dashed curves.
The  numbers $\ell_1,\ell_2$ are arbitrary real (positive) numbers; they are $\int_{\alpha}^{\alpha'} \sqrt{P}\rd x$ (where $\alpha,\alpha'$ denote the two critical points. 
The two numbers $\rho_i$ are instead complex (with  nonzero imaginary part) and correspond to $\int_{\alpha}^c\sqrt{P}\rd x$, where $c$ is the critical point in the middle.  
They satisfy the only constraint that $\Im(\rho_1) = \Im (\rho_2)$. Of course there could be more strips between the two connected components of $\mathfrak X_0$. (The normalization $\res{\infty} y\rd x=1$ imposes the constraint $\ell_1 +\ell_2 = 1$).

The topology of this example is the ``usual'' one in the {\em two-cut} potentials: there are two arcs supporting the asymptotic
 distributions of zeroes of the orthogonal polynomials and the differential $y\rd x$  has a critical  point between them. The degree of the potential $V$ is here quartic.

The second example (Fig. \ref{fig:examp2}) is ``unusual'' but it can happen for this class of semiclassical orthogonal polynomials: it is  a cubic  potential without double zero for the differential $y\rd x$ (similarly, one could have also a quartic potential without double zero). The support of the zeroes of the polynomials corresponds to the thick lines. There are three free real parameters $\ell_1,\ell_2,\ell_3$ (subject to $\ell_1+\ell_3 = 1$, which is the normalization condition).

The third  example (Fig. \ref{fig:examp3}) is the case of two double points and double support (something that cannot happen for ordinary orthogonal polynomials) 

The fourth example (Fig. \ref{fig:examp4}) shows that it can happen that not all Stokes sectors could be joined by curves $\Gamma$ satisfying the requirements for the steepest--descent method: 
here $\ell_1\in \R_+$ is arbitrary, and $\rho_1,\rho_2\in \mathcal H_+$ are also arbitrary, but with the condition $\Im(\rho_2)> \Im (\rho_1)$. The Stokes sector on 
the right (represented by the edge of the hexagon bordering the white area) cannot be joined to the other sectors because there is a river of $h<0$ ``underwater''
 in between (the critical value at the point $c$ of intersection of the dashed lines has $h(c)<0$).  

All these examples are admissible for our asymptotic study in the sense of Def. \ref{de:adm}; even without a formalization of the rules that make an admissible
 triple, the reader should have no difficulty in imagining and drawing on paper even very complicated decorated clock-diagrams that correspond to a situation as in Def. \ref{de:adm}.  

\section{Conclusion}
The first two parts (Sec. \ref{se:Jacobi} and \ref{se:relat}) have shown that any admissible triple (Def. \ref{de:adm}) is associated to the asymptotics with respect to  varying weights of certain pseudo--orthogonal polynomials. Some concluding remarks that follow from the construction are in order;
\begin{itemize}
\item  there might be different orthogonal polynomials that have the same asymptotic. This can happen if the potential is the same (necessarily) and the asymptotic Boutroux admissible curve has two Stokes sectors that belong to the same connected component of $\{h(x)>0\}$.  Indeed in this case the  jump on the contour joining them and remaining in the positive-$h$ region will be exponentially close (and uniformly) to the identity in the large $N$ limit, thus becoming irrelevant.
\item  If two Stokes sectors cannot be joined by a curve in $\mathcal Y_+\cup \mathcal B$ (which happens for example in Fig \ref{fig:examp4}) then the corresponding finite-$n$ RHP for the orthogonal polynomials cannot have a jump on a contour joining them.
\item In the asymptotic regime the change of Stokes' parameters (the $\varkappa$'s) is {\bf isospectral}; indeed it simply corresponds to choice of a different Stokes-Kirchoff's differential $\eta$, hence of a different twisting of the line--bundle of the
s. It is easy (but we don't do it here for brevity) to write explicitly the conjugating matrix for the spectral --problem associated to the spinors (eq. \ref{Laxmatrix})
 \end{itemize}

The last part (Sec. \ref{se:reconstr}) has shown that there can be admissible triples whose branch-cut structure $\mathcal B$ can be basically as complicated as one may wish and it is (topologically) a  forest of loop-free trivalent trees; the zeroes of the pseudo--orthogonal polynomials then accumulate on $\mathcal B$ and we have (not completely rigorously) specified the asymptotic density of zeroes along these arcs in Sec. \ref{se:density}.

\subsection{Outlook}
The definition of admissibility (Def. \ref{de:adm}) is forced upon us because the jump matrices (Stokes' matrices) in the Riemann--Hilbert problem for orthogonal polynomials are all {\bf upper triangular}.  In a more general setting, e.g. as  in the study of Painlev\'e\ II equations in the asymptotic regime \cite{its}, there can be also lower triangular Stokes matrices: in this case we could relax the notion of admissibility. The key fact  that drives the steepest-descent method is that  the upper-lower triangular form of the jump matrix should be related to the signs of $h$ on the two sides of the branch-cut (i.e. if $h<0$ on both sides then the jump should be upper triangular, if $h>0$ then the jump should be lower triangular). 

This means that one could use any Boutroux curve to model the asymptotics of {\em some} RHP by choosing the jump-matrices of the appropriate form according to the signs of $h$.
This seems a promising avenue of research that we intend to pursue in a different publication.
  
 Another extension of the present setting that we set out to explore is the inclusion of hard-edges and more general potentials with rational derivative; while the heuristic approach remains unchanged (and in fact the first part of this paper is already developed to the full extent of this generality) several changes need to be made in the steepest descent part and in the analysis of critical trajectories.
 
 It seems that the introduction of hard-edges is the simplest generalization; this requires the use of a different local parametrix near each hard-edge (built out of Bessel functions) and the modification of the construction of admissible curves (some ingredients for this generalization can be found in \cite{kenarno}). All this will be dealt with  in a forthcoming publication.

\end{document}

%% file: Introfig_bw_enhanced.pdf_t
\begin{picture}(0,0)%
\includegraphics{Introfig_bw_enhanced.pdf}%
\end{picture}%
\setlength{\unitlength}{3947sp}%
\begingroup\makeatletter\ifx\SetFigFont\undefined%
\gdef\SetFigFont#1#2#3#4#5{%
  \reset@font\fontsize{#1}{#2pt}%
  \fontfamily{#3}\fontseries{#4}\fontshape{#5}%
  \selectfont}%
\fi\endgroup%
\begin{picture}(3224,3223)(589,-2561)
\end{picture}%

%% file: Branchcute.pdf_t
\begin{picture}(0,0)%
\includegraphics{Branchcute.pdf}%
\end{picture}%
\setlength{\unitlength}{3947sp}%
\begingroup\makeatletter\ifx\SetFigFont\undefined%
\gdef\SetFigFont#1#2#3#4#5{%
  \reset@font\fontsize{#1}{#2pt}%
  \fontfamily{#3}\fontseries{#4}\fontshape{#5}%
  \selectfont}%
\fi\endgroup%
\begin{picture}(6549,1899)(289,-1723)
\end{picture}%

%% file: Admissyes.pdf_t
\begin{picture}(0,0)%
\includegraphics{Admissyes.pdf}%
\end{picture}%
\setlength{\unitlength}{3947sp}%
\begingroup\makeatletter\ifx\SetFigFont\undefined%
\gdef\SetFigFont#1#2#3#4#5{%
  \reset@font\fontsize{#1}{#2pt}%
  \fontfamily{#3}\fontseries{#4}\fontshape{#5}%
  \selectfont}%
\fi\endgroup%
\begin{picture}(19866,8105)(193,-7751)
\put(13774,-3114){\makebox(0,0)[lb]{\smash{{\SetFigFont{20}{24.0}{\familydefault}{\mddefault}{\updefault}{\color[rgb]{0,0,0}-}%
}}}}
\put(11987,-4283){\makebox(0,0)[lb]{\smash{{\SetFigFont{20}{24.0}{\familydefault}{\mddefault}{\updefault}{\color[rgb]{0,0,0}+}%
}}}}
\put(14117,-6275){\makebox(0,0)[lb]{\smash{{\SetFigFont{20}{24.0}{\familydefault}{\mddefault}{\updefault}{\color[rgb]{0,0,0}-}%
}}}}
\put(17071,-6206){\makebox(0,0)[lb]{\smash{{\SetFigFont{20}{24.0}{\familydefault}{\mddefault}{\updefault}{\color[rgb]{0,0,0}+}%
}}}}
\put(19133,-4558){\makebox(0,0)[lb]{\smash{{\SetFigFont{20}{24.0}{\familydefault}{\mddefault}{\updefault}{\color[rgb]{0,0,0}-}%
}}}}
\put(16934,-2909){\makebox(0,0)[lb]{\smash{{\SetFigFont{20}{24.0}{\familydefault}{\mddefault}{\updefault}{\color[rgb]{0,0,0}+}%
}}}}
\put(4801,-1861){\makebox(0,0)[lb]{\smash{{\SetFigFont{20}{24.0}{\familydefault}{\mddefault}{\updefault}{\color[rgb]{0,0,0}+}%
}}}}
\put(6751,-2986){\makebox(0,0)[lb]{\smash{{\SetFigFont{20}{24.0}{\familydefault}{\mddefault}{\updefault}{\color[rgb]{0,0,0}-}%
}}}}
\put(7426,-4861){\makebox(0,0)[lb]{\smash{{\SetFigFont{20}{24.0}{\familydefault}{\mddefault}{\updefault}{\color[rgb]{0,0,0}+}%
}}}}
\put(4801,-4636){\makebox(0,0)[lb]{\smash{{\SetFigFont{20}{24.0}{\familydefault}{\mddefault}{\updefault}{\color[rgb]{0,0,0}-}%
}}}}
\put(2176,-4861){\makebox(0,0)[lb]{\smash{{\SetFigFont{20}{24.0}{\familydefault}{\mddefault}{\updefault}{\color[rgb]{0,0,0}+}%
}}}}
\put(2401,-2911){\makebox(0,0)[lb]{\smash{{\SetFigFont{20}{24.0}{\familydefault}{\mddefault}{\updefault}{\color[rgb]{0,0,0}-}%
}}}}
\end{picture}%

%% file: Sigmacut.pdf_t
\begin{picture}(0,0)%
\includegraphics{Sigmacut.pdf}%
\end{picture}%
\setlength{\unitlength}{3947sp}%
\begingroup\makeatletter\ifx\SetFigFont\undefined%
\gdef\SetFigFont#1#2#3#4#5{%
  \reset@font\fontsize{#1}{#2pt}%
  \fontfamily{#3}\fontseries{#4}\fontshape{#5}%
  \selectfont}%
\fi\endgroup%
\begin{picture}(10002,2896)(1411,-4284)
\put(9751,-2686){\makebox(0,0)[lb]{\smash{{\SetFigFont{20}{24.0}{\familydefault}{\mddefault}{\updefault}{\color[rgb]{0,0,0}$\Sigma$}%
}}}}
\put(1426,-2086){\makebox(0,0)[lb]{\smash{{\SetFigFont{17}{20.4}{\familydefault}{\mddefault}{\updefault}{\color[rgb]{0,0,0}$\alpha_1$}%
}}}}
\end{picture}%

%% file: Kirchoff_bw.pdf_t
\begin{picture}(0,0)%
\includegraphics{Kirchoff_bw.pdf}%
\end{picture}%
\setlength{\unitlength}{3947sp}%
\begingroup\makeatletter\ifx\SetFigFont\undefined%
\gdef\SetFigFont#1#2#3#4#5{%
  \reset@font\fontsize{#1}{#2pt}%
  \fontfamily{#3}\fontseries{#4}\fontshape{#5}%
  \selectfont}%
\fi\endgroup%
\begin{picture}(4499,4079)(1839,-5048)
\put(4111,-3046){\makebox(0,0)[lb]{\smash{{\SetFigFont{17}{20.4}{\familydefault}{\mddefault}{\updefault}{\color[rgb]{0,0,0}$\rho_3$}%
}}}}
\put(5926,-2791){\makebox(0,0)[lb]{\smash{{\SetFigFont{17}{20.4}{\familydefault}{\mddefault}{\updefault}{\color[rgb]{0,0,0}$\rho_4$}%
}}}}
\put(5941,-3706){\makebox(0,0)[lb]{\smash{{\SetFigFont{17}{20.4}{\familydefault}{\mddefault}{\updefault}{\color[rgb]{0,0,0}$\rho_5$}%
}}}}
\put(3061,-2611){\makebox(0,0)[lb]{\smash{{\SetFigFont{17}{20.4}{\familydefault}{\mddefault}{\updefault}{\color[rgb]{0,0,0}$\rho_1$}%
}}}}
\put(3076,-4036){\makebox(0,0)[lb]{\smash{{\SetFigFont{17}{20.4}{\familydefault}{\mddefault}{\updefault}{\color[rgb]{0,0,0}$\rho_2$}%
}}}}
\end{picture}%

%% file: CrazyExample_1_bw.pdf_t
\begin{picture}(0,0)%
\includegraphics{CrazyExample_1_bw.pdf}%
\end{picture}%
\setlength{\unitlength}{3947sp}%
\begingroup\makeatletter\ifx\SetFigFont\undefined%
\gdef\SetFigFont#1#2#3#4#5{%
  \reset@font\fontsize{#1}{#2pt}%
  \fontfamily{#3}\fontseries{#4}\fontshape{#5}%
  \selectfont}%
\fi\endgroup%
\begin{picture}(21325,21612)(-4933,-13278)
\end{picture}%

%% file: normallens.pdf_t
\begin{picture}(0,0)%
\includegraphics{normallens.pdf}%
\end{picture}%
\setlength{\unitlength}{3947sp}%
\begingroup\makeatletter\ifx\SetFigFont\undefined%
\gdef\SetFigFont#1#2#3#4#5{%
  \reset@font\fontsize{#1}{#2pt}%
  \fontfamily{#3}\fontseries{#4}\fontshape{#5}%
  \selectfont}%
\fi\endgroup%
\begin{picture}(5349,3102)(2914,-7882)
\put(4726,-5911){\makebox(0,0)[lb]{\smash{{\SetFigFont{14}{16.8}{\rmdefault}{\mddefault}{\updefault}{\color[rgb]{0,0,0}$\Sigma_i$}%
}}}}
\put(4876,-5011){\makebox(0,0)[lb]{\smash{{\SetFigFont{14}{16.8}{\rmdefault}{\mddefault}{\updefault}{\color[rgb]{0,0,0}$\Gamma_{i}^L$}%
}}}}
\put(4801,-7786){\makebox(0,0)[lb]{\smash{{\SetFigFont{14}{16.8}{\rmdefault}{\mddefault}{\updefault}{\color[rgb]{0,0,0}$\Gamma_{i}^R$}%
}}}}
\end{picture}%

%% file: biglens0.pdf_t
\begin{picture}(0,0)%
\includegraphics{biglens0.pdf}%
\end{picture}%
\setlength{\unitlength}{3947sp}%
\begingroup\makeatletter\ifx\SetFigFont\undefined%
\gdef\SetFigFont#1#2#3#4#5{%
  \reset@font\fontsize{#1}{#2pt}%
  \fontfamily{#3}\fontseries{#4}\fontshape{#5}%
  \selectfont}%
\fi\endgroup%
\begin{picture}(6723,3290)(5321,-6664)
\put(5701,-6436){\makebox(0,0)[lb]{\smash{{\SetFigFont{12}{14.4}{\rmdefault}{\mddefault}{\updefault}{\color[rgb]{0,0,0}$\alpha_{\ell_i}$}%
}}}}
\put(7276,-6361){\makebox(0,0)[lb]{\smash{{\SetFigFont{12}{14.4}{\rmdefault}{\mddefault}{\updefault}{\color[rgb]{0,0,0}$\alpha_{\ell_{i+1}}$}%
}}}}
\put(6676,-6586){\makebox(0,0)[lb]{\smash{{\SetFigFont{12}{14.4}{\rmdefault}{\mddefault}{\updefault}{\color[rgb]{0,0,0}$r_i$}%
}}}}
\end{picture}%

%% file: biglens.pdf_t
\begin{picture}(0,0)%
\includegraphics{biglens.pdf}%
\end{picture}%
\setlength{\unitlength}{3947sp}%
\begingroup\makeatletter\ifx\SetFigFont\undefined%
\gdef\SetFigFont#1#2#3#4#5{%
  \reset@font\fontsize{#1}{#2pt}%
  \fontfamily{#3}\fontseries{#4}\fontshape{#5}%
  \selectfont}%
\fi\endgroup%
\begin{picture}(6723,3077)(5321,-6451)
\end{picture}%

%% file: biglenscut1.pdf_t
\begin{picture}(0,0)%
\includegraphics{biglenscut1.pdf}%
\end{picture}%
\setlength{\unitlength}{4144sp}%
\begingroup\makeatletter\ifx\SetFigFont\undefined%
\gdef\SetFigFont#1#2#3#4#5{%
  \reset@font\fontsize{#1}{#2pt}%
  \fontfamily{#3}\fontseries{#4}\fontshape{#5}%
  \selectfont}%
\fi\endgroup%
\begin{picture}(5025,4316)(3406,-6981)
\put(4692,-4249){\makebox(0,0)[lb]{\smash{{\SetFigFont{14}{16.8}{\rmdefault}{\mddefault}{\updefault}{\color[rgb]{0,0,0}$r_{km}$}%
}}}}
\put(6324,-3717){\makebox(0,0)[lb]{\smash{{\SetFigFont{14}{16.8}{\rmdefault}{\mddefault}{\updefault}{\color[rgb]{0,0,0}$\Sigma_l$}%
}}}}
\put(6057,-5046){\makebox(0,0)[lb]{\smash{{\SetFigFont{14}{16.8}{\rmdefault}{\mddefault}{\updefault}{\color[rgb]{0,0,0}$\Sigma_m$}%
}}}}
\put(6020,-3337){\makebox(0,0)[lb]{\smash{{\SetFigFont{14}{16.8}{\rmdefault}{\mddefault}{\updefault}{\color[rgb]{0,0,0}$r_{lk}$}%
}}}}
\put(7021,-4111){\makebox(0,0)[lb]{\smash{{\SetFigFont{14}{16.8}{\rmdefault}{\mddefault}{\updefault}{\color[rgb]{0,0,0}$r_{ml}$}%
}}}}
\put(5176,-3661){\makebox(0,0)[lb]{\smash{{\SetFigFont{14}{16.8}{\rmdefault}{\mddefault}{\updefault}{\color[rgb]{0,0,0}$\Sigma_k$}%
}}}}
\put(3421,-5281){\makebox(0,0)[lb]{\smash{{\SetFigFont{14}{16.8}{\familydefault}{\mddefault}{\updefault}{\color[rgb]{0,0,0}$\Gamma_m^L$}%
}}}}
\put(8416,-4966){\makebox(0,0)[lb]{\smash{{\SetFigFont{14}{16.8}{\familydefault}{\mddefault}{\updefault}{\color[rgb]{0,0,0}$\Gamma_m^R$}%
}}}}
\put(7606,-3661){\makebox(0,0)[lb]{\smash{{\SetFigFont{14}{16.8}{\familydefault}{\mddefault}{\updefault}{\color[rgb]{0,0,0}$\Gamma_l^R$}%
}}}}
\put(6436,-2896){\makebox(0,0)[lb]{\smash{{\SetFigFont{14}{16.8}{\familydefault}{\mddefault}{\updefault}{\color[rgb]{0,0,0}$\Gamma_l^L$}%
}}}}
\end{picture}%

%% file: parametrix1.pdf_t
\begin{picture}(0,0)%
\includegraphics{parametrix1.pdf}%
\end{picture}%
\setlength{\unitlength}{3947sp}%
\begingroup\makeatletter\ifx\SetFigFont\undefined%
\gdef\SetFigFont#1#2#3#4#5{%
  \reset@font\fontsize{#1}{#2pt}%
  \fontfamily{#3}\fontseries{#4}\fontshape{#5}%
  \selectfont}%
\fi\endgroup%
\begin{picture}(11202,4161)(964,-8014)
\put(9826,-4036){\makebox(0,0)[rb]{\smash{{\SetFigFont{12}{14.4}{\rmdefault}{\mddefault}{\updefault}{\color[rgb]{0,0,0}$\gamma_1$}%
}}}}
\put(10201,-6286){\makebox(0,0)[rb]{\smash{{\SetFigFont{12}{14.4}{\rmdefault}{\mddefault}{\updefault}{\color[rgb]{0,0,0}$\gamma_2$}%
}}}}
\put(10351,-7936){\makebox(0,0)[rb]{\smash{{\SetFigFont{12}{14.4}{\rmdefault}{\mddefault}{\updefault}{\color[rgb]{0,0,0}$\gamma_3$}%
}}}}
\put(12151,-6361){\makebox(0,0)[rb]{\smash{{\SetFigFont{12}{14.4}{\rmdefault}{\mddefault}{\updefault}{\color[rgb]{0,0,0}$\gamma_4$}%
}}}}
\put(2476,-6286){\makebox(0,0)[rb]{\smash{{\SetFigFont{12}{14.4}{\rmdefault}{\mddefault}{\updefault}{\color[rgb]{0,0,0}$\gamma_2$}%
}}}}
\put(3376,-7411){\makebox(0,0)[rb]{\smash{{\SetFigFont{12}{14.4}{\rmdefault}{\mddefault}{\updefault}{\color[rgb]{0,0,0}$\gamma_3$}%
}}}}
\put(2626,-4411){\makebox(0,0)[rb]{\smash{{\SetFigFont{12}{14.4}{\rmdefault}{\mddefault}{\updefault}{\color[rgb]{0,0,0}$\gamma_1$}%
}}}}
\put(6076,-6286){\makebox(0,0)[rb]{\smash{{\SetFigFont{12}{14.4}{\rmdefault}{\mddefault}{\updefault}{\color[rgb]{0,0,0}$\gamma_4$}%
}}}}
\put(5176,-5536){\makebox(0,0)[lb]{\smash{{\SetFigFont{12}{14.4}{\familydefault}{\mddefault}{\updefault}{\color[rgb]{0,0,0}I}%
}}}}
\put(5026,-6736){\makebox(0,0)[lb]{\smash{{\SetFigFont{12}{14.4}{\familydefault}{\mddefault}{\updefault}{\color[rgb]{0,0,0}IV}%
}}}}
\put(3601,-6511){\makebox(0,0)[lb]{\smash{{\SetFigFont{12}{14.4}{\familydefault}{\mddefault}{\updefault}{\color[rgb]{0,0,0}III}%
}}}}
\put(5716,-4441){\makebox(0,0)[rb]{\smash{{\SetFigFont{12}{14.4}{\familydefault}{\mddefault}{\updefault}{\color[rgb]{0,0,0}$O_{\alpha}$}%
}}}}
\end{picture}%

%% file: parametrix2.pdf_t
\begin{picture}(0,0)%
\includegraphics{parametrix2.pdf}%
\end{picture}%
\setlength{\unitlength}{3947sp}%
\begingroup\makeatletter\ifx\SetFigFont\undefined%
\gdef\SetFigFont#1#2#3#4#5{%
  \reset@font\fontsize{#1}{#2pt}%
  \fontfamily{#3}\fontseries{#4}\fontshape{#5}%
  \selectfont}%
\fi\endgroup%
\begin{picture}(11040,4781)(826,-7149)
\put(5251,-5986){\makebox(0,0)[lb]{\smash{{\SetFigFont{12}{14.4}{\rmdefault}{\mddefault}{\updefault}{\color[rgb]{0,0,0}$\gamma_6$}%
}}}}
\put(2011,-2551){\makebox(0,0)[lb]{\smash{{\SetFigFont{12}{14.4}{\familydefault}{\mddefault}{\updefault}{\color[rgb]{0,0,0}$O_{\alpha}$}%
}}}}
\put(3766,-7071){\makebox(0,0)[lb]{\smash{{\SetFigFont{12}{14.4}{\rmdefault}{\mddefault}{\updefault}{\color[rgb]{0,0,0}$\gamma_1$}%
}}}}
\put(1296,-6591){\makebox(0,0)[lb]{\smash{{\SetFigFont{12}{14.4}{\rmdefault}{\mddefault}{\updefault}{\color[rgb]{0,0,0}$\gamma_2$}%
}}}}
\put(841,-2731){\makebox(0,0)[lb]{\smash{{\SetFigFont{12}{14.4}{\rmdefault}{\mddefault}{\updefault}{\color[rgb]{0,0,0}$\gamma_3$}%
}}}}
\put(3636,-2566){\makebox(0,0)[lb]{\smash{{\SetFigFont{12}{14.4}{\rmdefault}{\mddefault}{\updefault}{\color[rgb]{0,0,0}$\gamma_4$}%
}}}}
\put(5826,-3276){\makebox(0,0)[lb]{\smash{{\SetFigFont{12}{14.4}{\rmdefault}{\mddefault}{\updefault}{\color[rgb]{0,0,0}$\gamma_5$}%
}}}}
\put(7426,-4711){\makebox(0,0)[lb]{\smash{{\SetFigFont{12}{14.4}{\rmdefault}{\mddefault}{\updefault}{\color[rgb]{0,0,0}$\gamma_6$}%
}}}}
\put(8701,-6511){\makebox(0,0)[lb]{\smash{{\SetFigFont{12}{14.4}{\rmdefault}{\mddefault}{\updefault}{\color[rgb]{0,0,0}$\gamma_5$}%
}}}}
\put(10801,-6511){\makebox(0,0)[lb]{\smash{{\SetFigFont{12}{14.4}{\rmdefault}{\mddefault}{\updefault}{\color[rgb]{0,0,0}$\gamma_4$}%
}}}}
\put(11851,-4636){\makebox(0,0)[lb]{\smash{{\SetFigFont{12}{14.4}{\rmdefault}{\mddefault}{\updefault}{\color[rgb]{0,0,0}$\gamma_3$}%
}}}}
\put(8701,-2761){\makebox(0,0)[lb]{\smash{{\SetFigFont{12}{14.4}{\rmdefault}{\mddefault}{\updefault}{\color[rgb]{0,0,0}$\gamma_1$}%
}}}}
\put(10801,-2836){\makebox(0,0)[lb]{\smash{{\SetFigFont{12}{14.4}{\rmdefault}{\mddefault}{\updefault}{\color[rgb]{0,0,0}$\gamma_2$}%
}}}}
\end{picture}%

%% file: Infinity.pdf_t
\begin{picture}(0,0)%
\includegraphics{Infinity.pdf}%
\end{picture}%
\setlength{\unitlength}{3947sp}%
\begingroup\makeatletter\ifx\SetFigFont\undefined%
\gdef\SetFigFont#1#2#3#4#5{%
  \reset@font\fontsize{#1}{#2pt}%
  \fontfamily{#3}\fontseries{#4}\fontshape{#5}%
  \selectfont}%
\fi\endgroup%
\begin{picture}(3024,3024)(4489,-6673)
\end{picture}%

%% file: Clock6.pdf_t
\begin{picture}(0,0)%
\includegraphics{Clock6.pdf}%
\end{picture}%
\setlength{\unitlength}{3947sp}%
\begingroup\makeatletter\ifx\SetFigFont\undefined%
\gdef\SetFigFont#1#2#3#4#5{%
  \reset@font\fontsize{#1}{#2pt}%
  \fontfamily{#3}\fontseries{#4}\fontshape{#5}%
  \selectfont}%
\fi\endgroup%
\begin{picture}(16302,7674)(514,-7273)
\end{picture}%

%% file: Examp3.pdf_t
\begin{picture}(0,0)%
\includegraphics{Examp3.pdf}%
\end{picture}%
\setlength{\unitlength}{3947sp}%
\begingroup\makeatletter\ifx\SetFigFont\undefined%
\gdef\SetFigFont#1#2#3#4#5{%
  \reset@font\fontsize{#1}{#2pt}%
  \fontfamily{#3}\fontseries{#4}\fontshape{#5}%
  \selectfont}%
\fi\endgroup%
\begin{picture}(8166,8052)(1468,-8107)
\put(5551,-3211){\makebox(0,0)[lb]{\smash{{\SetFigFont{14}{16.8}{\familydefault}{\mddefault}{\updefault}{\color[rgb]{0,0,0}$\Sigma_2$}%
}}}}
\put(4126,-3286){\makebox(0,0)[lb]{\smash{{\SetFigFont{14}{16.8}{\familydefault}{\mddefault}{\updefault}{\color[rgb]{0,0,0}A}%
}}}}
\put(5701,-4936){\makebox(0,0)[lb]{\smash{{\SetFigFont{14}{16.8}{\familydefault}{\mddefault}{\updefault}{\color[rgb]{0,0,0}B}%
}}}}
\put(6301,-2236){\makebox(0,0)[lb]{\smash{{\SetFigFont{14}{16.8}{\familydefault}{\mddefault}{\updefault}{\color[rgb]{0,0,0}C}%
}}}}
\put(7501,-3661){\makebox(0,0)[lb]{\smash{{\SetFigFont{14}{16.8}{\familydefault}{\mddefault}{\updefault}{\color[rgb]{0,0,0}D}%
}}}}
\put(4351,-4486){\makebox(0,0)[lb]{\smash{{\SetFigFont{14}{16.8}{\familydefault}{\mddefault}{\updefault}{\color[rgb]{0,0,0}$\Sigma_1$}%
}}}}
\put(6601,-4036){\makebox(0,0)[lb]{\smash{{\SetFigFont{14}{16.8}{\familydefault}{\mddefault}{\updefault}{\color[rgb]{0,0,0}$\Sigma_3$}%
}}}}
\put(1801,-2086){\makebox(0,0)[lb]{\smash{{\SetFigFont{14}{16.8}{\familydefault}{\mddefault}{\updefault}{\color[rgb]{0,0,0}$\Pi_9$}%
}}}}
\put(5176,-286){\makebox(0,0)[lb]{\smash{{\SetFigFont{14}{16.8}{\familydefault}{\mddefault}{\updefault}{\color[rgb]{0,0,0}$\Pi_4$}%
}}}}
\put(8851,-2086){\makebox(0,0)[lb]{\smash{{\SetFigFont{14}{16.8}{\familydefault}{\mddefault}{\updefault}{\color[rgb]{0,0,0}$\Pi_5$}%
}}}}
\put(9001,-5836){\makebox(0,0)[lb]{\smash{{\SetFigFont{14}{16.8}{\familydefault}{\mddefault}{\updefault}{\color[rgb]{0,0,0}$\Pi_6$}%
}}}}
\put(5101,-8011){\makebox(0,0)[lb]{\smash{{\SetFigFont{14}{16.8}{\familydefault}{\mddefault}{\updefault}{\color[rgb]{0,0,0}$\Pi_7$}%
}}}}
\put(1576,-6136){\makebox(0,0)[lb]{\smash{{\SetFigFont{14}{16.8}{\familydefault}{\mddefault}{\updefault}{\color[rgb]{0,0,0}$\Pi_8$}%
}}}}
\put(3901,-1336){\makebox(0,0)[lb]{\smash{{\SetFigFont{14}{16.8}{\familydefault}{\mddefault}{\updefault}{\color[rgb]{0,0,0}$V_1$}%
}}}}
\put(4576,-1486){\makebox(0,0)[lb]{\smash{{\SetFigFont{14}{16.8}{\familydefault}{\mddefault}{\updefault}{\color[rgb]{0,0,0}$V_2$}%
}}}}
\put(5551,-961){\makebox(0,0)[lb]{\smash{{\SetFigFont{14}{16.8}{\familydefault}{\mddefault}{\updefault}{\color[rgb]{0,0,0}$V_3$}%
}}}}
\put(7276,-1786){\makebox(0,0)[lb]{\smash{{\SetFigFont{14}{16.8}{\familydefault}{\mddefault}{\updefault}{\color[rgb]{0,0,0}$V_4$}%
}}}}
\put(8101,-2836){\makebox(0,0)[lb]{\smash{{\SetFigFont{14}{16.8}{\familydefault}{\mddefault}{\updefault}{\color[rgb]{0,0,0}$V_5$}%
}}}}
\put(8101,-5461){\makebox(0,0)[lb]{\smash{{\SetFigFont{14}{16.8}{\familydefault}{\mddefault}{\updefault}{\color[rgb]{0,0,0}$V_6$}%
}}}}
\put(7201,-6211){\makebox(0,0)[lb]{\smash{{\SetFigFont{14}{16.8}{\familydefault}{\mddefault}{\updefault}{\color[rgb]{0,0,0}$V_7$}%
}}}}
\put(6826,-6436){\makebox(0,0)[lb]{\smash{{\SetFigFont{14}{16.8}{\familydefault}{\mddefault}{\updefault}{\color[rgb]{0,0,0}$V_8$}%
}}}}
\put(4726,-7036){\makebox(0,0)[lb]{\smash{{\SetFigFont{14}{16.8}{\familydefault}{\mddefault}{\updefault}{\color[rgb]{0,0,0}$V_9$}%
}}}}
\put(3826,-6361){\makebox(0,0)[lb]{\smash{{\SetFigFont{14}{16.8}{\familydefault}{\mddefault}{\updefault}{\color[rgb]{0,0,0}$V_{10}$}%
}}}}
\put(2326,-4786){\makebox(0,0)[lb]{\smash{{\SetFigFont{14}{16.8}{\familydefault}{\mddefault}{\updefault}{\color[rgb]{0,0,0}$V_{11}$}%
}}}}
\put(2476,-2536){\makebox(0,0)[lb]{\smash{{\SetFigFont{14}{16.8}{\familydefault}{\mddefault}{\updefault}{\color[rgb]{0,0,0}$V_{12}$}%
}}}}
\end{picture}%

%% file: compact.pdf_t
\begin{picture}(0,0)%
\includegraphics{compact.pdf}%
\end{picture}%
\setlength{\unitlength}{3947sp}%
\begingroup\makeatletter\ifx\SetFigFont\undefined%
\gdef\SetFigFont#1#2#3#4#5{%
  \reset@font\fontsize{#1}{#2pt}%
  \fontfamily{#3}\fontseries{#4}\fontshape{#5}%
  \selectfont}%
\fi\endgroup%
\begin{picture}(3784,4404)(2959,-6233)
\put(4471,-3971){\makebox(0,0)[lb]{\smash{{\SetFigFont{17}{20.4}{\familydefault}{\mddefault}{\updefault}{\color[rgb]{0,0,0}$\infty$}%
}}}}
\end{picture}%

%% file: compact2.pdf_t
\begin{picture}(0,0)%
\includegraphics{compact2.pdf}%
\end{picture}%
\setlength{\unitlength}{3947sp}%
\begingroup\makeatletter\ifx\SetFigFont\undefined%
\gdef\SetFigFont#1#2#3#4#5{%
  \reset@font\fontsize{#1}{#2pt}%
  \fontfamily{#3}\fontseries{#4}\fontshape{#5}%
  \selectfont}%
\fi\endgroup%
\begin{picture}(11022,5122)(2086,-5182)
\put(2101,-5086){\makebox(0,0)[lb]{\smash{{\SetFigFont{17}{20.4}{\rmdefault}{\mddefault}{\updefault}{\color[rgb]{0,0,0}$w$--plane}%
}}}}
\put(12841,-1482){\makebox(0,0)[lb]{\smash{{\SetFigFont{17}{20.4}{\rmdefault}{\mddefault}{\updefault}{\color[rgb]{0,0,0}$\frac{2\pi}{n+2}$}%
}}}}
\put(10201,-3717){\makebox(0,0)[lb]{\smash{{\SetFigFont{17}{20.4}{\rmdefault}{\mddefault}{\updefault}{\color[rgb]{0,0,0}$\zeta$--plane}%
}}}}
\end{picture}%

%% file: Examp1_enhanced.pdf_t
\begin{picture}(0,0)%
\includegraphics{Examp1_enhanced.pdf}%
\end{picture}%
\setlength{\unitlength}{3947sp}%
\begingroup\makeatletter\ifx\SetFigFont\undefined%
\gdef\SetFigFont#1#2#3#4#5{%
  \reset@font\fontsize{#1}{#2pt}%
  \fontfamily{#3}\fontseries{#4}\fontshape{#5}%
  \selectfont}%
\fi\endgroup%
\begin{picture}(6086,6095)(110,-5212)
\put(5914,-2165){\makebox(0,0)[lb]{\smash{{\SetFigFont{17}{20.4}{\familydefault}{\mddefault}{\updefault}{\color[rgb]{0,0,0}+}%
}}}}
\put(5102,-293){\makebox(0,0)[lb]{\smash{{\SetFigFont{17}{20.4}{\familydefault}{\mddefault}{\updefault}{\color[rgb]{0,0,0}-}%
}}}}
\put(2980,546){\makebox(0,0)[lb]{\smash{{\SetFigFont{17}{20.4}{\familydefault}{\mddefault}{\updefault}{\color[rgb]{0,0,0}+}%
}}}}
\put(270,-2204){\makebox(0,0)[lb]{\smash{{\SetFigFont{17}{20.4}{\familydefault}{\mddefault}{\updefault}{\color[rgb]{0,0,0}+}%
}}}}
\put(3010,-4984){\makebox(0,0)[lb]{\smash{{\SetFigFont{17}{20.4}{\familydefault}{\mddefault}{\updefault}{\color[rgb]{0,0,0}+}%
}}}}
\put(1020,-4154){\makebox(0,0)[lb]{\smash{{\SetFigFont{17}{20.4}{\familydefault}{\mddefault}{\updefault}{\color[rgb]{0,0,0}-}%
}}}}
\put(1040,-314){\makebox(0,0)[lb]{\smash{{\SetFigFont{17}{20.4}{\familydefault}{\mddefault}{\updefault}{\color[rgb]{0,0,0}-}%
}}}}
\put(5084,-4140){\makebox(0,0)[lb]{\smash{{\SetFigFont{17}{20.4}{\familydefault}{\mddefault}{\updefault}{\color[rgb]{0,0,0}-}%
}}}}
\put(3238,-2016){\makebox(0,0)[lb]{\smash{{\SetFigFont{14}{16.8}{\familydefault}{\mddefault}{\updefault}{\color[rgb]{0,0,0}$\rho_2$}%
}}}}
\put(2939,-2010){\makebox(0,0)[lb]{\smash{{\SetFigFont{14}{16.8}{\familydefault}{\mddefault}{\updefault}{\color[rgb]{0,0,0}$\rho_1$}%
}}}}
\put(2345,-2090){\makebox(0,0)[lb]{\smash{{\SetFigFont{14}{16.8}{\familydefault}{\mddefault}{\updefault}{\color[rgb]{0,0,0}$\ell_1$}%
}}}}
\put(3805,-2121){\makebox(0,0)[lb]{\smash{{\SetFigFont{14}{16.8}{\familydefault}{\mddefault}{\updefault}{\color[rgb]{0,0,0}$\ell_2$}%
}}}}
\end{picture}%

%% file: Examp2_enhanced.pdf_t
\begin{picture}(0,0)%
\includegraphics{Examp2_enhanced.pdf}%
\end{picture}%
\setlength{\unitlength}{3947sp}%
\begingroup\makeatletter\ifx\SetFigFont\undefined%
\gdef\SetFigFont#1#2#3#4#5{%
  \reset@font\fontsize{#1}{#2pt}%
  \fontfamily{#3}\fontseries{#4}\fontshape{#5}%
  \selectfont}%
\fi\endgroup%
\begin{picture}(3224,3224)(1732,-3816)
\put(2682,-2132){\makebox(0,0)[lb]{\smash{{\SetFigFont{10}{12.0}{\familydefault}{\mddefault}{\updefault}{\color[rgb]{0,0,0}$\ell_1$}%
}}}}
\put(3223,-2296){\makebox(0,0)[lb]{\smash{{\SetFigFont{10}{12.0}{\familydefault}{\mddefault}{\updefault}{\color[rgb]{0,0,0}$\ell_2$}%
}}}}
\put(3814,-2109){\makebox(0,0)[lb]{\smash{{\SetFigFont{10}{12.0}{\familydefault}{\mddefault}{\updefault}{\color[rgb]{0,0,0}$\ell_3$}%
}}}}
\end{picture}%

%% file: VeryStrange_enhanced.pdf_t
\begin{picture}(0,0)%
\includegraphics{VeryStrange_enhanced.pdf}%
\end{picture}%
\setlength{\unitlength}{3947sp}%
\begingroup\makeatletter\ifx\SetFigFont\undefined%
\gdef\SetFigFont#1#2#3#4#5{%
  \reset@font\fontsize{#1}{#2pt}%
  \fontfamily{#3}\fontseries{#4}\fontshape{#5}%
  \selectfont}%
\fi\endgroup%
\begin{picture}(5443,5295)(449,-4726)
\put(3094,-4711){\makebox(0,0)[lb]{\smash{{\SetFigFont{12}{14.4}{\familydefault}{\mddefault}{\updefault}{\color[rgb]{0,0,0}+}%
}}}}
\put(5054,-4111){\makebox(0,0)[lb]{\smash{{\SetFigFont{12}{14.4}{\familydefault}{\mddefault}{\updefault}{\color[rgb]{0,0,0}-}%
}}}}
\put(5754,-2081){\makebox(0,0)[lb]{\smash{{\SetFigFont{12}{14.4}{\familydefault}{\mddefault}{\updefault}{\color[rgb]{0,0,0}+}%
}}}}
\put(4894,-331){\makebox(0,0)[lb]{\smash{{\SetFigFont{12}{14.4}{\familydefault}{\mddefault}{\updefault}{\color[rgb]{0,0,0}-}%
}}}}
\put(3074,449){\makebox(0,0)[lb]{\smash{{\SetFigFont{12}{14.4}{\familydefault}{\mddefault}{\updefault}{\color[rgb]{0,0,0}+}%
}}}}
\put(1074,-371){\makebox(0,0)[lb]{\smash{{\SetFigFont{12}{14.4}{\familydefault}{\mddefault}{\updefault}{\color[rgb]{0,0,0}-}%
}}}}
\put(464,-2111){\makebox(0,0)[lb]{\smash{{\SetFigFont{12}{14.4}{\familydefault}{\mddefault}{\updefault}{\color[rgb]{0,0,0}+}%
}}}}
\put(1224,-3971){\makebox(0,0)[lb]{\smash{{\SetFigFont{12}{14.4}{\familydefault}{\mddefault}{\updefault}{\color[rgb]{0,0,0}-}%
}}}}
\put(2448,-1967){\makebox(0,0)[lb]{\smash{{\SetFigFont{10}{12.0}{\familydefault}{\mddefault}{\updefault}{\color[rgb]{0,0,0}$\ell_1$}%
}}}}
\put(2896,-2069){\makebox(0,0)[lb]{\smash{{\SetFigFont{10}{12.0}{\familydefault}{\mddefault}{\updefault}{\color[rgb]{0,0,0}$\rho_1$}%
}}}}
\put(3570,-2046){\makebox(0,0)[lb]{\smash{{\SetFigFont{10}{12.0}{\familydefault}{\mddefault}{\updefault}{\color[rgb]{0,0,0}$\rho_2$}%
}}}}
\end{picture}%

%% file: Strange.pdf_t
\begin{picture}(0,0)%
\includegraphics{Strange.pdf}%
\end{picture}%
\setlength{\unitlength}{3947sp}%
\begingroup\makeatletter\ifx\SetFigFont\undefined%
\gdef\SetFigFont#1#2#3#4#5{%
  \reset@font\fontsize{#1}{#2pt}%
  \fontfamily{#3}\fontseries{#4}\fontshape{#5}%
  \selectfont}%
\fi\endgroup%
\begin{picture}(8305,7882)(3729,-9319)
\put(7687,-1788){\makebox(0,0)[lb]{\smash{{\SetFigFont{14}{16.8}{\familydefault}{\mddefault}{\updefault}{\color[rgb]{0,0,0}+}%
}}}}
\put(5537,-2512){\makebox(0,0)[lb]{\smash{{\SetFigFont{14}{16.8}{\familydefault}{\mddefault}{\updefault}{\color[rgb]{0,0,0}-}%
}}}}
\put(4247,-4342){\makebox(0,0)[lb]{\smash{{\SetFigFont{14}{16.8}{\familydefault}{\mddefault}{\updefault}{\color[rgb]{0,0,0}+}%
}}}}
\put(4360,-6597){\makebox(0,0)[lb]{\smash{{\SetFigFont{14}{16.8}{\familydefault}{\mddefault}{\updefault}{\color[rgb]{0,0,0}-}%
}}}}
\put(5593,-8444){\makebox(0,0)[lb]{\smash{{\SetFigFont{14}{16.8}{\familydefault}{\mddefault}{\updefault}{\color[rgb]{0,0,0}+}%
}}}}
\put(7857,-9164){\makebox(0,0)[lb]{\smash{{\SetFigFont{14}{16.8}{\familydefault}{\mddefault}{\updefault}{\color[rgb]{0,0,0}-}%
}}}}
\put(9927,-8484){\makebox(0,0)[lb]{\smash{{\SetFigFont{14}{16.8}{\familydefault}{\mddefault}{\updefault}{\color[rgb]{0,0,0}+}%
}}}}
\put(11233,-6701){\makebox(0,0)[lb]{\smash{{\SetFigFont{14}{16.8}{\familydefault}{\mddefault}{\updefault}{\color[rgb]{0,0,0}-}%
}}}}
\put(11197,-4321){\makebox(0,0)[lb]{\smash{{\SetFigFont{14}{16.8}{\familydefault}{\mddefault}{\updefault}{\color[rgb]{0,0,0}+}%
}}}}
\put(10020,-2402){\makebox(0,0)[lb]{\smash{{\SetFigFont{14}{16.8}{\familydefault}{\mddefault}{\updefault}{\color[rgb]{0,0,0}-}%
}}}}
\put(6826,-4486){\makebox(0,0)[lb]{\smash{{\SetFigFont{17}{20.4}{\familydefault}{\mddefault}{\updefault}$\rho_1$}}}}
\put(8551,-6061){\makebox(0,0)[lb]{\smash{{\SetFigFont{17}{20.4}{\familydefault}{\mddefault}{\updefault}{\color[rgb]{0,0,0}$\rho_0$}%
}}}}
\put(7801,-4261){\makebox(0,0)[lb]{\smash{{\SetFigFont{17}{20.4}{\familydefault}{\mddefault}{\updefault}$\rho_2$}}}}
\put(6301,-4561){\makebox(0,0)[lb]{\smash{{\SetFigFont{17}{20.4}{\familydefault}{\mddefault}{\updefault}{\color[rgb]{0,0,0}$\ell_1$}%
}}}}
\put(8926,-6736){\makebox(0,0)[lb]{\smash{{\SetFigFont{17}{20.4}{\familydefault}{\mddefault}{\updefault}{\color[rgb]{0,0,0}$\ell_2$}%
}}}}
\end{picture}%